\documentclass[longauth]{aa}

\usepackage{graphicx}
\usepackage{natbib}
\usepackage{scalerel}

\usepackage[table]{xcolor}

\usepackage[varg]{txfonts}
\usepackage[T1]{fontenc}

\usepackage{amsmath,amssymb,amsthm,mathtools}

\newcommand{\w}{\mathfrak{w}}
\newcommand{\U}{\hat{u}}
\DeclarePairedDelimiter{\ev}{\langle}{\rangle}
\newcommand{\D}{\mathrm{d}}
\newcommand{\E}{\mathrm{e}}
\newcommand{\I}{\mathrm{i}}

\makeatletter
\def\@endtheorem{\endtrivlist}
\makeatother
\theoremstyle{definition}
\newtheorem{assumption}{Assumption}

\makeatletter
\def\@threej(#1,#2,#3,#4,#5,#6){%
\begin{pmatrix}#1&#3&#5\\#2&#4&#6\end{pmatrix}}
\newcommand{\threej}[3]{\@threej(#1,#2,#3)}
\makeatother

\usepackage[pdfencoding=auto,psdextra]{hyperref}
\hypersetup{
    colorlinks=true,
    linkcolor=blue,
    filecolor=magenta,
    urlcolor=blue,
    citecolor=blue
}
\makeatletter
\renewcommand*\aa@pageof{, page \thepage{} of \pageref*{LastPage}}
\makeatother

\usepackage[utf8]{inputenc}

\usepackage[switch, modulo]{lineno}

\usepackage{euclid}

\usepackage{orcidlink}

\begin{document}

\title{\Euclid preparation}
\subtitle{LIX. Angular power spectra from discrete observations}

\newcommand{\orcid}[1]{\orcidlink{#1}} 
\author{Euclid Collaboration: N.~Tessore\orcid{0000-0002-9696-7931}\thanks{\email{n.tessore@ucl.ac.uk}}\inst{\ref{aff1}}
\and B.~Joachimi\orcid{0000-0001-7494-1303}\inst{\ref{aff1}}
\and A.~Loureiro\orcid{0000-0002-4371-0876}\inst{\ref{aff2},\ref{aff3}}
\and A.~Hall\orcid{0000-0002-3139-8651}\inst{\ref{aff4}}
\and G.~Ca\~nas-Herrera\orcid{0000-0003-2796-2149}\inst{\ref{aff5},\ref{aff6}}
\and I.~Tutusaus\orcid{0000-0002-3199-0399}\inst{\ref{aff7}}
\and N.~Jeffrey\orcid{0000-0003-2927-1800}\inst{\ref{aff1}}
\and K.~Naidoo\orcid{0000-0002-9182-1802}\inst{\ref{aff1}}
\and J.~D.~McEwen\inst{\ref{aff8}}
\and A.~Amara\inst{\ref{aff9}}
\and S.~Andreon\orcid{0000-0002-2041-8784}\inst{\ref{aff10}}
\and N.~Auricchio\orcid{0000-0003-4444-8651}\inst{\ref{aff11}}
\and C.~Baccigalupi\orcid{0000-0002-8211-1630}\inst{\ref{aff12},\ref{aff13},\ref{aff14},\ref{aff15}}
\and M.~Baldi\orcid{0000-0003-4145-1943}\inst{\ref{aff16},\ref{aff11},\ref{aff17}}
\and S.~Bardelli\orcid{0000-0002-8900-0298}\inst{\ref{aff11}}
\and F.~Bernardeau\inst{\ref{aff18},\ref{aff19}}
\and D.~Bonino\orcid{0000-0002-3336-9977}\inst{\ref{aff20}}
\and E.~Branchini\orcid{0000-0002-0808-6908}\inst{\ref{aff21},\ref{aff22},\ref{aff10}}
\and M.~Brescia\orcid{0000-0001-9506-5680}\inst{\ref{aff23},\ref{aff24},\ref{aff25}}
\and J.~Brinchmann\orcid{0000-0003-4359-8797}\inst{\ref{aff26},\ref{aff27}}
\and A.~Caillat\inst{\ref{aff28}}
\and S.~Camera\orcid{0000-0003-3399-3574}\inst{\ref{aff29},\ref{aff30},\ref{aff20}}
\and V.~Capobianco\orcid{0000-0002-3309-7692}\inst{\ref{aff20}}
\and C.~Carbone\orcid{0000-0003-0125-3563}\inst{\ref{aff31}}
\and V.~F.~Cardone\inst{\ref{aff32},\ref{aff33}}
\and J.~Carretero\orcid{0000-0002-3130-0204}\inst{\ref{aff34},\ref{aff35}}
\and S.~Casas\orcid{0000-0002-4751-5138}\inst{\ref{aff36},\ref{aff37}}
\and M.~Castellano\orcid{0000-0001-9875-8263}\inst{\ref{aff32}}
\and G.~Castignani\orcid{0000-0001-6831-0687}\inst{\ref{aff11}}
\and S.~Cavuoti\orcid{0000-0002-3787-4196}\inst{\ref{aff24},\ref{aff25}}
\and A.~Cimatti\inst{\ref{aff38}}
\and C.~Colodro-Conde\inst{\ref{aff39}}
\and G.~Congedo\orcid{0000-0003-2508-0046}\inst{\ref{aff4}}
\and C.~J.~Conselice\orcid{0000-0003-1949-7638}\inst{\ref{aff40}}
\and L.~Conversi\orcid{0000-0002-6710-8476}\inst{\ref{aff41},\ref{aff42}}
\and Y.~Copin\orcid{0000-0002-5317-7518}\inst{\ref{aff43}}
\and F.~Courbin\orcid{0000-0003-0758-6510}\inst{\ref{aff44},\ref{aff45},\ref{aff46}}
\and H.~M.~Courtois\orcid{0000-0003-0509-1776}\inst{\ref{aff47}}
\and M.~Cropper\orcid{0000-0003-4571-9468}\inst{\ref{aff8}}
\and A.~Da~Silva\orcid{0000-0002-6385-1609}\inst{\ref{aff48},\ref{aff49}}
\and H.~Degaudenzi\orcid{0000-0002-5887-6799}\inst{\ref{aff50}}
\and G.~De~Lucia\orcid{0000-0002-6220-9104}\inst{\ref{aff13}}
\and J.~Dinis\orcid{0000-0001-5075-1601}\inst{\ref{aff48},\ref{aff49}}
\and F.~Dubath\orcid{0000-0002-6533-2810}\inst{\ref{aff50}}
\and C.~A.~J.~Duncan\inst{\ref{aff40}}
\and X.~Dupac\inst{\ref{aff42}}
\and S.~Dusini\orcid{0000-0002-1128-0664}\inst{\ref{aff51}}
\and M.~Farina\orcid{0000-0002-3089-7846}\inst{\ref{aff52}}
\and S.~Farrens\orcid{0000-0002-9594-9387}\inst{\ref{aff53}}
\and F.~Faustini\orcid{0000-0001-6274-5145}\inst{\ref{aff54},\ref{aff32}}
\and S.~Ferriol\inst{\ref{aff43}}
\and M.~Frailis\orcid{0000-0002-7400-2135}\inst{\ref{aff13}}
\and E.~Franceschi\orcid{0000-0002-0585-6591}\inst{\ref{aff11}}
\and M.~Fumana\orcid{0000-0001-6787-5950}\inst{\ref{aff31}}
\and S.~Galeotta\orcid{0000-0002-3748-5115}\inst{\ref{aff13}}
\and W.~Gillard\orcid{0000-0003-4744-9748}\inst{\ref{aff55}}
\and B.~Gillis\orcid{0000-0002-4478-1270}\inst{\ref{aff4}}
\and C.~Giocoli\orcid{0000-0002-9590-7961}\inst{\ref{aff11},\ref{aff56}}
\and P.~G\'omez-Alvarez\orcid{0000-0002-8594-5358}\inst{\ref{aff57},\ref{aff42}}
\and A.~Grazian\orcid{0000-0002-5688-0663}\inst{\ref{aff58}}
\and F.~Grupp\inst{\ref{aff59},\ref{aff60}}
\and L.~Guzzo\orcid{0000-0001-8264-5192}\inst{\ref{aff61},\ref{aff10}}
\and S.~V.~H.~Haugan\orcid{0000-0001-9648-7260}\inst{\ref{aff62}}
\and H.~Hoekstra\orcid{0000-0002-0641-3231}\inst{\ref{aff63}}
\and W.~Holmes\inst{\ref{aff64}}
\and F.~Hormuth\inst{\ref{aff65}}
\and A.~Hornstrup\orcid{0000-0002-3363-0936}\inst{\ref{aff66},\ref{aff67}}
\and P.~Hudelot\inst{\ref{aff19}}
\and K.~Jahnke\orcid{0000-0003-3804-2137}\inst{\ref{aff68}}
\and M.~Jhabvala\inst{\ref{aff69}}
\and E.~Keih\"anen\orcid{0000-0003-1804-7715}\inst{\ref{aff70}}
\and S.~Kermiche\orcid{0000-0002-0302-5735}\inst{\ref{aff55}}
\and A.~Kiessling\orcid{0000-0002-2590-1273}\inst{\ref{aff64}}
\and B.~Kubik\orcid{0009-0006-5823-4880}\inst{\ref{aff43}}
\and M.~K\"ummel\orcid{0000-0003-2791-2117}\inst{\ref{aff60}}
\and M.~Kunz\orcid{0000-0002-3052-7394}\inst{\ref{aff71}}
\and H.~Kurki-Suonio\orcid{0000-0002-4618-3063}\inst{\ref{aff72},\ref{aff73}}
\and S.~Ligori\orcid{0000-0003-4172-4606}\inst{\ref{aff20}}
\and P.~B.~Lilje\orcid{0000-0003-4324-7794}\inst{\ref{aff62}}
\and V.~Lindholm\orcid{0000-0003-2317-5471}\inst{\ref{aff72},\ref{aff73}}
\and I.~Lloro\inst{\ref{aff74}}
\and G.~Mainetti\orcid{0000-0003-2384-2377}\inst{\ref{aff75}}
\and E.~Maiorano\orcid{0000-0003-2593-4355}\inst{\ref{aff11}}
\and O.~Mansutti\orcid{0000-0001-5758-4658}\inst{\ref{aff13}}
\and O.~Marggraf\orcid{0000-0001-7242-3852}\inst{\ref{aff76}}
\and M.~Martinelli\orcid{0000-0002-6943-7732}\inst{\ref{aff32},\ref{aff33}}
\and N.~Martinet\orcid{0000-0003-2786-7790}\inst{\ref{aff28}}
\and F.~Marulli\orcid{0000-0002-8850-0303}\inst{\ref{aff77},\ref{aff11},\ref{aff17}}
\and R.~Massey\orcid{0000-0002-6085-3780}\inst{\ref{aff78}}
\and E.~Medinaceli\orcid{0000-0002-4040-7783}\inst{\ref{aff11}}
\and S.~Mei\orcid{0000-0002-2849-559X}\inst{\ref{aff79}}
\and M.~Melchior\inst{\ref{aff80}}
\and Y.~Mellier\inst{\ref{aff81},\ref{aff19}}
\and M.~Meneghetti\orcid{0000-0003-1225-7084}\inst{\ref{aff11},\ref{aff17}}
\and E.~Merlin\orcid{0000-0001-6870-8900}\inst{\ref{aff32}}
\and G.~Meylan\inst{\ref{aff44}}
\and J.~J.~Mohr\orcid{0000-0002-6875-2087}\inst{\ref{aff60},\ref{aff59}}
\and M.~Moresco\orcid{0000-0002-7616-7136}\inst{\ref{aff77},\ref{aff11}}
\and B.~Morin\inst{\ref{aff53}}
\and L.~Moscardini\orcid{0000-0002-3473-6716}\inst{\ref{aff77},\ref{aff11},\ref{aff17}}
\and E.~Munari\orcid{0000-0002-1751-5946}\inst{\ref{aff13},\ref{aff12}}
\and R.~Nakajima\inst{\ref{aff76}}
\and S.-M.~Niemi\inst{\ref{aff5}}
\and C.~Padilla\orcid{0000-0001-7951-0166}\inst{\ref{aff82}}
\and S.~Paltani\orcid{0000-0002-8108-9179}\inst{\ref{aff50}}
\and F.~Pasian\orcid{0000-0002-4869-3227}\inst{\ref{aff13}}
\and K.~Pedersen\inst{\ref{aff83}}
\and W.~J.~Percival\orcid{0000-0002-0644-5727}\inst{\ref{aff84},\ref{aff85},\ref{aff86}}
\and V.~Pettorino\inst{\ref{aff5}}
\and S.~Pires\orcid{0000-0002-0249-2104}\inst{\ref{aff53}}
\and G.~Polenta\orcid{0000-0003-4067-9196}\inst{\ref{aff54}}
\and M.~Poncet\inst{\ref{aff87}}
\and L.~A.~Popa\inst{\ref{aff88}}
\and F.~Raison\orcid{0000-0002-7819-6918}\inst{\ref{aff59}}
\and A.~Renzi\orcid{0000-0001-9856-1970}\inst{\ref{aff89},\ref{aff51}}
\and J.~Rhodes\orcid{0000-0002-4485-8549}\inst{\ref{aff64}}
\and G.~Riccio\inst{\ref{aff24}}
\and E.~Romelli\orcid{0000-0003-3069-9222}\inst{\ref{aff13}}
\and M.~Roncarelli\orcid{0000-0001-9587-7822}\inst{\ref{aff11}}
\and E.~Rossetti\orcid{0000-0003-0238-4047}\inst{\ref{aff16}}
\and R.~Saglia\orcid{0000-0003-0378-7032}\inst{\ref{aff60},\ref{aff59}}
\and Z.~Sakr\orcid{0000-0002-4823-3757}\inst{\ref{aff90},\ref{aff7},\ref{aff91}}
\and A.~G.~S\'anchez\orcid{0000-0003-1198-831X}\inst{\ref{aff59}}
\and D.~Sapone\orcid{0000-0001-7089-4503}\inst{\ref{aff92}}
\and B.~Sartoris\orcid{0000-0003-1337-5269}\inst{\ref{aff60},\ref{aff13}}
\and M.~Schirmer\orcid{0000-0003-2568-9994}\inst{\ref{aff68}}
\and P.~Schneider\orcid{0000-0001-8561-2679}\inst{\ref{aff76}}
\and T.~Schrabback\orcid{0000-0002-6987-7834}\inst{\ref{aff93}}
\and A.~Secroun\orcid{0000-0003-0505-3710}\inst{\ref{aff55}}
\and G.~Seidel\orcid{0000-0003-2907-353X}\inst{\ref{aff68}}
\and M.~Seiffert\orcid{0000-0002-7536-9393}\inst{\ref{aff64}}
\and S.~Serrano\orcid{0000-0002-0211-2861}\inst{\ref{aff94},\ref{aff95},\ref{aff96}}
\and C.~Sirignano\orcid{0000-0002-0995-7146}\inst{\ref{aff89},\ref{aff51}}
\and G.~Sirri\orcid{0000-0003-2626-2853}\inst{\ref{aff17}}
\and L.~Stanco\orcid{0000-0002-9706-5104}\inst{\ref{aff51}}
\and J.~Steinwagner\orcid{0000-0001-7443-1047}\inst{\ref{aff59}}
\and P.~Tallada-Cresp\'{i}\orcid{0000-0002-1336-8328}\inst{\ref{aff34},\ref{aff35}}
\and A.~N.~Taylor\inst{\ref{aff4}}
\and I.~Tereno\inst{\ref{aff48},\ref{aff97}}
\and R.~Toledo-Moreo\orcid{0000-0002-2997-4859}\inst{\ref{aff98}}
\and F.~Torradeflot\orcid{0000-0003-1160-1517}\inst{\ref{aff35},\ref{aff34}}
\and L.~Valenziano\orcid{0000-0002-1170-0104}\inst{\ref{aff11},\ref{aff99}}
\and T.~Vassallo\orcid{0000-0001-6512-6358}\inst{\ref{aff60},\ref{aff13}}
\and Y.~Wang\orcid{0000-0002-4749-2984}\inst{\ref{aff100}}
\and J.~Weller\orcid{0000-0002-8282-2010}\inst{\ref{aff60},\ref{aff59}}
\and G.~Zamorani\orcid{0000-0002-2318-301X}\inst{\ref{aff11}}
\and E.~Zucca\orcid{0000-0002-5845-8132}\inst{\ref{aff11}}
\and A.~Biviano\orcid{0000-0002-0857-0732}\inst{\ref{aff13},\ref{aff12}}
\and M.~Bolzonella\orcid{0000-0003-3278-4607}\inst{\ref{aff11}}
\and A.~Boucaud\orcid{0000-0001-7387-2633}\inst{\ref{aff79}}
\and E.~Bozzo\orcid{0000-0002-8201-1525}\inst{\ref{aff50}}
\and C.~Burigana\orcid{0000-0002-3005-5796}\inst{\ref{aff101},\ref{aff99}}
\and M.~Calabrese\orcid{0000-0002-2637-2422}\inst{\ref{aff102},\ref{aff31}}
\and D.~Di~Ferdinando\inst{\ref{aff17}}
\and J.~A.~Escartin~Vigo\inst{\ref{aff59}}
\and F.~Finelli\orcid{0000-0002-6694-3269}\inst{\ref{aff11},\ref{aff99}}
\and J.~Gracia-Carpio\inst{\ref{aff59}}
\and S.~Matthew\orcid{0000-0001-8448-1697}\inst{\ref{aff4}}
\and N.~Mauri\orcid{0000-0001-8196-1548}\inst{\ref{aff38},\ref{aff17}}
\and A.~Pezzotta\orcid{0000-0003-0726-2268}\inst{\ref{aff59}}
\and M.~P\"ontinen\orcid{0000-0001-5442-2530}\inst{\ref{aff72}}
\and V.~Scottez\inst{\ref{aff81},\ref{aff103}}
\and A.~Spurio~Mancini\orcid{0000-0001-5698-0990}\inst{\ref{aff104},\ref{aff8}}
\and M.~Tenti\orcid{0000-0002-4254-5901}\inst{\ref{aff17}}
\and M.~Viel\orcid{0000-0002-2642-5707}\inst{\ref{aff12},\ref{aff13},\ref{aff15},\ref{aff14},\ref{aff105}}
\and M.~Wiesmann\orcid{0009-0000-8199-5860}\inst{\ref{aff62}}
\and Y.~Akrami\orcid{0000-0002-2407-7956}\inst{\ref{aff106},\ref{aff107}}
\and S.~Anselmi\orcid{0000-0002-3579-9583}\inst{\ref{aff51},\ref{aff89},\ref{aff108}}
\and M.~Archidiacono\orcid{0000-0003-4952-9012}\inst{\ref{aff61},\ref{aff109}}
\and F.~Atrio-Barandela\orcid{0000-0002-2130-2513}\inst{\ref{aff110}}
\and A.~Balaguera-Antolinez\orcid{0000-0001-5028-3035}\inst{\ref{aff39},\ref{aff111}}
\and M.~Ballardini\orcid{0000-0003-4481-3559}\inst{\ref{aff112},\ref{aff11},\ref{aff113}}
\and D.~Benielli\inst{\ref{aff55}}
\and A.~Blanchard\orcid{0000-0001-8555-9003}\inst{\ref{aff7}}
\and L.~Blot\orcid{0000-0002-9622-7167}\inst{\ref{aff114},\ref{aff108}}
\and H.~B\"ohringer\orcid{0000-0001-8241-4204}\inst{\ref{aff59},\ref{aff115},\ref{aff116}}
\and S.~Borgani\orcid{0000-0001-6151-6439}\inst{\ref{aff117},\ref{aff12},\ref{aff13},\ref{aff14}}
\and S.~Bruton\orcid{0000-0002-6503-5218}\inst{\ref{aff118}}
\and R.~Cabanac\orcid{0000-0001-6679-2600}\inst{\ref{aff7}}
\and A.~Calabro\orcid{0000-0003-2536-1614}\inst{\ref{aff32}}
\and B.~Camacho~Quevedo\orcid{0000-0002-8789-4232}\inst{\ref{aff94},\ref{aff96}}
\and A.~Cappi\inst{\ref{aff11},\ref{aff119}}
\and F.~Caro\inst{\ref{aff32}}
\and C.~S.~Carvalho\inst{\ref{aff97}}
\and T.~Castro\orcid{0000-0002-6292-3228}\inst{\ref{aff13},\ref{aff14},\ref{aff12},\ref{aff105}}
\and K.~C.~Chambers\orcid{0000-0001-6965-7789}\inst{\ref{aff120}}
\and A.~R.~Cooray\orcid{0000-0002-3892-0190}\inst{\ref{aff121}}
\and S.~de~la~Torre\inst{\ref{aff28}}
\and G.~Desprez\orcid{0000-0001-8325-1742}\inst{\ref{aff122}}
\and A.~D\'iaz-S\'anchez\orcid{0000-0003-0748-4768}\inst{\ref{aff123}}
\and S.~Di~Domizio\orcid{0000-0003-2863-5895}\inst{\ref{aff21},\ref{aff22}}
\and H.~Dole\orcid{0000-0002-9767-3839}\inst{\ref{aff124}}
\and S.~Escoffier\orcid{0000-0002-2847-7498}\inst{\ref{aff55}}
\and A.~G.~Ferrari\orcid{0009-0005-5266-4110}\inst{\ref{aff17}}
\and P.~G.~Ferreira\orcid{0000-0002-3021-2851}\inst{\ref{aff125}}
\and I.~Ferrero\orcid{0000-0002-1295-1132}\inst{\ref{aff62}}
\and A.~Finoguenov\orcid{0000-0002-4606-5403}\inst{\ref{aff72}}
\and A.~Fontana\orcid{0000-0003-3820-2823}\inst{\ref{aff32}}
\and F.~Fornari\orcid{0000-0003-2979-6738}\inst{\ref{aff99}}
\and L.~Gabarra\orcid{0000-0002-8486-8856}\inst{\ref{aff125}}
\and K.~Ganga\orcid{0000-0001-8159-8208}\inst{\ref{aff79}}
\and J.~Garc\'ia-Bellido\orcid{0000-0002-9370-8360}\inst{\ref{aff106}}
\and T.~Gasparetto\orcid{0000-0002-7913-4866}\inst{\ref{aff13}}
\and E.~Gaztanaga\orcid{0000-0001-9632-0815}\inst{\ref{aff96},\ref{aff94},\ref{aff37}}
\and F.~Giacomini\orcid{0000-0002-3129-2814}\inst{\ref{aff17}}
\and F.~Gianotti\orcid{0000-0003-4666-119X}\inst{\ref{aff11}}
\and G.~Gozaliasl\orcid{0000-0002-0236-919X}\inst{\ref{aff126}}
\and C.~M.~Gutierrez\orcid{0000-0001-7854-783X}\inst{\ref{aff127}}
\and W.~G.~Hartley\inst{\ref{aff50}}
\and H.~Hildebrandt\orcid{0000-0002-9814-3338}\inst{\ref{aff128}}
\and J.~Hjorth\orcid{0000-0002-4571-2306}\inst{\ref{aff83}}
\and A.~Jimenez~Mu\~noz\orcid{0009-0004-5252-185X}\inst{\ref{aff129}}
\and S.~Joudaki\orcid{0000-0001-8820-673X}\inst{\ref{aff37}}
\and J.~J.~E.~Kajava\orcid{0000-0002-3010-8333}\inst{\ref{aff130},\ref{aff131}}
\and V.~Kansal\orcid{0000-0002-4008-6078}\inst{\ref{aff132},\ref{aff133}}
\and D.~Karagiannis\orcid{0000-0002-4927-0816}\inst{\ref{aff134},\ref{aff135}}
\and C.~C.~Kirkpatrick\inst{\ref{aff70}}
\and S.~Kruk\orcid{0000-0001-8010-8879}\inst{\ref{aff42}}
\and F.~Lacasa\orcid{0000-0002-7268-3440}\inst{\ref{aff136},\ref{aff124}}
\and M.~Lattanzi\inst{\ref{aff113}}
\and A.~M.~C.~Le~Brun\orcid{0000-0002-0936-4594}\inst{\ref{aff108}}
\and J.~Le~Graet\orcid{0000-0001-6523-7971}\inst{\ref{aff55}}
\and L.~Legrand\orcid{0000-0003-0610-5252}\inst{\ref{aff137}}
\and J.~Lesgourgues\orcid{0000-0001-7627-353X}\inst{\ref{aff36}}
\and T.~I.~Liaudat\orcid{0000-0002-9104-314X}\inst{\ref{aff138}}
\and J.~Macias-Perez\orcid{0000-0002-5385-2763}\inst{\ref{aff129}}
\and M.~Magliocchetti\orcid{0000-0001-9158-4838}\inst{\ref{aff52}}
\and F.~Mannucci\orcid{0000-0002-4803-2381}\inst{\ref{aff139}}
\and R.~Maoli\orcid{0000-0002-6065-3025}\inst{\ref{aff140},\ref{aff32}}
\and J.~Mart\'{i}n-Fleitas\orcid{0000-0002-8594-569X}\inst{\ref{aff141}}
\and C.~J.~A.~P.~Martins\orcid{0000-0002-4886-9261}\inst{\ref{aff142},\ref{aff26}}
\and L.~Maurin\orcid{0000-0002-8406-0857}\inst{\ref{aff124}}
\and R.~B.~Metcalf\orcid{0000-0003-3167-2574}\inst{\ref{aff77},\ref{aff11}}
\and M.~Miluzio\inst{\ref{aff42},\ref{aff143}}
\and P.~Monaco\orcid{0000-0003-2083-7564}\inst{\ref{aff117},\ref{aff13},\ref{aff14},\ref{aff12}}
\and A.~Montoro\orcid{0000-0003-4730-8590}\inst{\ref{aff96},\ref{aff94}}
\and C.~Moretti\orcid{0000-0003-3314-8936}\inst{\ref{aff15},\ref{aff105},\ref{aff13},\ref{aff12},\ref{aff14}}
\and G.~Morgante\inst{\ref{aff11}}
\and C.~Murray\inst{\ref{aff79}}
\and S.~Nadathur\orcid{0000-0001-9070-3102}\inst{\ref{aff37}}
\and N.~A.~Walton\orcid{0000-0003-3983-8778}\inst{\ref{aff144}}
\and L.~Patrizii\inst{\ref{aff17}}
\and V.~Popa\orcid{0000-0002-9118-8330}\inst{\ref{aff88}}
\and D.~Potter\orcid{0000-0002-0757-5195}\inst{\ref{aff145}}
\and P.~Reimberg\orcid{0000-0003-3410-0280}\inst{\ref{aff81}}
\and I.~Risso\orcid{0000-0003-2525-7761}\inst{\ref{aff146}}
\and P.-F.~Rocci\inst{\ref{aff124}}
\and R.~P.~Rollins\orcid{0000-0003-1291-1023}\inst{\ref{aff4}}
\and M.~Sahl\'en\orcid{0000-0003-0973-4804}\inst{\ref{aff147}}
\and E.~Sarpa\orcid{0000-0002-1256-655X}\inst{\ref{aff15},\ref{aff105},\ref{aff14}}
\and A.~Schneider\orcid{0000-0001-7055-8104}\inst{\ref{aff145}}
\and M.~Sereno\orcid{0000-0003-0302-0325}\inst{\ref{aff11},\ref{aff17}}
\and P.~Simon\inst{\ref{aff76}}
\and K.~Tanidis\inst{\ref{aff125}}
\and C.~Tao\orcid{0000-0001-7961-8177}\inst{\ref{aff55}}
\and G.~Testera\inst{\ref{aff22}}
\and R.~Teyssier\orcid{0000-0001-7689-0933}\inst{\ref{aff148}}
\and S.~Toft\orcid{0000-0003-3631-7176}\inst{\ref{aff149},\ref{aff150}}
\and S.~Tosi\orcid{0000-0002-7275-9193}\inst{\ref{aff21},\ref{aff22}}
\and A.~Troja\orcid{0000-0003-0239-4595}\inst{\ref{aff89},\ref{aff51}}
\and M.~Tucci\inst{\ref{aff50}}
\and C.~Valieri\inst{\ref{aff17}}
\and J.~Valiviita\orcid{0000-0001-6225-3693}\inst{\ref{aff72},\ref{aff73}}
\and D.~Vergani\orcid{0000-0003-0898-2216}\inst{\ref{aff11}}
\and G.~Verza\orcid{0000-0002-1886-8348}\inst{\ref{aff151},\ref{aff152}}
\and P.~Vielzeuf\orcid{0000-0003-2035-9339}\inst{\ref{aff55}}
\and M.~L.~Brown\orcid{0000-0002-0370-8077}\inst{\ref{aff40}}
\and E.~Sellentin\inst{\ref{aff153},\ref{aff63}}}

\institute{Department of Physics and Astronomy, University College London, Gower Street, London WC1E 6BT, UK\label{aff1}
\and
Oskar Klein Centre for Cosmoparticle Physics, Department of Physics, Stockholm University, Stockholm, SE-106 91, Sweden\label{aff2}
\and
Astrophysics Group, Blackett Laboratory, Imperial College London, London SW7 2AZ, UK\label{aff3}
\and
Institute for Astronomy, University of Edinburgh, Royal Observatory, Blackford Hill, Edinburgh EH9 3HJ, UK\label{aff4}
\and
European Space Agency/ESTEC, Keplerlaan 1, 2201 AZ Noordwijk, The Netherlands\label{aff5}
\and
Institute Lorentz, Leiden University, Niels Bohrweg 2, 2333 CA Leiden, The Netherlands\label{aff6}
\and
Institut de Recherche en Astrophysique et Plan\'etologie (IRAP), Universit\'e de Toulouse, CNRS, UPS, CNES, 14 Av. Edouard Belin, 31400 Toulouse, France\label{aff7}
\and
Mullard Space Science Laboratory, University College London, Holmbury St Mary, Dorking, Surrey RH5 6NT, UK\label{aff8}
\and
School of Mathematics and Physics, University of Surrey, Guildford, Surrey, GU2 7XH, UK\label{aff9}
\and
INAF-Osservatorio Astronomico di Brera, Via Brera 28, 20122 Milano, Italy\label{aff10}
\and
INAF-Osservatorio di Astrofisica e Scienza dello Spazio di Bologna, Via Piero Gobetti 93/3, 40129 Bologna, Italy\label{aff11}
\and
IFPU, Institute for Fundamental Physics of the Universe, via Beirut 2, 34151 Trieste, Italy\label{aff12}
\and
INAF-Osservatorio Astronomico di Trieste, Via G. B. Tiepolo 11, 34143 Trieste, Italy\label{aff13}
\and
INFN, Sezione di Trieste, Via Valerio 2, 34127 Trieste TS, Italy\label{aff14}
\and
SISSA, International School for Advanced Studies, Via Bonomea 265, 34136 Trieste TS, Italy\label{aff15}
\and
Dipartimento di Fisica e Astronomia, Universit\`a di Bologna, Via Gobetti 93/2, 40129 Bologna, Italy\label{aff16}
\and
INFN-Sezione di Bologna, Viale Berti Pichat 6/2, 40127 Bologna, Italy\label{aff17}
\and
Institut de Physique Th\'eorique, CEA, CNRS, Universit\'e Paris-Saclay 91191 Gif-sur-Yvette Cedex, France\label{aff18}
\and
Institut d'Astrophysique de Paris, UMR 7095, CNRS, and Sorbonne Universit\'e, 98 bis boulevard Arago, 75014 Paris, France\label{aff19}
\and
INAF-Osservatorio Astrofisico di Torino, Via Osservatorio 20, 10025 Pino Torinese (TO), Italy\label{aff20}
\and
Dipartimento di Fisica, Universit\`a di Genova, Via Dodecaneso 33, 16146, Genova, Italy\label{aff21}
\and
INFN-Sezione di Genova, Via Dodecaneso 33, 16146, Genova, Italy\label{aff22}
\and
Department of Physics "E. Pancini", University Federico II, Via Cinthia 6, 80126, Napoli, Italy\label{aff23}
\and
INAF-Osservatorio Astronomico di Capodimonte, Via Moiariello 16, 80131 Napoli, Italy\label{aff24}
\and
INFN section of Naples, Via Cinthia 6, 80126, Napoli, Italy\label{aff25}
\and
Instituto de Astrof\'isica e Ci\^encias do Espa\c{c}o, Universidade do Porto, CAUP, Rua das Estrelas, PT4150-762 Porto, Portugal\label{aff26}
\and
Faculdade de Ci\^encias da Universidade do Porto, Rua do Campo de Alegre, 4150-007 Porto, Portugal\label{aff27}
\and
Aix-Marseille Universit\'e, CNRS, CNES, LAM, Marseille, France\label{aff28}
\and
Dipartimento di Fisica, Universit\`a degli Studi di Torino, Via P. Giuria 1, 10125 Torino, Italy\label{aff29}
\and
INFN-Sezione di Torino, Via P. Giuria 1, 10125 Torino, Italy\label{aff30}
\and
INAF-IASF Milano, Via Alfonso Corti 12, 20133 Milano, Italy\label{aff31}
\and
INAF-Osservatorio Astronomico di Roma, Via Frascati 33, 00078 Monteporzio Catone, Italy\label{aff32}
\and
INFN-Sezione di Roma, Piazzale Aldo Moro, 2 - c/o Dipartimento di Fisica, Edificio G. Marconi, 00185 Roma, Italy\label{aff33}
\and
Centro de Investigaciones Energ\'eticas, Medioambientales y Tecnol\'ogicas (CIEMAT), Avenida Complutense 40, 28040 Madrid, Spain\label{aff34}
\and
Port d'Informaci\'{o} Cient\'{i}fica, Campus UAB, C. Albareda s/n, 08193 Bellaterra (Barcelona), Spain\label{aff35}
\and
Institute for Theoretical Particle Physics and Cosmology (TTK), RWTH Aachen University, 52056 Aachen, Germany\label{aff36}
\and
Institute of Cosmology and Gravitation, University of Portsmouth, Portsmouth PO1 3FX, UK\label{aff37}
\and
Dipartimento di Fisica e Astronomia "Augusto Righi" - Alma Mater Studiorum Universit\`a di Bologna, Viale Berti Pichat 6/2, 40127 Bologna, Italy\label{aff38}
\and
Instituto de Astrof\'isica de Canarias, Calle V\'ia L\'actea s/n, 38204, San Crist\'obal de La Laguna, Tenerife, Spain\label{aff39}
\and
Jodrell Bank Centre for Astrophysics, Department of Physics and Astronomy, University of Manchester, Oxford Road, Manchester M13 9PL, UK\label{aff40}
\and
European Space Agency/ESRIN, Largo Galileo Galilei 1, 00044 Frascati, Roma, Italy\label{aff41}
\and
ESAC/ESA, Camino Bajo del Castillo, s/n., Urb. Villafranca del Castillo, 28692 Villanueva de la Ca\~nada, Madrid, Spain\label{aff42}
\and
Universit\'e Claude Bernard Lyon 1, CNRS/IN2P3, IP2I Lyon, UMR 5822, Villeurbanne, F-69100, France\label{aff43}
\and
Institute of Physics, Laboratory of Astrophysics, Ecole Polytechnique F\'ed\'erale de Lausanne (EPFL), Observatoire de Sauverny, 1290 Versoix, Switzerland\label{aff44}
\and
Institut de Ci\`{e}ncies del Cosmos (ICCUB), Universitat de Barcelona (IEEC-UB), Mart\'{i} i Franqu\`{e}s 1, 08028 Barcelona, Spain\label{aff45}
\and
Instituci\'o Catalana de Recerca i Estudis Avan\c{c}ats (ICREA), Passeig de Llu\'{\i}s Companys 23, 08010 Barcelona, Spain\label{aff46}
\and
UCB Lyon 1, CNRS/IN2P3, IUF, IP2I Lyon, 4 rue Enrico Fermi, 69622 Villeurbanne, France\label{aff47}
\and
Departamento de F\'isica, Faculdade de Ci\^encias, Universidade de Lisboa, Edif\'icio C8, Campo Grande, PT1749-016 Lisboa, Portugal\label{aff48}
\and
Instituto de Astrof\'isica e Ci\^encias do Espa\c{c}o, Faculdade de Ci\^encias, Universidade de Lisboa, Campo Grande, 1749-016 Lisboa, Portugal\label{aff49}
\and
Department of Astronomy, University of Geneva, ch. d'Ecogia 16, 1290 Versoix, Switzerland\label{aff50}
\and
INFN-Padova, Via Marzolo 8, 35131 Padova, Italy\label{aff51}
\and
INAF-Istituto di Astrofisica e Planetologia Spaziali, via del Fosso del Cavaliere, 100, 00100 Roma, Italy\label{aff52}
\and
Universit\'e Paris-Saclay, Universit\'e Paris Cit\'e, CEA, CNRS, AIM, 91191, Gif-sur-Yvette, France\label{aff53}
\and
Space Science Data Center, Italian Space Agency, via del Politecnico snc, 00133 Roma, Italy\label{aff54}
\and
Aix-Marseille Universit\'e, CNRS/IN2P3, CPPM, Marseille, France\label{aff55}
\and
Istituto Nazionale di Fisica Nucleare, Sezione di Bologna, Via Irnerio 46, 40126 Bologna, Italy\label{aff56}
\and
FRACTAL S.L.N.E., calle Tulip\'an 2, Portal 13 1A, 28231, Las Rozas de Madrid, Spain\label{aff57}
\and
INAF-Osservatorio Astronomico di Padova, Via dell'Osservatorio 5, 35122 Padova, Italy\label{aff58}
\and
Max Planck Institute for Extraterrestrial Physics, Giessenbachstr. 1, 85748 Garching, Germany\label{aff59}
\and
Universit\"ats-Sternwarte M\"unchen, Fakult\"at f\"ur Physik, Ludwig-Maximilians-Universit\"at M\"unchen, Scheinerstrasse 1, 81679 M\"unchen, Germany\label{aff60}
\and
Dipartimento di Fisica "Aldo Pontremoli", Universit\`a degli Studi di Milano, Via Celoria 16, 20133 Milano, Italy\label{aff61}
\and
Institute of Theoretical Astrophysics, University of Oslo, P.O. Box 1029 Blindern, 0315 Oslo, Norway\label{aff62}
\and
Leiden Observatory, Leiden University, Einsteinweg 55, 2333 CC Leiden, The Netherlands\label{aff63}
\and
Jet Propulsion Laboratory, California Institute of Technology, 4800 Oak Grove Drive, Pasadena, CA, 91109, USA\label{aff64}
\and
Felix Hormuth Engineering, Goethestr. 17, 69181 Leimen, Germany\label{aff65}
\and
Technical University of Denmark, Elektrovej 327, 2800 Kgs. Lyngby, Denmark\label{aff66}
\and
Cosmic Dawn Center (DAWN), Denmark\label{aff67}
\and
Max-Planck-Institut f\"ur Astronomie, K\"onigstuhl 17, 69117 Heidelberg, Germany\label{aff68}
\and
NASA Goddard Space Flight Center, Greenbelt, MD 20771, USA\label{aff69}
\and
Department of Physics and Helsinki Institute of Physics, Gustaf H\"allstr\"omin katu 2, 00014 University of Helsinki, Finland\label{aff70}
\and
Universit\'e de Gen\`eve, D\'epartement de Physique Th\'eorique and Centre for Astroparticle Physics, 24 quai Ernest-Ansermet, CH-1211 Gen\`eve 4, Switzerland\label{aff71}
\and
Department of Physics, P.O. Box 64, 00014 University of Helsinki, Finland\label{aff72}
\and
Helsinki Institute of Physics, Gustaf H{\"a}llstr{\"o}min katu 2, University of Helsinki, Helsinki, Finland\label{aff73}
\and
NOVA optical infrared instrumentation group at ASTRON, Oude Hoogeveensedijk 4, 7991PD, Dwingeloo, The Netherlands\label{aff74}
\and
Centre de Calcul de l'IN2P3/CNRS, 21 avenue Pierre de Coubertin 69627 Villeurbanne Cedex, France\label{aff75}
\and
Universit\"at Bonn, Argelander-Institut f\"ur Astronomie, Auf dem H\"ugel 71, 53121 Bonn, Germany\label{aff76}
\and
Dipartimento di Fisica e Astronomia "Augusto Righi" - Alma Mater Studiorum Universit\`a di Bologna, via Piero Gobetti 93/2, 40129 Bologna, Italy\label{aff77}
\and
Department of Physics, Institute for Computational Cosmology, Durham University, South Road, DH1 3LE, UK\label{aff78}
\and
Universit\'e Paris Cit\'e, CNRS, Astroparticule et Cosmologie, 75013 Paris, France\label{aff79}
\and
University of Applied Sciences and Arts of Northwestern Switzerland, School of Engineering, 5210 Windisch, Switzerland\label{aff80}
\and
Institut d'Astrophysique de Paris, 98bis Boulevard Arago, 75014, Paris, France\label{aff81}
\and
Institut de F\'{i}sica d'Altes Energies (IFAE), The Barcelona Institute of Science and Technology, Campus UAB, 08193 Bellaterra (Barcelona), Spain\label{aff82}
\and
DARK, Niels Bohr Institute, University of Copenhagen, Jagtvej 155, 2200 Copenhagen, Denmark\label{aff83}
\and
Waterloo Centre for Astrophysics, University of Waterloo, Waterloo, Ontario N2L 3G1, Canada\label{aff84}
\and
Department of Physics and Astronomy, University of Waterloo, Waterloo, Ontario N2L 3G1, Canada\label{aff85}
\and
Perimeter Institute for Theoretical Physics, Waterloo, Ontario N2L 2Y5, Canada\label{aff86}
\and
Centre National d'Etudes Spatiales -- Centre spatial de Toulouse, 18 avenue Edouard Belin, 31401 Toulouse Cedex 9, France\label{aff87}
\and
Institute of Space Science, Str. Atomistilor, nr. 409 M\u{a}gurele, Ilfov, 077125, Romania\label{aff88}
\and
Dipartimento di Fisica e Astronomia "G. Galilei", Universit\`a di Padova, Via Marzolo 8, 35131 Padova, Italy\label{aff89}
\and
Institut f\"ur Theoretische Physik, University of Heidelberg, Philosophenweg 16, 69120 Heidelberg, Germany\label{aff90}
\and
Universit\'e St Joseph; Faculty of Sciences, Beirut, Lebanon\label{aff91}
\and
Departamento de F\'isica, FCFM, Universidad de Chile, Blanco Encalada 2008, Santiago, Chile\label{aff92}
\and
Universit\"at Innsbruck, Institut f\"ur Astro- und Teilchenphysik, Technikerstr. 25/8, 6020 Innsbruck, Austria\label{aff93}
\and
Institut d'Estudis Espacials de Catalunya (IEEC),  Edifici RDIT, Campus UPC, 08860 Castelldefels, Barcelona, Spain\label{aff94}
\and
Satlantis, University Science Park, Sede Bld 48940, Leioa-Bilbao, Spain\label{aff95}
\and
Institute of Space Sciences (ICE, CSIC), Campus UAB, Carrer de Can Magrans, s/n, 08193 Barcelona, Spain\label{aff96}
\and
Instituto de Astrof\'isica e Ci\^encias do Espa\c{c}o, Faculdade de Ci\^encias, Universidade de Lisboa, Tapada da Ajuda, 1349-018 Lisboa, Portugal\label{aff97}
\and
Universidad Polit\'ecnica de Cartagena, Departamento de Electr\'onica y Tecnolog\'ia de Computadoras,  Plaza del Hospital 1, 30202 Cartagena, Spain\label{aff98}
\and
INFN-Bologna, Via Irnerio 46, 40126 Bologna, Italy\label{aff99}
\and
Infrared Processing and Analysis Center, California Institute of Technology, Pasadena, CA 91125, USA\label{aff100}
\and
INAF, Istituto di Radioastronomia, Via Piero Gobetti 101, 40129 Bologna, Italy\label{aff101}
\and
Astronomical Observatory of the Autonomous Region of the Aosta Valley (OAVdA), Loc. Lignan 39, I-11020, Nus (Aosta Valley), Italy\label{aff102}
\and
Junia, EPA department, 41 Bd Vauban, 59800 Lille, France\label{aff103}
\and
Department of Physics, Royal Holloway, University of London, TW20 0EX, UK\label{aff104}
\and
ICSC - Centro Nazionale di Ricerca in High Performance Computing, Big Data e Quantum Computing, Via Magnanelli 2, Bologna, Italy\label{aff105}
\and
Instituto de F\'isica Te\'orica UAM-CSIC, Campus de Cantoblanco, 28049 Madrid, Spain\label{aff106}
\and
CERCA/ISO, Department of Physics, Case Western Reserve University, 10900 Euclid Avenue, Cleveland, OH 44106, USA\label{aff107}
\and
Laboratoire Univers et Th\'eorie, Observatoire de Paris, Universit\'e PSL, Universit\'e Paris Cit\'e, CNRS, 92190 Meudon, France\label{aff108}
\and
INFN-Sezione di Milano, Via Celoria 16, 20133 Milano, Italy\label{aff109}
\and
Departamento de F{\'\i}sica Fundamental. Universidad de Salamanca. Plaza de la Merced s/n. 37008 Salamanca, Spain\label{aff110}
\and
Departamento de Astrof\'isica, Universidad de La Laguna, 38206, La Laguna, Tenerife, Spain\label{aff111}
\and
Dipartimento di Fisica e Scienze della Terra, Universit\`a degli Studi di Ferrara, Via Giuseppe Saragat 1, 44122 Ferrara, Italy\label{aff112}
\and
Istituto Nazionale di Fisica Nucleare, Sezione di Ferrara, Via Giuseppe Saragat 1, 44122 Ferrara, Italy\label{aff113}
\and
Center for Data-Driven Discovery, Kavli IPMU (WPI), UTIAS, The University of Tokyo, Kashiwa, Chiba 277-8583, Japan\label{aff114}
\and
Ludwig-Maximilians-University, Schellingstrasse 4, 80799 Munich, Germany\label{aff115}
\and
Max-Planck-Institut f\"ur Physik, Boltzmannstr. 8, 85748 Garching, Germany\label{aff116}
\and
Dipartimento di Fisica - Sezione di Astronomia, Universit\`a di Trieste, Via Tiepolo 11, 34131 Trieste, Italy\label{aff117}
\and
Minnesota Institute for Astrophysics, University of Minnesota, 116 Church St SE, Minneapolis, MN 55455, USA\label{aff118}
\and
Universit\'e C\^{o}te d'Azur, Observatoire de la C\^{o}te d'Azur, CNRS, Laboratoire Lagrange, Bd de l'Observatoire, CS 34229, 06304 Nice cedex 4, France\label{aff119}
\and
Institute for Astronomy, University of Hawaii, 2680 Woodlawn Drive, Honolulu, HI 96822, USA\label{aff120}
\and
Department of Physics \& Astronomy, University of California Irvine, Irvine CA 92697, USA\label{aff121}
\and
Department of Astronomy \& Physics and Institute for Computational Astrophysics, Saint Mary's University, 923 Robie Street, Halifax, Nova Scotia, B3H 3C3, Canada\label{aff122}
\and
Departamento F\'isica Aplicada, Universidad Polit\'ecnica de Cartagena, Campus Muralla del Mar, 30202 Cartagena, Murcia, Spain\label{aff123}
\and
Universit\'e Paris-Saclay, CNRS, Institut d'astrophysique spatiale, 91405, Orsay, France\label{aff124}
\and
Department of Physics, Oxford University, Keble Road, Oxford OX1 3RH, UK\label{aff125}
\and
Department of Computer Science, Aalto University, PO Box 15400, Espoo, FI-00 076, Finland\label{aff126}
\and
Instituto de Astrof\'\i sica de Canarias, c/ Via Lactea s/n, La Laguna E-38200, Spain. Departamento de Astrof\'\i sica de la Universidad de La Laguna, Avda. Francisco Sanchez, La Laguna, E-38200, Spain\label{aff127}
\and
Ruhr University Bochum, Faculty of Physics and Astronomy, Astronomical Institute (AIRUB), German Centre for Cosmological Lensing (GCCL), 44780 Bochum, Germany\label{aff128}
\and
Univ. Grenoble Alpes, CNRS, Grenoble INP, LPSC-IN2P3, 53, Avenue des Martyrs, 38000, Grenoble, France\label{aff129}
\and
Department of Physics and Astronomy, Vesilinnantie 5, 20014 University of Turku, Finland\label{aff130}
\and
Serco for European Space Agency (ESA), Camino bajo del Castillo, s/n, Urbanizacion Villafranca del Castillo, Villanueva de la Ca\~nada, 28692 Madrid, Spain\label{aff131}
\and
ARC Centre of Excellence for Dark Matter Particle Physics, Melbourne, Australia\label{aff132}
\and
Centre for Astrophysics \& Supercomputing, Swinburne University of Technology,  Hawthorn, Victoria 3122, Australia\label{aff133}
\and
School of Physics and Astronomy, Queen Mary University of London, Mile End Road, London E1 4NS, UK\label{aff134}
\and
Department of Physics and Astronomy, University of the Western Cape, Bellville, Cape Town, 7535, South Africa\label{aff135}
\and
Universit\'e Libre de Bruxelles (ULB), Service de Physique Th\'eorique CP225, Boulevard du Triophe, 1050 Bruxelles, Belgium\label{aff136}
\and
ICTP South American Institute for Fundamental Research, Instituto de F\'{\i}sica Te\'orica, Universidade Estadual Paulista, S\~ao Paulo, Brazil\label{aff137}
\and
IRFU, CEA, Universit\'e Paris-Saclay 91191 Gif-sur-Yvette Cedex, France\label{aff138}
\and
INAF-Osservatorio Astrofisico di Arcetri, Largo E. Fermi 5, 50125, Firenze, Italy\label{aff139}
\and
Dipartimento di Fisica, Sapienza Universit\`a di Roma, Piazzale Aldo Moro 2, 00185 Roma, Italy\label{aff140}
\and
Aurora Technology for European Space Agency (ESA), Camino bajo del Castillo, s/n, Urbanizacion Villafranca del Castillo, Villanueva de la Ca\~nada, 28692 Madrid, Spain\label{aff141}
\and
Centro de Astrof\'{\i}sica da Universidade do Porto, Rua das Estrelas, 4150-762 Porto, Portugal\label{aff142}
\and
HE Space for European Space Agency (ESA), Camino bajo del Castillo, s/n, Urbanizacion Villafranca del Castillo, Villanueva de la Ca\~nada, 28692 Madrid, Spain\label{aff143}
\and
Institute of Astronomy, University of Cambridge, Madingley Road, Cambridge CB3 0HA, UK\label{aff144}
\and
Department of Astrophysics, University of Zurich, Winterthurerstrasse 190, 8057 Zurich, Switzerland\label{aff145}
\and
INAF-Osservatorio Astronomico di Brera, Via Brera 28, 20122 Milano, Italy, and INFN-Sezione di Genova, Via Dodecaneso 33, 16146, Genova, Italy\label{aff146}
\and
Theoretical astrophysics, Department of Physics and Astronomy, Uppsala University, Box 515, 751 20 Uppsala, Sweden\label{aff147}
\and
Department of Astrophysical Sciences, Peyton Hall, Princeton University, Princeton, NJ 08544, USA\label{aff148}
\and
Cosmic Dawn Center (DAWN)\label{aff149}
\and
Niels Bohr Institute, University of Copenhagen, Jagtvej 128, 2200 Copenhagen, Denmark\label{aff150}
\and
Center for Cosmology and Particle Physics, Department of Physics, New York University, New York, NY 10003, USA\label{aff151}
\and
Center for Computational Astrophysics, Flatiron Institute, 162 5th Avenue, 10010, New York, NY, USA\label{aff152}
\and
Mathematical Institute, University of Leiden, Niels Bohrweg 1, 2333 CA Leiden, The Netherlands\label{aff153}}

\date{8 November 2024}

\abstract{%
    We present the framework for measuring angular power spectra in the \Euclid
    mission.  The observables in galaxy surveys, such as galaxy clustering and
    cosmic shear, are not continuous fields, but discrete sets of data,
    obtained only at the positions of galaxies.  We show how to compute the
    angular power spectra of such discrete data sets, without treating
    observations as maps of an underlying continuous field that is overlaid
    with a noise component.  This formalism allows us to compute exact
    theoretical expectations for our measured spectra, under a number of
    assumptions that we track explicitly.  In particular, we obtain exact
    expressions for the additive biases (``shot noise'') in angular galaxy
    clustering and cosmic shear.  For efficient practical computations, we
    introduce a spin-weighted spherical convolution with a well-defined
    convolution theorem, which allows us to apply exact theoretical predictions
    to finite-resolution maps, including \emph{HEALPix}.  When validating our
    methodology, we find that our measurements are biased by less than 1\% of
    their statistical uncertainty in simulations of \Euclid's first data
    release.
}

\keywords{%
    Methods: statistical;
    Surveys;
    Cosmology: observations;
    large-scale structure of Universe;
    Gravitational lensing: weak
}

\titlerunning{\Euclid preparation: LIX. Angular power spectra}
\authorrunning{N.~Tessore et al.}

\maketitle

\section{Introduction}

\begin{figure*}%
\centering%
\includegraphics[width=0.9\textwidth]{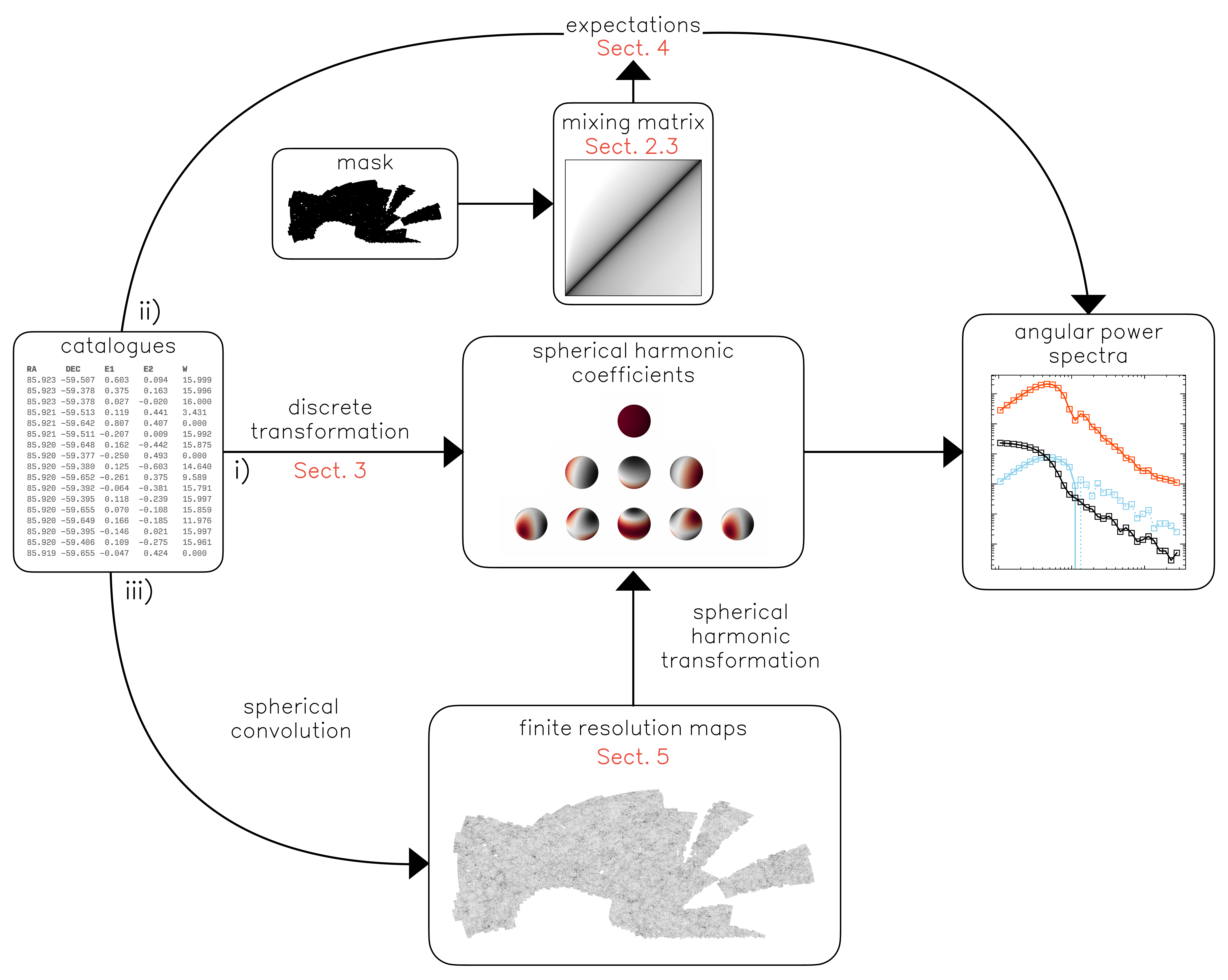}%
\caption{%
    Overview of the methodology presented below.  We apply the formalism for
    discrete angular power spectra in three distinct ways: \emph{i)} Exact
    spherical harmonic coefficients can be computed from the discrete data,
    without the use of maps.  In turn, angular power spectra can be computed
    from combinations of spherical harmonic coefficients.  \emph{ii)} The
    angular power spectra themselves can be computed from the discrete data.
    This is inefficient for practical computation, but makes it possible to
    obtain exact expressions for the expected spectra.  \emph{iii)} The
    discrete data can be turned into maps, and subsequently into spherical
    harmonic coefficients by means of a spherical harmonic transform.  This can
    yield the same results as the discrete transformation.
}%
\label{fig:overview}%
\end{figure*}

The photometric survey of the \Euclid mission
\citep{2011arXiv1110.3193L,2024arXiv240513491E} will infer cosmology using
correlations between the observed angular positions of galaxies (angular galaxy
clustering), their observed shapes (cosmic shear), and the cross-correlation
between positions and shapes (galaxy--galaxy lensing).  These so-called
two-point statistics are powerful probes of the late-time evolution of the
Universe, both on their own and in a joint ``3\texttimes2pt'' analysis.  As a
result, two-point statistics have become the de facto standard observable for
cosmological analysis in Stage-III galaxy surveys such as the Kilo-Degree
Survey \citep{2021A&A...646A.140H}, the Dark Energy Survey
\citep{2022PhRvD.105b3520A}, and the Subaru Hyper Suprime-Cam Survey
\citep{2023PhRvD.108l3520M}.

Angular correlations can be quantified and measured in a variety of ways.  In
so-called real-space methods, correlations are measured in terms of real
angular separation on the sky.  Conversely, in harmonic-space methods,
observations first undergo a spherical harmonic transform before two-point
statistics are extracted.  Examples of real-space methods include angular
correlation functions \citep{1973ApJ...185..413P,2002A&A...396....1S}, COSEBIs
\citep{2010A&A...520A.116S}, and band powers \citep{2002A&A...396....1S}, while
examples of harmonic-space methods include various flavours of angular power
spectra.  As we will show below, there are exact mathematical relations to
transform between real-space and harmonic-space observables.  In practice,
however, these transformations usually cannot be applied to measured data, so
that real-space and harmonic-space methods are effectively slightly different
probes of the same underlying information.  For that reason,  \Euclid will
deliver data products for all of the aforementioned methods.  In what follows,
we describe the harmonic-space measurement, whereas the real-space methods will
be described separately (Euclid Collaboration: Kilbinger et al. in prep.).

Most current methodology for the measurement of angular power spectra for
3\texttimes2pt cosmology comes from the analysis of the cosmic microwave
background \citep[CMB; e.g.,][]{1997PhRvD..55.7368K, 1997PhRvD..55.1830Z,
2001PhRvD..64h3003W, 2002ApJ...567....2H}.  However, the observables of the CMB
are continuous temperature and polarisation fields, of which maps are created
by carefully planned observations that are slightly oversampled with respect to
the instrument's beam size \citep{2005A&A...430..363D}.  The same is not true
for the observables in galaxy surveys such as \Euclid:  galaxy clustering
observes the individual, discrete positions of galaxies, and cosmic shear
probes the gravitational lensing fields through the ellipticities of galaxies
at whatever positions these may be located.

To extract angular power spectra from a photometric galaxy survey, the typical
approach is then to treat observations as if they were sampling continuous
fields, much like the CMB \citep{2019MNRAS.484.4127A, 2021JCAP...03..067N}.
For galaxy clustering, this requires an assumption that galaxies are discrete
``tracers'' of an underlying galaxy density field.  By making pixelated maps of
observed galaxy number counts, the idea is to create a fair representation of
this underlying field, up to a ``shot noise'' contribution in each pixel.
Similarly, for cosmic shear, observed ellipticities of galaxies are considered
tracers of the weak lensing signal.  By averaging all observed ellipticities in
each pixel of a cosmic shear map, the assumption is that one recovers the
underlying field, up to a ``shape noise'' contribution due to the distribution
of intrinsic galaxy shapes.

The approximation of having a smooth, continuous map of an underlying field
overlaid with noise starts to break down when observations are sparse with
respect to the map resolution.  For example, at the angular resolution required
for \Euclid's ambitious science goals \citep{2024arXiv240513491E}, we expect
about half of the observed pixels in the resulting maps to be empty.  Our
approach for \Euclid is therefore to consider the angular power spectra of the
discrete data itself, similar to the traditional analysis of spectroscopic
galaxy catalogues \citep{1995MNRAS.275..483H, 1999MNRAS.305..527T,
2004MNRAS.353.1201P}.  As we will show, this is possible without assuming that
observations recover an underlying continuous field.  In particular, the
angular power spectra from discrete data points are essentially the spherical
harmonics evaluated in said points, and can hence be calculated in practice.
During implementation of the \Euclid pipeline following this approach, the same
idea has been independently developed in two other recent works, first by
\citet{2024JCAP...05..010B}, and subsequently by \citet*{2024arXiv240721013W}.

Besides practical computation, the discrete angular power spectra offer an
additional advantage on the theoretical side:  not having to assume the
existence of intermediary maps with resolution-dependent ``noise'' greatly
simplifies theoretical predictions for the measured spectra.  Apart from a
number of scientific assumptions, which we track and call out explicitly, this
approach allows us to obtain an exact theory for the expectations of our
measurements.  For angular galaxy clustering, we find that the additive ``shot
noise'' bias is not random but a known number, and we obtain an expected galaxy
clustering signal that depends directly on the angular correlation
function~$\w(\theta)$ as originally defined by \cite{1973ApJ...185..413P},
instead of the two-point statistics of an ancillary galaxy density field.  For
cosmic shear, we obtain an expression for the additive ``shape noise'' bias
that correctly treats the interplay between reduced shear and intrinsic galaxy
shapes, as well as a novel method to remove the residual additive bias from the
intrinsic variance of the cosmic shear field.  In light of the stringent
requirements on admissible biases in the \Euclid data processing pipeline,
these results allow us to validate our measurements to unprecedented levels of
accuracy, which would otherwise be impossible due to uncertainty in the
expectation values.

Directly measuring angular power spectra from maps is fast, which makes it the
de facto standard approach for obtaining spectra, despite the emergence of
competing harmonic-space methods such as Quadratic Maximum Likelihood (QML)
estimators \citep{1997PhRvD..55.5895T, 2001PhRvD..64f3001T,
2023MNRAS.520.4836M} or Bayesian Hierarchical Model (BHM) estimators
\citep{2016MNRAS.455.4452A, 2023OJAp....6E...6L, 2023OJAp....6E..31S}.
Compared to discrete angular power spectra, the overall computational cost of
map-based spectra is essentially a function of map resolution, and largely
independent of the number of objects in the input catalogues.  Map-based
angular power spectra therefore remain an attractive computational option,
particularly in the context of a large galaxy survey such as \Euclid, where we
eventually expect more than 1.5 billion galaxies to be observed.

For this reason, we investigate ways to obtain spectra from finite-resolution
maps, while keeping the theoretical benefits of the discrete methodology.  We
can achieve this by using a formalism for spin-weighted spherical convolution
with an exact convolution theorem.  In principle, we are able to recover the
discrete angular power spectra up to a resolution-dependent band limit, and
hence apply the exact theoretical predictions to map-based spectra.  To improve
performance even further, we also show how this approach can be approximated
using \emph{HEALPix} maps \citep{2005ApJ...622..759G}, which do not have an
exact convolution theorem, and require special handling of the additive bias.

Overall, a schema of our approach is shown in Fig.~\ref{fig:overview}.  The
text is organised similarly.  In Sect.~\ref{sec:angular-power-spectra}, we
review the theory of angular power spectra, and develop results on which we
rely later.  In Sect.~\ref{sec:observations}, we compute the angular power
spectra of discrete sets of observations.  In Sect.~\ref{sec:expectations}, we
obtain the expectations of said angular power spectra, for the cases in which
the observations are generated by point processes or random fields.  In
Sect.~\ref{sec:maps}, we show how to obtain the angular power spectra of
discrete observations from the usual maps.  In Sect.~\ref{sec:validation}, we
validate our results against simulations, and show that our methodology can be
applied to \Euclid's first data release.  We conclude with a brief discussion
of our method in Sect.~\ref{sec:discussion}.

The methodology presented here, for both the discrete and map-based spectra, is
implemented in a publicly available code called \texttt{Heracles}.\footnote{%
    \label{fn:url}\url{https://github.com/heracles-ec/heracles}}
This code is used for data processing in the 3\texttimes2pt pipeline within the
\Euclid Science Ground Segment.  However, it was designed from the ground up as
a modular, adaptable, and user-friendly general-purpose utility that can be
used for a multitude of probes and surveys.

\section{Angular power spectra}
\label{sec:angular-power-spectra}

In this section, we state key results and theorems regarding the two-point
statistics of arbitrary spherical functions (i.e., functions on the sphere).
The crucial point here will be that the concepts of angular power spectra and
angular correlation functions are well-defined not only for the particular case
of homogeneous random fields, but for any function on the sphere, such as,
e.g., an individual realisation of a random field.

In the following, we will always deal with spin-weighted spherical functions,
which sometimes have spin weight zero, and we follow the definitions of
\citet{2016JMP....57i2504B}.  We parametrise the sphere using unit vectors,
which we denote~$\U$ and~$\U'$.  A spherical function~$f$ has spin weight~$s$
if the function value~$f(\U)$ transforms under a rotation~$\gamma$ of the
coordinate frame in~$\U$ as
\begin{equation}
\label{eq:spinweight}
    f(\U) \mapsto \E^{-\I s \gamma} \, f(\U) \,.
\end{equation}
It follows that a spherical function with non-zero spin weight is necessarily
complex-valued.  Examples of a spin-weighted spherical functions are, e.g., the
global surface temperature on Earth ($s = 0$), wind speed and direction ($s =
1$), or the polarisation of the CMB ($s = 2$).

We generally only consider spherical functions~$f$ with spin weight~$s$ that
have an expansion into spin-weighted spherical harmonics~$_sY_{lm}$,
\begin{equation}
\label{eq:def-f-ylm}
    f(\U)
    = \sum_{lm} f_{lm} \, {}_sY_{lm}(\U) \,,
\end{equation}
where, here and in the following, sums always extend over all admissible
values~$l \ge |s|$ and~$-l \le m \le l$.  The coefficients~$f_{lm}$ of the
expansion are obtained by integration against the spherical harmonics,
\begin{equation}
\label{eq:def-flm}
    f_{lm}
    = \int \! f(\U) \, \bigl[{}_sY_{lm}(\U)\bigr]^* \, \D\U \,,
\end{equation}
where the integral extends over the entire sphere, and an asterisk denotes
complex conjugation.  For~$s = 0$, i.e., no spin weight, the expansion is in
the classical spherical harmonics~$Y_{lm} \equiv {}_0Y_{lm}$.  In practice, we
always have~$s = 0$ or~$s = 2$, but we will treat~$s$ as an arbitrary integer
spin weight as much as possible.

\subsection{Two-point statistics}

For any pair of spherical functions~$f$ and~$f'$ with respective spin
weights~$s$ and~$s'$, where $f = f'$ is allowed, we can define the angular
correlation~$C^{ff'}(\theta)$ as the correlation of~$f(\U)$ and~$f'(\U')$ over
all points~$\U, \U'$ on the sphere separated by the angle~$\theta$,
\begin{equation}
\label{eq:cf}
    C^{ff'}(\theta)
    = \frac{1}{8\pi^2} \smashoperator{\iint\limits_{\U \cdot \U' = \cos\theta}}
        \! \E^{\I s \alpha} \, f^*(\U) \, f'(\U') \, \E^{-\I s' \alpha'} \,
            \D\U \, \D\U' \,.
\end{equation}
Here, if the spin weights~$s$ and~$s'$ are non-zero, the angles~$\alpha$
and~$\alpha'$, defined in Appendix~\ref{sec:angles}, describe a rotation of the
respective coordinate frames in~$\U$ and~$\U'$ such that the resulting
correlation is frame-independent.  It is clear that the
definition~\eqref{eq:cf} of the angular correlation function does not
require~$f$ or~$f'$ to be random fields, or possess any kind of symmetry.
If~$f$ and~$f'$ are complex-valued, the two-point statistics are not fully
characterised by~\eqref{eq:cf} alone, but also by the correlations
between~$f^*$ and~$f'$, as described in Appendix~\ref{sec:e-b-modes}.

For any angular correlation function~$C^{ff'}$, we can define an associated
angular power spectrum~$C_l^{ff'}$ as the coefficients of the expansion
of~$C^{ff'}$ into the Wigner~$d$ functions~$d^l_{ss'}$
\citep{1960amqm.book.....E},
\begin{equation}
\label{eq:cf-cl}
    C^{ff'}(\theta)
    = \sum_{l} \frac{2l+1}{4\pi} \, C_l^{ff'} \, d^l_{ss'}(\theta) \,.
\end{equation}
As usual, the coefficients are obtained by projection against the basis
functions,
\begin{equation}
\label{eq:cl-cf}
    C_l^{ff'}
    = 2\pi \int_{0}^{\pi} \!
            C^{ff'}(\theta) \, d^l_{ss'}(\theta) \sin(\theta) \, \D\theta \,.
\end{equation}
The definition~\eqref{eq:cf-cl} of the angular power spectrum in terms of the
angular correlation function makes once again no reference to random fields or
symmetries.

To express the angular power spectrum~$C_l^{ff'}$ directly in terms of the
functions~$f$ and~$f'$, we can replace~$C^{ff'}(\theta)$ in the angular power
spectrum~\eqref{eq:cl-cf} by its definition~\eqref{eq:cf},
\begin{equation}
\label{eq:def-cl-ff}
    C_l^{ff'}
    = \frac{1}{4\pi} \iint f^*(\U) \, f'(\U') \,
        \E^{\I s \alpha} \, d^l_{ss'}(\theta) \, \E^{-\I s' \alpha'} \,
        \D\U \, \D\U' \,,
\end{equation}
where the angles~$\alpha, \theta, \alpha'$ still depend on~$\U$ and~$\U'$, but
we now have two unrestricted integrals over the sphere.  Inserting the
spherical harmonic addition theorem
\begin{equation}
\label{eq:addthm}
    \sum_{m} {}_sY_{lm}(\U) \, {}^{}_{s'}Y_{lm}^*(\U')
    = \frac{2l+1}{4\pi} \,
        \E^{\I s \alpha} \, d^l_{ss'}(\theta) \, \E^{-\I s' \alpha'}
\end{equation}
into the integrand in definition~\eqref{eq:def-cl-ff}, the two integrals
decouple, and reduce to the spherical harmonic coefficients~$f_{lm}$
and~$f'_{lm}$ given by definition~\eqref{eq:def-flm}.  The angular power
spectrum of spherical functions~$f$ and~$f'$ is hence equivalently defined in
terms of their spherical harmonic coefficients,
\begin{equation}
\label{eq:cl}
    C_l^{ff'}
    = \frac{1}{2l+1} \sum_{m} f^*_{lm} \, f'_{lm} \,.
\end{equation}
This expression is sometimes called ``the estimator of the angular power
spectrum'', for reasons that are given below.  However,
expression~\eqref{eq:cl} is in fact the true angular power spectrum of the
particular spherical functions~$f$ and~$f'$ (which, in cosmology, are a
particular realisation from a random process), as~$C_l^{ff'}$ contains exactly
the same information as the angular correlation function~\eqref{eq:cf}.  To
fully describe the two-point statistics of complex-valued functions~$f$
and~$f'$, we hence require both~$C_l^{ff'}$ and the
pseudo-spectrum~$C_l^{f^*f'}$ (see Appendix~\ref{sec:e-b-modes}).

\subsection{Homogeneous random fields}

The angular correlation function~\eqref{eq:cf} is obtained by averaging a
spherical function over all pairs of points at a given angular
separation~$\theta$.  There is an important class of fields where this
averaging over direction does not remove information from the two-point
statistics.  These are the random fields that are invariant under rotations,
which we call ``homogeneous'' on the sphere.\footnote{%
    Generally speaking, homogeneity is invariance under translations, whereas
    invariance under rotations is isotropy.  The translations of the sphere are
    rotations, and any rotation of the sphere can be achieved by three
    translations, so that homogeneity and isotropy are equivalent.}

Under a rotation~$R$ of the sphere the coefficients of the spherical harmonic
expansion~\eqref{eq:def-f-ylm} transform as \citep{2016JMP....57i2504B}
\begin{equation}
\label{eq:flm-rot}
    f_{lm}
    \mapsto \sum_{\mu} f_{l\mu} \, D^l_{\mu m}(R) \,,
\end{equation}
where~$D^l_{\mu m}$ is the Wigner $D$ function.  The importance of homogeneous
random fields on the sphere is closely related to this transformation:  if~$f$
is a realisation of such a field, all of its statistical properties are by
definition invariant under rotations, and both sides of
transformation~\eqref{eq:flm-rot} have the same distribution.  The distinction
between the random field itself and its realisations is important here: the
random field is invariant, but any given realisation that we may observe is a
fixed non-random spherical function.

Homogeneity is a powerful tool:  for example, consider the product~$f^*_{lm} \,
f'_{l'm'}$ of modes from the spherical harmonic expansion~\eqref{eq:def-f-ylm}
of a pair of functions~$f$ and~$f'$.  Under the rotation~\eqref{eq:flm-rot},
the product transforms as
\begin{equation}
\label{eq:flm-prod-rot}
    f^*_{lm} \, f'_{l'm'}
    \mapsto \sum_{\mu\mu'} f^*_{l\mu} \, f'_{l'\mu'} \,
        D^{l*}_{\mu m}(R) \, D^{l'}_{\mu' m'}(R) \,.
\end{equation}
If~$f$ and~$f'$ are realisations of jointly homogeneous random fields, both
sides of transformation~\eqref{eq:flm-prod-rot} must be equal in distribution.
Taking the expectation over realisations, denoted~$\ev{\:\cdot\:}$, we find
\begin{equation}
    \ev{f^*_{lm} \, f'_{l'm'}}
    = \sum_{\mu\mu'} \ev{f^*_{l\mu} \, f'_{l'\mu'}} \,
                        D^{l*}_{\mu m}(R) \, D^{l'}_{\mu' m'}(R) \,.
\end{equation}
Integrating out the rotation~$R$ on both sides using the orthogonality of the
$D$ functions \citep[][Eq.\ 4.6.1]{1960amqm.book.....E}, we recover
expression~\eqref{eq:cl} and thus obtain the well-known expectation,
\begin{equation}
\label{eq:ev-flm-prod}
    \ev{f^*_{lm} \, f'_{l'm'}}
    = \delta^{\rm K}_{ll'} \, \delta^{\rm K}_{mm'} \, \ev{C_l^{ff'}} \,,
\end{equation}
where~$\delta^{\rm K}$ is the Kronecker delta symbol.  In other words, the
modes of homogeneous random fields on the sphere are uncorrelated, unless their
modes numbers coincide.

Note that the expectation~\eqref{eq:ev-flm-prod} is sometimes used to define
the angular power spectrum~$\ev{C_l}$ of random fields, in which case the
expression~\eqref{eq:cl} is used as an estimator for~$\ev{C_l}$.  However, we
prefer to think of the sum~\eqref{eq:cl} as the actual, realised, observable
angular power spectrum, and~$\ev{C_l}$ as its expectation over realisations.

Having obtained the two-point expectation~\eqref{eq:ev-flm-prod} in harmonic
space, its equivalent~$\ev{f^*(\U) \, f'(\U')}$ in real-space can be obtained
by computing the spherical harmonic expansion~\eqref{eq:def-f-ylm} of the
product, substituting expectation~\eqref{eq:ev-flm-prod}, and using the complex
conjugate of the spherical harmonic addition theorem~\eqref{eq:addthm},
\begin{equation}
\label{eq:ev-ff-cl}
    \ev{f^*(\U) \, f'(\U')}
    = \frac{2l+1}{4\pi} \sum_{l} \ev{C_l^{ff'}} \,
        \E^{-\I s \alpha} \, d^l_{ss'}(\theta) \, \E^{\I s' \alpha'} \,.
\end{equation}
Factoring out the exponentials, the remaining sum is precisely the expectation
of the relation~\eqref{eq:cf-cl} between angular power spectrum and angular
correlation function,
\begin{equation}
\label{eq:ev-cf-cl}
    \ev{C^{ff'}(\theta)}
    = \frac{2l+1}{4\pi} \sum_{l} \ev{C_l^{ff'}} \, d^l_{ss'}(\theta) \,,
\end{equation}
and expectation~\eqref{eq:ev-ff-cl} thus yields the expected two-point
statistics in real space,
\begin{equation}
\label{eq:ev-ff-cf}
    \ev{f^*(\U) \, f'(\U')}
    = \E^{-\I s \alpha} \, \ev{C^{ff'}(\theta)} \, \E^{\I s' \alpha'} \,.
\end{equation}
Naturally, the inverse relation~\eqref{eq:cl-cf} to~\eqref{eq:ev-cf-cl} holds
in expectation as well,
\begin{equation}
\label{eq:ev-cl-cf}
    \ev{C_l^{ff'}}
    = 2\pi \int_{0}^{\pi} \!
        \ev{C^{ff'}(\theta)} \, d^l_{ss'}(\theta) \sin(\theta) \, \D\theta \,,
\end{equation}
and is consistent with expectations~\eqref{eq:ev-ff-cf}
and~\eqref{eq:ev-ff-cl}.

\subsection{Mixing matrices}

An important special case is a random field~$f$ that is the product of a
homogeneous random field~$g$ and a non-stochastic spherical function~$w$,
\begin{equation}
\label{eq:prod}
    f(\U) = g(\U) \, w(\U) \,.
\end{equation}
We usually call~$w$ a ``weight'' function, but it is in fact arbitrary, and
could in principle encode systematic effects such as, e.g., position-dependent
multiplicative biases, including higher-order biases with non-zero spin weights
\citep{2021OJAp....4E..17K, 2022OJAp....5E...6K}.  The functions in the
product~\eqref{eq:prod} can each have an associated spin weight; if~$s, s_1,
s_2$ are the respective spin weights of~$f, g, w$, it follows that~$s = s_1 +
s_2$ by the definition of the spin weight~\eqref{eq:spinweight}.

The angular correlation function~\eqref{eq:cf} of~$f$ and a second such
field~$f'$ can be expressed in terms of~$g, g'$ and~$w, w'$,
\begin{equation}
\begin{split}
\label{eq:cf-prod}
    C^{ff'}(\theta)
    = \frac{1}{8\pi^2}
        \smashoperator{\iint\limits_{\U \cdot \U' = \cos\theta}}
        &\Bigl[
            \E^{\I s_1 \alpha} \, g^*(\U) \, g'(\U') \, \E^{-\I s_1' \alpha'}
        \Bigr] \\[-10pt] &\quad \times \Bigl[
            \E^{\I s_2 \alpha} \, w^*(\U) \, w'(\U') \, \E^{-\I s_2' \alpha'}
        \Bigr] \, \D\U \, \D\U' \,.
\end{split}
\end{equation}
To compute the expectation of expression~\eqref{eq:cf-prod}, we assume that~$g$
and~$g'$ are independent of~$w$ and~$w'$.  The expectation can then be moved
into the integral, and we recover the angular correlation
function~\eqref{eq:ev-ff-cf} of~$g$ and~$g'$,
\begin{equation}
\label{eq:mm-crit}
    \E^{\I s_1 \alpha} \, \ev{g^*(\U) \, g'(\U')} \, \E^{-\I s_1' \alpha'}
    = \ev{C^{gg'}(\theta)} \,.
\end{equation}
We can factor~$\ev{C^{gg'}(\theta)}$ out of the integral, which reduces to the
angular correlation function~\eqref{eq:cf} of~$w$ and~$w'$,
\begin{equation}
\label{eq:ev-cf-prod}
    \ev{C^{ff'}(\theta)}
    = \ev{C^{gg'}(\theta)} \, C^{ww'}(\theta) \,.
\end{equation}
We thus find that the expected angular correlation of products of homogeneous
random fields and weight functions is the product of their (expected) angular
correlations.

Given expectation~\eqref{eq:ev-cf-prod}, we can also compute the expected
angular power spectrum using relation~\eqref{eq:cl-cf},
\begin{equation}
\label{eq:ev-cl-cf-prod}
    \ev{C_l^{ff'}}
    = 2\pi \int_{0}^{\pi} \! \ev{C^{gg'}(\theta)} \, C^{ww'}(\theta) \,
        d^l_{ss'}(\theta) \sin(\theta) \, \D\theta \,.
\end{equation}
We then expand the angular correlations~$\ev{C^{gg'}(\theta)}$
and~$C^{ww'}(\theta)$ in the integral using relation~\eqref{eq:cf-cl}.  Since
$s = s_1 + s_2$ and $s' = s_1' + s_2'$, we obtain Gaunt's integral for the $d$
functions \citep{1960amqm.book.....E},
\begin{equation}
\begin{split}
    &\frac{1}{2} \int_{0}^{\pi} \! d^{l_1}_{s_1s_1'}(\theta) \,
        d^{l_2}_{s_2s_2'}(\theta) \, d^l_{ss'}(\theta)
        \sin(\theta) \, \D\theta \\
    &\qquad
    = (-1)^{s-s'} \, \threej{l_1,s_1}{l_2,s_2}{l,-s} \,
        \threej{l_1,s_1'}{l_2,s_2'}{l,-s'} \,,
\end{split}
\end{equation}
where the right-hand side contains the Wigner~$3j$ symbols.  The result
expresses the expected angular power spectrum~$\ev{C_l^{ff'}}$ in terms of the
angular power spectra~$\ev{C_l^{gg'}}$ and~$C_l^{ww'}$,
\begin{equation}
\label{eq:ev-cl-prod}
\begin{split}
    \ev{C_l^{ff'}}
    = (-1)^{s-s'} \smash[b]{\sum_{l_1l_2}}
        & \frac{(2l_1+1) \, (2l_2+1)}{4\pi} \,
        \ev{C_{l_1}^{gg'}} \, C_{l_2}^{ww'}
        \\ & \; \times
        \threej{l_1,s_1}{l_2,s_2}{l,-s} \,
        \threej{l_1,s_1'}{l_2,s_2'}{l,-s'} \,.
\end{split}
\end{equation}
There is hence a convolution theorem for angular power spectra and angular
correlation functions, which more generally holds for expansions in Wigner~$d$
functions:  the product of functions in the real-space
expectation~\eqref{eq:ev-cf-prod} corresponds to a convolution in the
harmonic-space expectation~\eqref{eq:ev-cl-prod}.

In practice, we usually want to keep the weight functions fixed, and compute
the expectation~$\ev{C_l^{ff'}}$ as a function of the expected angular power
spectrum~$\ev{C_l^{gg'}}$ of the underlying random fields.  In that case, the
convolution~\eqref{eq:ev-cl-prod} can be separated into a linear operator
containing the sum over~$l_2$,
\begin{equation}
\begin{split}
\label{eq:mm}
    M_{ll_1}^{ww'}
    = (-1)^{s-s'} \smash[b]{\sum_{l_2}} & \frac{(2l_1+1) \, (2l_2+1)}{4\pi}
        \, C_{l_2}^{ww'}
        \\ & \; \times
        \threej{l_1,s_1}{l_2,s_2}{l,-s} \,
        \threej{l_1,s_1'}{l_2,s_2'}{l,-s'} \,,
\end{split}
\end{equation}
which can subsequently be applied to any given~$\ev{C_{l_1}^{gg'}}$,
\begin{equation}
\label{eq:ev-cl-mm}
    \ev{C_l^{ff'}}
    = \sum_{l_1} M_{ll_1}^{ww'} \, \ev{C_{l_1}^{gg'}} \,.
\end{equation}
We call the operator~$M^{ww'}$ the `mixing matrix' of the weights~$w, w'$
applied to the fields~$f, f'$ and~$g, g'$.  This is slightly misleading, since
expression~\eqref{eq:mm} is merely a formal ``matrix'' with infinitely many
rows and columns.  However, in practice, it is always truncated to a finite
size, and hence indeed a matrix.  Note that the mixing matrix~\eqref{eq:mm} not
only depends on~$w$ and~$w'$, but also on the full set of spin weights.

There is an important, non-trivial consequence of the above derivation:  the
mixing matrix only maps the expected angular power spectrum of a homogeneous
random field to the expected angular power spectrum of its product with another
function.  The critical step occurs in the expectation~\eqref{eq:mm-crit},
which only holds \emph{i)} in expectation and \emph{ii)} for homogeneous random
fields.  If either condition is not fulfilled, the mixing matrix formalism
breaks down.  In particular, it follows that mixing matrices for random fields
cannot in general be multiplied: If the function~$w = w_1 \, w_2$ is the
product of spherical functions~$w_1$ and~$w_2$, then
\begin{equation}
    M^{ww'} \ne M^{w_1w_1'} \, M^{w_2w_2'}
\end{equation}
except for special cases.  The reason is a lack of homogeneity:  the random
field~$w_2 \, g$ that yields the mixing matrix~$M^{w_2w_2'}$ is no longer
homogeneous, and the product~$w_1 \, (w_2 \, g)$ is hence not described by a
second mixing matrix application.  For example, consider the respective
footprint of the northern and southern hemisphere.  Individually, both
footprints have the same angular correlation function, same angular power
spectrum, and same non-vanishing mixing matrix.  But since the product of the
footprints is identically zero, so is their combined mixing matrix.

\section{Discrete observations}
\label{sec:observations}

Having reviewed the theory of angular power spectra, we now turn our attention
towards creating the necessary spherical functions from sets of discrete
observations.  To this end, we consider two distinct types of observations:
\begin{itemize}
\item \emph{Points}.  The information lies in the distribution of the observed
    positions themselves, which have no further data attached.
\vskip3pt
\item \emph{Fields}.  The information comes from the observed values of some
    underlying spherical function, which is observed in a discrete set of
    points.
\end{itemize}
Depending on which kind of data we wish to analyse, we must proceed in slightly
different ways.

\subsection{Points}

We first consider the case where we observe a number of points~$\U_k$, $k = 1,
2, \ldots$, on the sphere, as well as a set of weights~$w_k$.\footnote{%
    In what follows, we always implicitly assume that points have spin-$0$
    weights, since that is the only practically relevant case.  However, our
    results generalise straightforwardly to the spin-weighted case.  For
    unweighted observations, the weights are set to unity.}
In the specific case of \Euclid, this might be the observed angular positions
of galaxies.  We can represent the set of observed points as a sum of ``point
masses'' using the Dirac delta function~$\delta^{\rm D}$,
\begin{equation}
\label{eq:n}
    n(\U)
    = \sum_{k} w_k \, \delta^{\rm D}(\U - \U_k) \,,
\end{equation}
where the sum extends over the observed points.  This turns the discrete
observations into a function defined over the entire sphere.  The spherical
function~$n$ has spin weight~$s = 0$ and is a true (weighted) number density,
since the integral of the definition~\eqref{eq:n} over any given area of the
sphere produces the contained (weighted) number of observed points.

The spherical harmonic expansion~\eqref{eq:def-f-ylm} of the observed number
density~$n$ is readily obtained:  inserting the function~\eqref{eq:n} into the
definition~\eqref{eq:def-flm} of the spherical harmonic coefficients, we can
use the defining property of the delta function,
\begin{equation}
\label{eq:nlm}
    n_{lm}
    = \sum_{k} w_k \, Y_{lm}^*(\U_k) \,.
\end{equation}
The spherical harmonic coefficients of the number density~$n$ are hence simply
the weighted, complex-conjugated values of the spherical harmonics in the
observed points.

To compute the angular power spectrum~\eqref{eq:cl} of~$n$ and a second set of
points~$\U'_{k'}$ with weights~$w'_{k'}$ and associated number density~$n'$,
where the two observed sets of points can be one and the same, it suffices to
insert the sum~\eqref{eq:n} of delta functions for~$n$ and~$n'$ into
definition~\eqref{eq:def-cl-ff}, set the spin weights to zero, and carry out
the integration.  The result is
\begin{equation}
\label{eq:cl-nn}
    C_l^{nn'}
    = \frac{1}{4\pi} \sum_{kk'} w_k w'_{k'} \, P_l(\cos\theta_{kk'}) \,,
\end{equation}
where~$P_l = d^l_{00}$ is the Legendre polynomial, and~$\theta_{kk'}$ is the
angle between~$\U_k$ and~$\U'_{k'}$.  This is the exact angular power spectrum
for any two sets of points.

\subsection{Fields}

Next, we consider observations of a set of (complex) function values~$g_k$, $k
= 1, 2, \ldots$, which are observed at points~$\U_k$ on the sphere, and given
weights~$w_k$.  As in the case of the number density~\eqref{eq:n}, we can
construct a spherical function~$f$ from the discrete observations using the
Dirac delta function~$\delta^{\rm D}$,
\begin{equation}
\label{eq:f}
    f(\U)
    = \sum_{k} g_k \, w_k \, \delta^{\rm D}(\U - \U_k) \,,
\end{equation}
where the sum extends over all observed values.  As before, we obtain a
function which is defined over the entire sphere.  The spin weight~$s$ of~$f$
is the sum of the respective spin weights~$s_1$ and~$s_2$ of~$g$ and~$w$:  if a
rotation~$\gamma$ of the sphere in~$\U_k$ transforms~$g_k$ into~$\E^{-\I s_1
\gamma} \, g_k$ and~$w_k$ into~$\E^{-\I s_2 \gamma} \, w_k$, the function
value~$f(\U_k)$ transforms into~$\E^{-\I s \gamma} \, f(\U_k)$.

To compute the spherical harmonic expansion~\eqref{eq:def-f-ylm} of~$f$, it
once again suffices to insert the function~\eqref{eq:f} into the
definition~\eqref{eq:def-flm} of the spherical harmonic coefficient and use the
defining property of the delta function,
\begin{equation}
\label{eq:flm}
    f_{lm}
    = \sum_{k} g_k \, w_k \, {}^{}_sY_{lm}^*(\U_k) \,.
\end{equation}
The spherical harmonic coefficients are therefore the complex conjugate values
of the spin-weighted spherical harmonics in the observed points, multiplied by
the observed values and their weights.

To compute the angular power spectrum of~$f$ and a second, similarly-defined
function~$f'$, where both functions can be one and the same, we proceed as
above, inserting the sum~\eqref{eq:f} of delta functions for~$f$ and~$f'$ into
definition~\eqref{eq:def-cl-ff} and carrying out the integration.  The
resulting angular power spectrum for~$f$ and~$f'$ is
\begin{equation}
\label{eq:cl-ff}
    C_l^{ff'}
    = \frac{1}{4\pi} \sum_{kk'} g_k^* \, g'_{k'} \, w_k \, w'_{k'} \,
        \E^{\I s \alpha_{kk'}} \, d^l_{ss'}(\theta_{kk'}) \,
        \E^{-\I s' \alpha'_{kk'}} \,,
\end{equation}
where the angles~$\alpha_{kk'}, \theta_{kk'}, \alpha'_{kk'}$ are defined for
each pair of points~$\U_k, \U'_{k'}$ as in definition~\eqref{eq:cf}.  This is
the exact angular power spectrum given two discrete sets of observed values on
the sphere.

The discrete angular power spectrum~\eqref{eq:cl-ff} demonstrates the
equivalence between harmonic and real space nicely:  it is equivalent to the
well-known real-space estimator \citep{2002A&A...396....1S}, transformed pair
by pair to harmonic space using the transformation~\eqref{eq:cl-cf}.  In
practice, however, the two measurements do contain slightly different
information, since we cannot obtain them over all angular scales, which would
be required to carry out the transformation mathematically.

\subsection{Cross-correlations}

Finally, we can consider the case where we wish to obtain the two-point
statistics between discrete sets of measured points and measured function
values.  Following the preceding sections, we can construct spherical
functions~$n$ and~$f'$ using the sums~\eqref{eq:n} and~\eqref{eq:f} of delta
functions, respectively.  The angular power spectrum is once again obtained by
inserting~$n$ and~$f'$ into definition~\eqref{eq:def-cl-ff} and integrating out
the delta functions,
\begin{equation}
\label{eq:cl-nf}
    C_l^{nf'}
    = \frac{1}{4\pi} \sum_{kk'} g'_{k'} \, w_k \, w'_{k'} \,
        d^l_{0s'}(\theta_{kk'}) \,
        \E^{-\I s' \alpha'_{kk'}} \,,
\end{equation}
where~$s'$ is the spin weight of~$f'$, and the angles~$\theta_{kk'},
\alpha'_{kk'}$ are defined as above.  Naturally, the result~\eqref{eq:cl-nf} is
merely the special case of the angular power spectrum~\eqref{eq:cl-ff} when
setting~$f = n$, and hence $g_k \equiv 1$ and $s = 0$.

\section{Expectations}
\label{sec:expectations}

We are now able to compute expectations of the angular power
spectra~\eqref{eq:cl-nn}, \eqref{eq:cl-ff}, and~\eqref{eq:cl-nf} when the
observations are random variates, such as the cosmological data observed by
\Euclid.  We once again have to distinguish the cases where we observe points
(e.g., galaxy positions) and fields (e.g., cosmic shear).  In the first case,
we have two-point statistics from observed points, which are generated by point
processes on the sphere.  In the second case, we have two-point statistics from
observed function values, which are generated by random fields.

There is a subtle difference between point processes and random fields beyond
the fact that we observe positions for one and function values for the other.
It is encoded in what will be called Assumption~\ref{asm-0} and
Assumption~\ref{asm-4} below:  to compute an expectation for point processes,
we must allow the random positions to vary.  This means that we require a
priori information about the probability of observing a point anywhere on the
sphere.  For random fields, we are instead able to compute expectations
conditional on the observed positions and weights.

\subsection{Point processes, angular clustering}
\label{sec:point-processes}

For observations generated by point processes, we compute the expectation of
the angular power spectrum~\eqref{eq:cl-nn} for the observed number
densities~$n$ and~$n'$.  To do so, the sum in expression~\eqref{eq:cl-nn} is
split into separate sums over the set of true pairs of distinct points (denoted
here by~$k \nequiv k'$, meaning~$\U_k$ and~$\U'_{k'}$ are not the same observed
point) and over the set of degenerate pairs of identical points (denoted by~$k
\equiv k'$, meaning~$\U_k$ and~$\U'_{k'}$ are the same observed point),
\begin{equation}
\label{eq:cl-nn:1}
    C_l^{nn'}
    = \frac{1}{4\pi} \, \Biggl\{\,
        \sum_{k \nequiv k'} + \sum_{k \equiv k'}
    \,\Biggr\} \, w_k w'_{k'} \, P_l(\cos\theta_{kk'}) \,.
\end{equation}
The second sum is sometimes empty, but not always, e.g., when computing an
auto-correlation, where~$n$ and~$n'$ describe the same observation.
Since~$\theta_{kk'} = 0$ for~$k \equiv k'$, the second sum contains only
$P_l(1) = 1$, and reduces to the total weight of degenerate pairs of points
in~$n$ and~$n'$, for which we define
\begin{equation}
\label{eq:A-nn}
    A^{nn'}
    = \frac{1}{4\pi} \sum_{k \equiv k'} w_k w'_{k'} \,.
\end{equation}
For an auto-correlation, $A^{nn'}$ is simply the total squared weight.
Overall, we thus find that the angular power spectrum~\eqref{eq:cl-nn} can be
written as
\begin{equation}
\label{eq:cl-nn-split}
    C_l^{nn'}
    = \frac{1}{4\pi} \sum_{k \nequiv k'} w_k w'_{k'} \, P_l(\cos\theta_{kk'})
    + A^{nn'} \,,
\end{equation}
where the remaining sum contains the two-point statistics from true pairs of
distinct points.  The term~$A^{nn'}$ is an additive bias from degenerate pairs
of identical points, which is often called the ``noise bias''.  However, even
though~$A^{nn'}$ is a stochastic quantity over realisations of the point
processes, for any given realisation of points, the bias~\eqref{eq:A-nn} is
evidently a known number that we can compute exactly.

We thus subtract~$A^{nn'}$ from both sides of expression~\eqref{eq:cl-nn-split}
and compute the expectation of the bias-subtracted angular power spectrum,
\begin{equation}
\label{eq:ev-cl-nn:0}
    \ev{C_l^{nn'} - A^{nn'}}
    = \frac{1}{4\pi} \, \ev[\Bigg]{
        \,\sum_{k \nequiv k'} w_k w'_{k'} \, P_l(\cos\theta_{kk'})
    } \,.
\end{equation}
Our goal is to express this expectation in terms of the intrinsic two-point
statistics of the point process.  The main difficulty lies in the fact that we
may not have a complete sample of observations;  for example, because we were
only able to observe part of the sphere, as happens in any galaxy survey such
as \Euclid.  In addition, there may be complicated observational effects at
play, which result in some random points being missed even within the survey
footprint.  Any systematic removal of points from our sample affects the
observed two-point statistics, and must hence be carefully taken into account.

To compute the expectation~\eqref{eq:ev-cl-nn:0} with missing observations and
systematic effects, we set~$w_k = 0$ and~$w'_{k'} = 0$ for all unobserved
points~$\U_k$ and~$\U'_{k'}$ in the (unknown) complete sample.  We can then
extend the sum in expression~\eqref{eq:ev-cl-nn:0} to all points generated by
the point process, both observed and unobserved, without changing its value,
\begin{equation}
\label{eq:ev-cl-nn:1}
    \ev{C_l^{nn'} - A^{nn'}}
    = \frac{1}{4\pi} \, \ev[\Bigg]{
        \,\smashoperator[r]{\sum_{\text{all } k \nequiv k'}}
            w_k w'_{k'} \, P_l(\cos\theta_{kk'})
    } \,.
\end{equation}
Naturally, we have no knowledge about the unobserved points~$\U_k$ in the sum,
but that will not be a problem for computing the expectation.

Since each realisation of the point process yields a different set of observed
points, the weights~$w_k$ and~$w'_{k'}$ in the
expectation~\eqref{eq:ev-cl-nn:1} are themselves random variables.  We can use
the law of total expectation to compute the expectation of~$w_k$ conditional
on~$\U_k$, by making
\begin{assumption}
\label{asm-0}
    There exist functions~$v$ and~$v'$ that describe the expected weight
    conditional on the observed position,
    \begin{equation}
    \label{eq:wk-v}
        \ev{w_k \mid \U_k} = v(\U_k) \,,
    \end{equation}
    and similarly $\ev{w'_{k'} \mid \U'_{k'}} = v'(\U'_{k'})$.
\end{assumption}
We call~$v$ and~$v'$ the (weighted) `visibility' of the respective observation;
for unit weights, the value~$v(\U)$ is a number between~$0$ and~$1$ that
describes the a priori probability that a point sampled in a given
position~$\U_k$ is observed.  For general weights, the expectation is also
taken over realisations of their values.  In practice, estimating the
visibility of a galaxy imaging survey is an open problem, and the subject of
ongoing research \citep{2021A&A...648A..98J, 2022MNRAS.511.2665R}.

Using the visibility~\eqref{eq:wk-v}, the expectation~\eqref{eq:ev-cl-nn:1} no
longer depends on the exact set of observed points,
\begin{equation}
\label{eq:ev-cl-nn:2}
    \ev{C_l^{nn'} - A^{nn'}}
    = \frac{1}{4\pi} \, \ev[\Bigg]{
        \,\smashoperator[r]{\sum_{\text{all } k \nequiv k'}}
            v(\U_k) \, v'(\U'_{k'}) \, P_l(\cos\theta_{kk'})
        } \,.
\end{equation}
In fact, the expectation of the sum in expression~\eqref{eq:ev-cl-nn:2} depends
solely on the pairs of points~$\U_k, \U'_{k'}$ in a given realisation.  We can
hence make
\begin{assumption}
\label{asm-1}
    All observed pairs of points have the same a priori distribution.
\end{assumption}
This is a weak assumption, since it is difficult to imagine how any specific
pair of points in a realisation might be a priori distinguishable from the
rest.

Under Assumption~\ref{asm-1}, all terms in the sum in
expression~\eqref{eq:ev-cl-nn:2} have the same expectation.  If~$N$ and~$N'$
are the respective total number of points for each point process, there
are~$NN'$ pairs of points,\footnote{%
    For simplicity, we use~$NN'$ for the number of pairs here, while the true
    number of pairs might be slightly different, e.g., $N(N-1)$ for an
    auto-correlation.  If the difference is significant, one can introduce a
    pair count correction factor.}
and hence terms in the sum.  Introducing functions~$\bar{n}$ and~$\bar{n}'$
with
\begin{equation}
\label{eq:nbar}
    \bar{n}(\U)
    = \frac{N}{4\pi} \, v(\U) \,,
\end{equation}
and similarly $\bar{n}'(\U') = (4\pi)^{-1} N' \, v'(\U')$, the
expectation~\eqref{eq:ev-cl-nn:2} is
\begin{equation}
\label{eq:ev-cl-nn:3}
    \ev{C_l^{nn'} - A^{nn'}}
    = 4\pi \, \ev[\big]{
        \bar{n}(\U) \, \bar{n}'(\U') \, P_l(\cos\theta)
    } \,,
\end{equation}
where~$\U, \U'$ is a pair of random points, and~$\theta$ is the angular
separation between them.  The functions~$\bar{n}$ and~$\bar{n}'$ can be
understood as the position-dependent mean density of the observed points,
taking the visibility into account.  This has the conceptual advantage that we
never have to define the exact sample of points to which~$N$ and~$v(\U)$ refer,
which would be difficult for \Euclid with its complicated coverage from
different ground-based surveys.

The remaining expectation on the right-hand side of
expression~\eqref{eq:ev-cl-nn:3} contains two random effects: one is the
angular distribution of points, and the other is the random realisation of the
mean densities.  Here, we are only interested in the former, and we therefore
make
\begin{assumption}
\label{asm-2}
    The expected angular power spectrum is conditional on the observed
    densities of points.
\end{assumption}
To see why the expectation over realisations with varying density is not very
interesting, one can imagine a point process where the distribution of points
is smoother or clumpier depending on the realised density.  In that case, the
expected two-point statistics over all densities can be arbitrarily different
from the expectation conditional on the observed density, and we can extract
essentially no information from our measurement.  We hence want to compute an
expectation that is ``close'' to our observation, except for the angular
distribution of the points.  This also agrees with intuition, since the
(conditional) expectation of the observed density~\eqref{eq:n} over
realisations of positions is then equal to the mean density~\eqref{eq:nbar},
\begin{equation}
\label{eq:ev-n}
    \ev[\big]{n(\U)}
    = \bar{n}(\U) \,.
\end{equation}
However, our assumption comes with two important caveats:  firstly, for galaxy
clustering, the number of galaxies (as well as their weights, if given) will
depend to some degree on the underlying realisation of the universe, and we are
hence assuming that this correlation can be neglected.  Secondly, in practice,
we have no a priori knowledge about the mean density~$\bar{n}$, and we must
hence estimate it from the observations themselves.  We will check the impact
of the latter point in Sect.~\ref{sec:validation}.

Using Assumption~\ref{asm-2}, only the expectation over positions remains in
expression~\eqref{eq:ev-cl-nn:3}, which is a double integral over the sphere,
\begin{equation}
\label{eq:ev-twopos}
\begin{split}
    &\ev[\big]{\bar{n}(\U) \, \bar{n}'(\U') \, P_l(\cos\theta)} \\
    &\qquad = \iint \! \bar{n}(\U) \, \bar{n}'(\U') \, P_l(\cos\theta)
                        \, p(\U, \U') \, \D\U \, \D\U' \,,
\end{split}
\end{equation}
with~$p(\U, \U') \, \D\U \, \D\U'$ the a priori probability of the point
process to generate a pair of points in~$\D\U \, \D\U'$.  In the general case,
this integral must be evaluated explicitly.  But for the point processes in
which we are interested here, we can make
\begin{assumption}
\label{asm-3}
    The point processes are homogeneous on the sphere, i.e., their distribution
    is unchanged under rotations of the sphere.
\end{assumption}
For galaxy clustering, this assumption is usually granted by the ``cosmological
principle''.

Under Assumption~\ref{asm-3}, the joint probability density~$p(\U, \U')$ in the
integral~\eqref{eq:ev-twopos} depends only on the angular distance~$\theta$
between the pair of points~$\U$ and~$\U'$.  It can be written as
\citep{1993ApJ...412...64L, 1973ApJ...185..413P}
\begin{equation}
\label{eq:p-nn}
    p(\U, \U')
    = \frac{1 + \w(\theta)}{(4\pi)^2} \,,
\end{equation}
where~$\w$ is the expected angular correlation function of density fluctuations
in the observed point processes, which describes the clustering of
points.\footnote{%
    The angular correlation function~$\w$ is not to be confused with the
    weight~$w$.}

Inserting the integral~\eqref{eq:ev-twopos} and joint probability
density~\eqref{eq:p-nn} into expectation~\eqref{eq:ev-cl-nn:3}, we find that
only~$\bar{n}$ and~$\bar{n}'$ depend explicitly on the positions~$\U$
and~$\U'$, while everything else depends on the angular separation~$\theta$
alone,
\begin{equation}
\label{eq:ev-cl-nn:4}
    \ev{C_l^{nn'} - A^{nn'}}
    = \frac{1}{4\pi}
        \iint \! \bar{n}(\U) \, \bar{n}'(\U') \,
        \bigl[1 + \w(\theta)\bigr] \, P_l(\cos\theta) \, \D\U \, \D\U' \,.
\end{equation}
Writing the double integral over the sphere in terms of the angular
separation~$\theta$ recovers precisely the definition~\eqref{eq:cf} of the
angular correlation function~$C^{\bar{n}\bar{n}'}$,
\begin{equation}
\label{eq:ev-cl-nn:5}
    \ev{C_l^{nn'} - A^{nn'}}
    = 2\pi \int_{0}^{\pi} \! C^{\bar{n}\bar{n}'}(\theta) \,
        \bigl[1 + \w(\theta)\bigr] \, P_l(\cos\theta) \sin(\theta) \,
        \D\theta \,.
\end{equation}
Integrating the two terms of~$1 + \w(\theta)$ separately, the former is the
transformation~\eqref{eq:cl-cf} from~$C^{\bar{n}\bar{n}'}(\theta)$
to~$C_l^{\bar{n}\bar{n}'}$, while the latter is the
convolution~\eqref{eq:ev-cl-cf-prod} of~$C^{\bar{n}\bar{n}'}(\theta)$
and~$\w(\theta)$, which we can write in the form of a mixing matrix
product~\eqref{eq:ev-cl-mm}.  Overall, we can hence write the
expectation~\eqref{eq:ev-cl-nn:5} as
\begin{equation}
\label{eq:ev-cl-nn}
    \ev{C_l^{nn'} - A^{nn'}}
    = C_l^{\bar{n}\bar{n}'}
        + \sum_{l_1} M_{ll_1}^{\bar{n}\bar{n}'} \, \w_{l_1} \,,
\end{equation}
where~$M_{ll_1}^{\bar{n}\bar{n}'}$ is the mixing matrix~\eqref{eq:mm} due to
the mean density functions~$\bar{n}$ and~$\bar{n}'$, and~$\w_l$ is the angular
power spectrum of the point processes, obtained from the intrinsic angular
correlation function~$\w$ using relation~\eqref{eq:cl-cf}.

We hence find that the expectation~\eqref{eq:ev-cl-nn} contains the desired
intrinsic two-point statistics of the point processes, in the form of~$\w_l$.
However, the signal is doubly contaminated when the mean densities~$\bar{n}$
and~$\bar{n}'$ contain systematic variations, by both the angular power
spectrum~$C_l^{\bar{n}\bar{n}'}$ and by the associated mixing
matrix~$M_{ll_1}^{\bar{n}\bar{n}'}$.  To remove these contaminations, we can
directly manipulate expression~\eqref{eq:ev-cl-nn} until it yields an estimator
for~$\w_l$.  While this approach is somewhat unusual in harmonic space, we show
in Appendix~\ref{sec:alt-ang-clus} that it recovers well-known results from
real space, such as the estimator of \citet{1993ApJ...412...64L}.

In what follows, we focus instead on the more traditional approach for
isolating the signal~$\w_l$ in the expectation~\eqref{eq:ev-cl-nn}.  We
directly construct spherical functions~$\delta$ and~$\delta'$ for the number
density contrast of the observed points,
\begin{equation}
\label{eq:delta}
    \delta(\U)
    = \frac{n(\U) - \bar{n}(\U)}{\bar{n}_0} \,,
\end{equation}
and equivalently for~$\delta'(\U')$, where~$\bar{n}_0 = N/(4\pi)$ denotes the
total mean density over the sphere.  Note that we divide here by a constant,
and not by the function~$\bar{n}$;  for the alternative case, see
Appendix~\ref{sec:alt-ang-clus}.  The density contrast~\eqref{eq:delta} is
hence a linear combination of the spherical functions~$n$ and~$\bar{n}$, and
it follows that the angular power spectrum of~$\delta$ and~$\delta'$ is
\begin{equation}
\label{eq:cl-dd}
    C_l^{\delta\delta'}
    = \frac{
        C_l^{nn'} - C_l^{\bar{n}n'} - C_l^{n\bar{n}'} + C_l^{\bar{n}\bar{n}'}
    }{
        \bar{n}_0 \bar{n}_0'
    } \,.
\end{equation}
We therefore find that measuring the angular power spectrum of the number
density contrast~\eqref{eq:delta} yields a result that is equivalent to the
partial-sky harmonic-space Landy--Szalay estimator~\eqref{eq:est-ls-part}.  The
expectation~$\ev{C_l^{\delta\delta'}}$ of the angular power
spectrum~\eqref{eq:cl-dd} is readily computed using
expressions~\eqref{eq:nbar}, \eqref{eq:ev-n}, and~\eqref{eq:ev-cl-nn},
\begin{equation}
\label{eq:ev-cl-dd}
    \ev{C_l^{\delta\delta'} - A^{\delta\delta'}}
    = \sum_{l_1} M_{ll_1}^{vv'} \, \w_{l_1} \,,
\end{equation}
where~$M^{vv'}$ is the mixing matrix for the visibilities~$v$ and~$v'$,
and~$A^{\delta\delta'} = (\bar{n}_0 \bar{n}_0')^{-1} A^{nn'}$ is the rescaled
additive bias.

\subsection{Random fields, cosmic shear}
\label{sec:random-fields}

The second case of interest is where we observe values~$g_k$ which are the
variates of an underlying random field, and use them to construct a spherical
function~$f$ using definition~\eqref{eq:f}.  If there is a spherical
function~$g$ such that the observations are the function values~$g_k = g(\U_k)$
of~$g$ in the observed points, we can use the defining property of the delta
function to factor~$g(\U)$ out of the sum in definition~\eqref{eq:f},
\begin{equation}
    f(\U)
    = g(\U) \sum_{k} w_k \, \delta^{\rm D}(\U - \U_k) \,.
\end{equation}
For the remaining sum, we introduce a spherical function~$w$, which we call the
`weight function' of the random field,
\begin{equation}
\label{eq:w}
    w(\U)
    = \sum_{k} w_k \, \delta^{\rm D}(\U - \U_k) \,.
\end{equation}
We can therefore write~$f(\U) = g(\U) \, w(\U)$, and understand our constructed
function~$f$ as the product of the function~$g$ under observation and a weight
function~$w$ that encodes where and how well~$g$ has been observed.  While the
visibility~\eqref{eq:wk-v} of a point process is, firstly, an expectation and,
secondly, usually a relatively smooth function over the sphere, the weight
function~\eqref{eq:w} of a random field consists of the given weights~$w_k$ in
the observed positions~$\U_k$.

If the function~$g$ is the realisation of a random field, we want to use the
mixing matrix formalism~\eqref{eq:ev-cl-mm} to compute the expectation of the
angular power spectrum~\eqref{eq:cl-ff} of~$f$ and a second such function~$f'$
with~$f'(\U') = g'(\U') \, w'(\U')$.  To this end, we firstly require
\begin{assumption}
\label{asm-7}
    The functions~$g$ and~$g'$ are realisations of jointly homogeneous random
    fields.
\end{assumption}
In the case of cosmic shear, this is once again a reasonable assumption by the
cosmological principle.  To apply the mixing matrix formalism, we further
require
\begin{assumption}
\label{asm-4}
    The distribution of observed values~$g_k$ is conditional on the observed
    positions~$\U_k$ and weights~$w_k$.
\end{assumption}
For cosmic shear, this assumption implies two approximations.  Firstly, it
ignores that the positions of galaxies are slightly correlated with their
shears \citep[source--lens clustering,][]{2024arXiv240709810L}, since both
positions and shears are ultimately connected to the large-scale structure of
the universe.  Secondly, the weights and values of shear observations are
generally also slightly correlated, since more extreme galaxy shapes are harder
to measure accurately, and thus given lower weights.

Under Assumptions~\ref{asm-7} and~\ref{asm-4}, only the functions~$g$ and~$g'$
are considered realisations of (homogeneous) random fields when computing the
expectation of the angular power spectrum~\eqref{eq:cl-ff} for~$f = g \, w$
and~$f' = g' \, w'$, while~$w$ and~$w'$ are considered fixed functions.  We can
hence use the mixing matrix formalism~\eqref{eq:ev-cl-mm} to obtain the
expected angular power spectrum of~$f$ and~$f'$,
\begin{equation}
\label{eq:ev-cl-ff}
    \ev{C_l^{ff'}}
    = \sum_{l_1} M_{ll_1}^{ww'} \, \ev{C_{l_1}^{gg'}} \,,
\end{equation}
where the mixing matrix is computed for the weight functions~$w$ and~$w'$ of
point masses following definition~\eqref{eq:w}.

The situation is slightly more complicated if we observe the field~$g$ only
indirectly via some intermediary observable.  For cosmic shear, that observable
is the galaxy ellipticity~$\epsilon_k$, which probes the cosmic shear field
through the effect of weak gravitational lensing on the intrinsic galaxy shapes
\citep[e.g.,][]{2001PhR...340..291B},
\begin{equation}
\label{eq:shear}
    \epsilon_k
    = \frac{\epsilon^{\rm i}_k + g_k}{1 + g_k^* \epsilon^{\rm i}_k} \,,
\end{equation}
where~$\epsilon^{\rm i}_k$ is the intrinsic galaxy ellipticity that would have
been observed without gravitational lensing.  We say that the
ellipticity~$\epsilon_k$ traces the cosmic shear field~$g$, because the
conditional expectation of~$\epsilon_k$ for a fixed value~$g_k$ and random
orientations of the galaxy is \citep{1997A&A...318..687S}
\begin{equation}
    \ev[\big]{\epsilon_k \bigm\vert g_k} = g_k \,.
\end{equation}
However, even though the observed ellipticity is an unbiased estimate of the
cosmic shear field, the intrinsic variability of galaxy shapes leads to an
increase in variance compared to the pure cosmic shear signal,
\begin{equation}
\label{eq:ev-eps2}
    \ev[\big]{|\epsilon_k|^2}
    = \ev[\big]{|g_k|^2}
    + \ev[\Bigg]{
        |\epsilon^{\rm i}_k|^2 \,
        \frac{\bigl(1 - |g_k|^2\bigr)^2}{1 - |g_k|^2 \, |\epsilon^{\rm i}_k|^2}
    } \,.
\end{equation}
The second term in expectation~\eqref{eq:ev-eps2} is an additional variance
commonly called ``shape noise'', and we see that the effect depends on both the
variance of the intrinsic galaxy ellipticity and the one-point statistics of
the cosmic shear field.  In practice, there is a further contribution to shape
noise due to the variance from imperfect shape measurement.

To understand the impact of noise on the expected angular power spectrum of a
random field~$g$, we make
\begin{assumption}
\label{asm-5}
    Observed values of the random field~$g$ have independent noise
    contributions.
\end{assumption}
Taken in isolation, this is not a good approximation for the shape noise of
cosmic shear, since galaxies have intrinsic alignments
\citep{2015SSRv..193....1J, 2015SSRv..193...67K, 2015SSRv..193..139K,
2015PhR...558....1T}.  However, intrinsic alignments are generally absorbed
into the theoretical prediction of the cosmic shear signal, so that our
assumption is effectively a statement about our capability to model this
effect.

Under Assumption~\ref{asm-5}, the expectation~\eqref{eq:ev-cl-ff} of the
angular power spectrum does not change its signal content, but picks up an
additional variance term,
\begin{equation}
\label{eq:ev-cl-ff-noi}
    \ev{C_l^{ff'}}
    = \sum_{l_1} M_{ll_1}^{ww'} \, \ev{C_{l_1}^{gg'}}
    + A^{ff'} \,,
\end{equation}
where~$A^{ff'}$ is the additive bias due to the noise variance~$\sigma^2_{kk'}$
from degenerate pairs of identical objects (denoted as before by $k \equiv
k'$),\footnote{%
    Here, degenerate pairs refer to observations of the same random field
    value~$g_k \equiv g'_{k'}$.  Apart from auto-correlations, such pairs also
    arise, e.g., for cosmic shear when one set of galaxies is observed with two
    different shape measurement methods, where it may be the case that~$g_k
    \equiv g'_{k'}$ but~$\epsilon_k \ne \epsilon'_{k'}$ and~$w_k \ne w'_{k'}$.}
\begin{equation}
\label{eq:A-ff}
    A^{ff'}
    = \delta^{\rm K}_{ss'} \,
        \frac{1}{4\pi} \sum_{k \equiv k'} w_k w'_{k'} \, \sigma^2_{kk'} \,,
\end{equation}
where~$s$ and~$s'$ are the spin weights of~$f$ and~$f'$, respectively, as
before.  For random fields, the additive bias~$A^{ff'}$ is therefore a true
``noise bias'', in the sense that it is the expectation of a stochastic noise
contribution, unlike the additive bias~$A^{nn'}$ of the point process, which is
a known number for each realisation.

For cosmic shear, we do not know, a priori, the additional
variance~$\sigma^2_{kk'}$ due to shape noise for each observed value~$g_k$
or~$g'_{k'}$.  In that situation, we can construct an
estimate~$\mathcal{A}^{ff'}$ of the additive bias from the variance of the
noisy observations \citep{2021JCAP...03..067N},
\begin{equation}
\label{eq:est-A-ff}
    \mathcal{A}^{ff'}
    = \delta^{\rm K}_{ss'} \, \frac{1}{4\pi}
        \sum_{k \equiv k'} w_k w'_{k'} \, \epsilon_k^* \epsilon'_{k'} \,.
\end{equation}
By expectation~\eqref{eq:ev-eps2}, this is a biased estimator for a
non-vanishing~$A^{ff}$, since it contains not only the variance due to shape
noise, but the sum of intrinsic and noise variance,
\begin{equation}
\label{eq:ev-est-A-ff}
    \ev{\mathcal{A}^{ff'}}
    = A^{ff'}
    + \delta^{\rm K}_{ss'} \,
        \frac{1}{4\pi} \sum_{k \equiv k'} w_k w'_{k'} \, \ev{C^{gg'}(0)} \,,
\end{equation}
where the expected zero-lag angular correlation~$\ev{C^{gg'}(0)}$ is the
intrinsic variance of the random fields~$g$ and~$g'$.
Subtracting~$\mathcal{A}^{ff'}$ from the measured angular power
spectrum~$C_l^{ff'}$ and taking the expectation using
expressions~\eqref{eq:ev-cl-ff-noi} and~\eqref{eq:ev-est-A-ff} , we obtain
\begin{equation}
\begin{split}
\label{eq:ev-cl-ff-debias}
    \ev{C_l^{ff'} - \mathcal{A}^{ff'}}
    &= \sum_{l_1} M_{ll_1}^{ww'} \, \ev{C_{l_1}^{gg'}}
    \\ & \quad
    - \delta^{\rm K}_{ss'} \,
        \frac{1}{4\pi} \sum_{k \equiv k'} w_k w'_{k'} \, \ev{C^{gg'}(0)} \,.
\end{split}
\end{equation}
Noting that the two-point statistics of~$g$ and~$g'$ enter both terms on the
right-hand side of the expectation, we use relation~\eqref{eq:cf-cl} and the
properties of the Wigner $d$~function to replace~$\ev{C^{gg'}(0)}$ by a sum
over the expected angular power spectrum,
\begin{equation}
\label{eq:ev-cf-gg-0}
    \ev{C^{gg'}(0)}
    = \delta^{\rm K}_{s_1s_1'}
        \sum_{l} \frac{2l + 1}{4\pi} \, \ev{C_l^{gg'}} \,,
\end{equation}
with~$s_1$ and~$s_1'$ the respective spin weights of~$g$ and~$g'$, as above.
The expectation~\eqref{eq:ev-cl-ff-debias} is therefore equivalent to
\begin{equation}
\label{eq:ev-cl-ff-debias-mm}
    \ev{C_l^{ff'} - \mathcal{A}^{ff'}}
    = \sum_{l_1} \mathcal{M}^{ww'}_{ll_1} \, \ev{C_{l_1}^{gg'}} \,,
\end{equation}
where we have introduced the reduced mixing matrix
\begin{equation}
\label{eq:reduced-mm-def}
    \mathcal{M}_{ll_1}^{ww'}
    = M_{ll_1}^{ww'}
      - \delta^{\rm K}_{ss'} \, \delta^{\rm K}_{s_1s_1'} \,
        \frac{2l_1 + 1}{4\pi} \,
        \frac{1}{4\pi} \sum_{k \equiv k'} w_k w'_{k'} \,.
\end{equation}
In the expectation~\eqref{eq:ev-cl-ff-debias-mm}, the bias introduced
by~$\mathcal{A}^{ff'}$ is thus completely absorbed
into~$\mathcal{M}_{ll_1}^{ww'}$.

As it turns out, the reduced mixing matrix has a much simpler interpretation
than the definition~\eqref{eq:reduced-mm-def} suggests.  Consider the angular
power spectrum~\eqref{eq:cl-ff} for the pair of weight functions~$w$ and~$w'$
with respective spin weights~$s_2$ and~$s_2'$.  Following
expression~\eqref{eq:cl-nn:1}, we split~$C_l^{ww'}$ into contributions from
true pairs of distinct points ($k \nequiv k'$) and degenerate pairs of
identical points ($k \equiv k'$), so that we may define the known additive
bias~$A^{ww'}$ for the weight functions~$w$ and~$w'$,
\begin{equation}
    A^{ww'}
    = \delta^{\rm K}_{s_2s_2'} \,
        \frac{1}{4\pi} \sum_{k \equiv k'} w_k w'_{k'} \,.
\end{equation}
Since~$s = s_1 + s_2$ and~$s' = s_1' + s_2$, we can substitute~$A^{ww'}$ for
the sum in expression~\eqref{eq:reduced-mm-def},
\begin{equation}
\label{eq:reduced-mm:0}
    \mathcal{M}_{ll_1}^{ww'}
    = M_{ll_1}^{ww'}
      - \delta^{\rm K}_{ss'} \, \delta^{\rm K}_{s_1s_1'} \,
        \frac{2l_1 + 1}{4\pi} \, A^{ww'} \,.
\end{equation}
Furthermore, we can substitute the Kronecker symbols by an identity for the
Wigner $3j$ symbols,
\begin{equation}
    \delta^{\rm K}_{ss'} \, \delta^{\rm K}_{s_1s_1'}
    = \sum_{l_2\sigma} (2l_2 + 1) \, \threej{l,-s}{l_1,s_1}{l_2,\sigma} \,
        \threej{l,-s'}{l_1,s_1'}{l_2,\sigma} \,.
\end{equation}
Using the fact that~$A^{ww'}$ vanishes unless~$s_2 = s_2'$, an equivalent way
to write expression~\eqref{eq:reduced-mm:0} is therefore
\begin{equation}
\begin{split}
\label{eq:reduced-mm:1}
    \mathcal{M}_{ll_1}^{ww'}
    = M_{ll_1}^{ww'}
    - \delta^{\rm K}_{ss'}
        \smash[b]{\sum_{l_2}}
        & \frac{(2l_1 + 1) \, (2l_2 + 1)}{4\pi} \, A^{ww'} \,
        \\ & \; \times
            \threej{l_1,s_1}{l_2,s_2}{l,-s} \,
            \threej{l_1,s_1'}{l_2,s_2'}{l,-s'} \,.
\end{split}
\end{equation}
Comparing the result to the definition~\eqref{eq:mm} of the mixing matrix, we
indeed obtain a straightforward interpretation of the reduced mixing matrix,
\begin{equation}
\begin{split}
\label{eq:reduced-mm}
    \mathcal{M}_{ll_1}^{ww'}
    = (-1)^{s-s'} \smash[b]{\sum_{l_2}} & \frac{(2l_1+1) \, (2l_2+1)}{4\pi}
        \, \Bigl(C_{l_2}^{ww'} - \delta^{\rm K}_{ss'} \, A^{ww'}\Bigr)
        \\ & \; \times
        \threej{l_1,s_1}{l_2,s_2}{l,-s} \,
        \threej{l_1,s_1'}{l_2,s_2'}{l,-s'} \,.
\end{split}
\end{equation}
In other words, the reduced mixing matrix is the mixing matrix of the angular
power spectrum~$C_l^{ww'}$ with its additive bias~$A^{ww'}$ subtracted.

To summarise, we obtain the following four key results.  For noisy observations
where the additive bias to the angular power spectrum is not known, which is
the case for cosmic shear, we can construct the estimate~\eqref{eq:est-A-ff}
using the variance of the noisy observations.  Subtracting the estimated
additive bias from the measured angular power spectrum leads to a biased
expectation~\eqref{eq:ev-cl-ff-debias} with respect to the mixing matrix
formalism, since the estimate contains not only the additional variance due to
noise, but also the intrinsic variance of the fields.  However, we can return
the expectation~\eqref{eq:ev-cl-ff-debias-mm} to standard form by introducing a
reduced mixing matrix, which implicitly removes the intrinsic variance of the
random fields from the expected angular power spectrum.  Finally, the reduced
mixing matrix~\eqref{eq:reduced-mm} is simply the mixing matrix with the
additive bias of the weight functions, which is a known number, subtracted.

The nature of this correction becomes clear in real space.  The unknown noise
variance~$\sigma^2$ is a delta-like contribution to the expected angular
correlation function of the random fields,
\begin{equation}
    \ev{C^{gg'}(\theta)}
    \mapsto \ev{C^{gg'}(\theta)}
    + \sigma^2 \, \delta^{\rm D}(\cos\theta - \cos0) \,.
\end{equation}
Subtracting the additive bias from the angular power spectrum is equivalent to
subtracting the variance, which is the zero-lag correlation, from the angular
correlation function.  There is hence a correspondence
\begin{equation}
    C_l^{ff'} - \mathcal{A}^{ff'}
    \Longleftrightarrow
    C^{ff'}(\theta)
    - C^{ff'}(0) \, \delta^{\rm D}(\cos\theta - \cos0)
\end{equation}
for the random fields, and
\begin{equation}
    C_l^{ww'} - A^{ww'}
    \Longleftrightarrow
    C^{ww'}(\theta)
    - C^{ww'}(0) \, \delta^{\rm D}(\cos\theta - \cos0)
\end{equation}
for the weight functions.  By expectation~\eqref{eq:ev-cf-prod}, the real-space
equivalent of the reduced mixing matrix
expectation~\eqref{eq:ev-cl-ff-debias-mm} is hence
\begin{multline}
    \ev[\Big]{
        C^{ff'}(\theta)
        - C^{ff'}(0) \, \delta^{\rm D}(\cos\theta - \cos0)
    } \\
    \quad= \Bigl[
        \ev{C^{gg'}(\theta)} + \sigma^2 \, \delta^{\rm D}(\cos\theta - \cos0)
    \Bigr] \\
    \times \Bigl[
        C^{ww'}(\theta) - C^{ww'}(0) \, \delta^{\rm D}(\cos\theta - \cos0)
    \Bigr] \,,
\end{multline}
where we can evaluate the right-hand side for all~$\theta \ge 0$ without
knowing the value of~$\sigma^2$.

\subsection{Cross-correlations, galaxy--galaxy lensing}
\label{sec:cross-correlations}

The final case of interest is the cross-correlation of points~$\U_k$ generated
by a point process, and observed values~$g'_{k'} = g(\U'_{k'})$ from the
realisation~$g$ of a random field.  The two observations define the spherical
functions~$n$ and~$f'$ as above.

For the expectation of the angular power spectrum~\eqref{eq:cl-nf} of~$n$
and~$f'$, we again fundamentally rely on Assumption~\ref{asm-4}:  the
distribution of observed values~$g'_{k'}$ is conditional on the observed
points~$\U_k$ and weights~$w'_{k'}$, which are held fixed.  We assume that this
remains true even when correlating positions and values from a single
observation, in which case the observed positions are both random variates
(within~$n$) and fixed (within~$w'$ and hence~$f'$).  For galaxy--galaxy
lensing, the approximation performs worse than for cosmic shear; this is seen
in Sect.~\ref{sec:validation}.  As in the case of intrinsic alignments, the
assumption is therefore effectively a statement about our ability to model the
effect of source--lens clustering in the theory part of the expectation.

To treat the point process in the expectation of the angular power
spectrum~\eqref{eq:cl-nf}, we proceed as before.  We extend the sum over~$k$ to
all points using the visibility~\eqref{eq:wk-v}, and replace~$w_k$ by~$v(\U_k)$
under Assumption~\ref{asm-0},
\begin{equation}
\label{eq:ev-cl-nf:0}
    \ev{C_l^{nf'}}
    = \frac{1}{4\pi} \, \ev[\Bigg]{\,\sum_{\text{all } k} \sum_{k'}
        g'_{k'} \, v(\U_k) \, w'_{k'} \,
        d^l_{0s'}(\theta_{kk'}) \, \E^{-\I s' \alpha'_{kk'}}
    } \,.
\end{equation}
While Assumption~\ref{asm-1} considers pairs of points, here we only have a
single set, and hence make
\begin{assumption}
\label{asm-6}
    All random points in the cross-correlation have the same a priori
    distribution.
\end{assumption}
As in the case of pairs of points, this seems a weak assumption, since it is
difficult to imagine how individual points might be a priori distinguishable
from each other.

Under Assumption~\ref{asm-6}, the sum over~$k$ in
expectation~\eqref{eq:ev-cl-nf:0} reduces to~$N$ identically distributed terms.
Using definition~\eqref{eq:nbar}, we can replace the product of~$N$ and
visibility~$v$ by the mean number density~$\bar{n}$.  Using the
definition~\eqref{eq:w} of the weight function~$w'$, we may also replace the
remaining sum over~$k'$ by an integral over~$\U'$,
\begin{equation}
\label{eq:ev-cl-nf:1}
    \ev{C_l^{nf'}}
    = \int \ev[\Big]{
        g'(\U') \, \bar{n}(\U) \, w'(\U') \,
        d^l_{0s'}(\theta) \, \E^{-\I s' \alpha'}
    } \, \D\U' \,,
\end{equation}
where the angles~$\theta$ and~$\alpha'$ now describe the relative orientation
between the random point~$\U$ and~$\U'$.

Using Assumptions~\ref{asm-4}, we can factor the weight~$w'(\U')$ out of the
integral in expectation~\eqref{eq:ev-cl-nf:1}.  Furthermore, using
Assumption~\ref{asm-2}, the expectation is conditional on the mean number
density~$\bar{n}$, and only the position~$\U$ in $\bar{n}(\U)$ is random.  The
remaining expectation in~\eqref{eq:ev-cl-nf:1} therefore reduces to the random
point~$\U$ and the random field~$g'$.  It can be computed in two steps using
the law of total expectation.  Using the angular
correlation~\eqref{eq:ev-ff-cf}, the expectation of a homogeneous (by
Assumption~\ref{asm-7}) random field~$g'$ conditional on~$\U$ is
\begin{equation}
\label{eq:ev-gamma-theta}
    \ev{g'(\U') \, \E^{-\I s' \alpha'} \mid \U}
    = \gamma(\theta) \,,
\end{equation}
where~$\gamma$ is the expected angular cross-correlation function.\footnote{%
    Not to be confused with the shear in gravitational lensing.}
For galaxy--galaxy lensing, the expected correlation is more commonly written
in terms of a tangential component~$\gamma_{\rm t}$ and
cross-component~$\gamma_{\times}$ as~$\gamma(\theta) = \gamma_{\rm t}(\theta) +
\I \, \gamma_{\times}(\theta)$.

Combining expectations~\eqref{eq:ev-gamma-theta} and~\eqref{eq:ev-cl-nf:1}, it
remains to compute the expectation over random positions~$\U$.  Under
Assumption~\ref{asm-3}, the point process is homogeneous, and hence
\begin{equation}
\label{eq:ev-cl-nf:2}
    \ev{C_l^{nf'}}
    = \frac{1}{4\pi} \iint \! \bar{n}(\U) \, w'(\U') \,
        \gamma(\theta) \, d^l_{0s'}(\theta) \, \D\U \, \D\U' \,.
\end{equation}
As before, the double integral recovers the definition~\eqref{eq:cf} of the
angular correlation function for~$\bar{n}$ and~$w'$,
\begin{equation}
\label{eq:ev-cl-nf:3}
    \ev{C_l^{nf'}}
    = 2\pi \int_{0}^{\pi} \! C^{\bar{n} w'}(\theta) \, \gamma(\theta) \,
        d^l_{0s'}(\theta) \sin(\theta) \, \D\theta \,,
\end{equation}
which in turn is the convolution~\eqref{eq:ev-cl-cf-prod} of~$C^{\bar{n} w'}$
and~$\gamma$ that yields the mixing matrix~\eqref{eq:ev-cl-mm},
\begin{equation}
\label{eq:ev-cl-nf}
    \ev{C_l^{nf'}}
    = \sum_{l_1} M^{\bar{n} w'}_{ll_1} \, \gamma_{l_1} \,,
\end{equation}
where~$\gamma_{l}$ is the angular power spectrum associated with the angular
cross-correlation function~$\gamma$.

Overall, we therefore obtain the intuitively clear result that the expected
angular power spectrum is given by the intrinsic spectrum~$\gamma_l$ for point
process and random field, modulated by a mixing matrix coming from the mean
number density~$\bar{n}$ (due to the point process) and weight function~$w'$
(due to the random field).

In the case of two point processes, it was advantageous to correlate the
density contrast~$\delta$ instead of the number density~$n$.  Since the density
contrast~\eqref{eq:delta} is linear in~$n$, the angular power spectrum
of~$\delta$ and~$f'$ is
\begin{equation}
\label{eq:cl-deltaf}
    C_l^{\delta f'}
    = \frac{C_l^{nf'} - C_l^{\bar{n} f'}}{\bar{n}_0} \,.
\end{equation}
The expectation~$\ev{C_l^{\delta f'}}$ then follows immediately from the
definition~\eqref{eq:nbar} of the mean density and
expectation~\eqref{eq:ev-cl-nf},
\begin{equation}
\label{eq:ev-cl-deltaf}
    \ev{C_l^{\delta f'}}
    = \sum_{l_1} M^{vw'}_{ll_1} \, \gamma_{l_1} - \ev{C_l^{vf'}}
    = \sum_{l_1} M^{vw'}_{ll_1} \, \gamma_{l_1} \,,
\end{equation}
where the second equality assumes that the expectation of~$f'$ vanishes.

Whether to correlate~$n$ or~$\delta$ is a well-known question for real-space
estimators of galaxy--galaxy lensing \citep{2021A&A...646A.129J}.  Formally, we
can construct a direct estimator of the signal from either the number
density~$n$ using expectation~\eqref{eq:ev-cl-nf},
\begin{equation}
\label{eq:est-gamma-n}
    \hat{\gamma}_l^{n}
    = \sum_{l_1} (M^{\bar{n} w'})^{-1}_{ll_1} \, C_l^{nf'} \,,
\end{equation}
or from the density contrast~$\delta$ using expectation~\eqref{eq:ev-cl-deltaf}
and the definition~\eqref{eq:nbar} of the mean density,
\begin{equation}
\label{eq:est-gamma-delta}
\begin{split}
    \hat{\gamma}_l^{\delta}
    &= \sum_{l_1} (M^{vw'})^{-1}_{ll_1} \,
        \frac{C_l^{nf'} - C_l^{\bar{n} f'}}{\bar{n}_0} \\
    &= \sum_{l_1} (M^{\bar{n}w'})^{-1}_{ll_1} \, C_l^{nf'}
    - \sum_{l_1} (M^{\bar{n}w'})^{-1}_{ll_1} \, C_l^{\bar{n} f'} \,.
\end{split}
\end{equation}
In real space, the mean number density~$\bar{n}$ corresponds to an equivalent
distribution of uniform random points (``randoms''); the inverse mixing matrix
in both~$\hat{\gamma}_l^{n}$ and~$\hat{\gamma}_l^{\delta}$ thus corresponds to
a normalisation by weighted pairs of randoms and the observed positions of the
field.  The difference between~$\hat{\gamma}_l^{n}$
and~$\hat{\gamma}_l^{\delta}$ is the second term in the
estimator~\eqref{eq:est-gamma-delta}, which corresponds to correlations between
randoms and field values.  It was shown by \citet{2017MNRAS.471.3827S} that the
estimator~\eqref{eq:est-gamma-delta} has lower variance relative to the
estimator~\eqref{eq:est-gamma-n}, particularly on large scales, because it
suppresses covariance terms that couple with the survey mask. This suppression
also increases the accuracy of covariance estimation via resampling techniques,
as these modify the effective survey window in the subsampling compared to the
original survey.  Moreover, the subtraction of correlations around random
points can subtract residual additive systematics in the signal.  As in the
case of angular clustering, we therefore generally prefer the density
contrast~$\delta$ instead of the number density~$n$ to measure
cross-correlations.

\section{Finite resolution maps}
\label{sec:maps}

The preceding sections demonstrate how we can obtain angular power spectra from
discrete observations, and how we can relate their expectations to the
intrinsic two-point statistics of the observed point processes or random
fields.  We now turn to the practical task of computing the angular power
spectra.  Retracing our steps, we find that this can be done in one of two
ways:
\begin{itemize}
\item[i)] Compute~$C_l^{nn'}$, $C_l^{ff'}$, and~$C_l^{nf'}$ directly using
    their respective expressions~\eqref{eq:cl-nn}, \eqref{eq:cl-ff},
    and~\eqref{eq:cl-nf}.
\item[ii)] Compute~$n_{lm}$ and~$f_{lm}$ from their analytical
    expansions~\eqref{eq:nlm} and~\eqref{eq:flm}, then compute the angular
    power spectra~\eqref{eq:cl} from the spherical harmonic coefficients.
\end{itemize}
If~$N$ is the number of observations, and~$l_{\max}$ is the highest angular
mode number of interest, then the former method has a runtime complexity
of~$O(N^2 l_{\max})$, i.e., quadratic in the number of observations, which is
the same as for real-space estimators.  The complexity of the latter method,
however, is~$O(N l_{\max}^2)$, and it is hence favourable when~$l_{\max} \ll
N$.  We therefore generally want to obtain angular power spectra~$C_l$ from
their constituent coefficients~$a_{lm}$.

The $a_{lm}$ computed from the sums~\eqref{eq:nlm} and~\eqref{eq:flm}
potentially still contain more information than we need:  if the observed
points are sufficiently dense, they probe scales beyond our desired scale
of~$l_{\max}$.  We can then reduce the computational complexity further by
introducing a spatial binning of the points -- or, in other words, by making a
map.

Map-making consists of two separate but related parts.  The first is sampling,
so that spherical functions are represented by their values in a finite set of
basis points on the sphere.  The number and locations of the basis points are
chosen such that it is possible to accurately recover angular modes up to some
given~$l_{\max}$ from the spherical harmonic expansion~\eqref{eq:def-f-ylm}.
Several sampling schemes for that purpose have been proposed; commonly used in
astronomy are, e.g., the scheme of \citet{1994AdApM..15..202D}, schemes with
exact spherical harmonic transforms for band-limited functions
\citep{2010ApJS..189..255H, 2011ITSP...59.5876M}, and \emph{HEALPix}
\citep{2005ApJ...622..759G}.

Sampling a random set of point masses, such as the spherical
functions~\eqref{eq:n} and~\eqref{eq:f} we construct from our discrete
observations, with a fixed set of basis points will result in a map that is
almost surely zero everywhere.  The second part of map-making is hence the
collection of function values (i.e., observed points) over a finite region
around each sampling point.  This is achieved using spherical convolution,
which we define below.  The area over which observations are collected is, at
least in principle, entirely independent of the sampling scheme.  Naturally, we
want every observation to be counted, in which case this area must be large
enough to cover the spaces between sampling points.  On the other hand, the
area should also not be much larger than necessary, or we needlessly degrade
the angular power spectra that we wish to measure.  In practice, there is hence
always a close match between the convolution and the sampling scheme.

\subsection{Spherical convolution}

Convolution is a mathematical operation that produces a new function~$F$ from a
given function~$f$ and convolution kernel~$K$.  The value of the convolution in
a point is obtained by making said point the origin of a local copy of~$K$ and
computing the integral of~$f$ weighted by that kernel.  Convolution is
therefore not a local operation.  And since spin-weighted spherical functions
are always defined relative to a local coordinate frame
\citep{2016JMP....57i2504B}, it follows that convolution on the sphere has to
explicitly take this non-local nature into account.

As an illustrative example, consider the following situation, where the dot
marks the centre of a small, essentially flat patch of the sphere, and the
arrows indicate the complex argument of a local spin-weighted function of
constant magnitude:
\begin{equation*}
\renewcommand*{\arraystretch}{1}
\setlength{\arraycolsep}{2pt}
\begin{array}{ccc}
    &\downarrow& \\
    \rightarrow&\boldsymbol{\cdot}&\leftarrow \\
    &\uparrow&
\end{array}
\end{equation*}
Intuitively, the sum of the function values should be zero.  Not accounting for
coordinate frame effects, this is indeed the case if the dot is near the
equator.  But if the dot marks the north pole, all arrows point north, and
naive summation produces an incorrect result.

For a spherical convolution that treats non-zero spin weights in the correct
manner, we define the convolution~$F$ of a spherical function~$f$ and a
symmetric convolution kernel~$K$ as
\begin{equation}
\label{eq:conv}
    F(\U)
    = \int \! f(\U') \,
        \E^{\I S\! \alpha} \, K(\theta) \, \E^{-\I s \alpha'} \, \D\U' \,,
\end{equation}
where the angles~$\theta, \alpha, \alpha'$ are the separation and relative
orientation of~$\U$ and~$\U'$ as in the spherical harmonic addition
theorem~\eqref{eq:addthm}.  Here, $s$ is the spin weight of the convolved
function~$f$, and~$S$ is the spin weight of the convolution~$F$.  We can
choose~$S$ freely:  under a rotation of~$\gamma$ in~$\U$, the angle~$\alpha$ in
the convolution~\eqref{eq:conv} transforms as $\alpha \mapsto \alpha - \gamma$,
so that $F$ indeed picks up the phase factor~$\E^{-\I S\! \gamma}$ of a
function with spin weight~$S$.  Overall, the convolution~\eqref{eq:conv} is
equivalent to the directional spin-weighted spherical convolution of
\citet{2015arXiv150906749M} with a symmetric kernel, and reduces to the usual
spherical convolution of scalar functions when the spin weight is zero
\citep[e.g.,][]{2001PhRvD..63l3002W}.

Most importantly, the definition~\eqref{eq:conv} of spherical convolution
yields a useful convolution theorem for spherical harmonic expansions.  The
convolution kernel~$K$ is a function of separation, similar to an angular
correlation function, so that we can apply the expansion~\eqref{eq:cf-cl} into
Wigner $d$ functions,
\begin{equation}
\label{eq:Kf-Kl}
    K(\theta)
    = \sum_{l} \frac{2l+1}{4\pi} \, K_l \, d^l_{S\!s}(\theta) \,,
\end{equation}
where the coefficients~$K_l$ of the expansion are given by~\eqref{eq:cl-cf},
\begin{equation}
\label{eq:Kl-Kf}
    K_l
    = 2\pi \int_{0}^{\pi} \! K(\theta) \, d^l_{S\!s}(\theta)
        \sin(\theta) \, \D \theta \,.
\end{equation}
Inserting the expansion~\eqref{eq:Kf-Kl} into definition~\eqref{eq:conv} and
substituting the addition theorem~\eqref{eq:addthm} yields the
integral~\eqref{eq:def-flm} for the coefficients~$f_{lm}$ in the spherical
harmonic expansion~\eqref{eq:def-f-ylm} of~$f$.  We thus obtain the desired
spherical harmonic convolution theorem,
\begin{equation}
\label{eq:convthm}
    F_{lm} = K_l \, f_{lm} \,,
\end{equation}
where the coefficients~$F_{lm}$ of the convolution~$F$ are the product of the
coefficients~$K_l$ of the convolution kernel~$K$ and the coefficients~$f_{lm}$
of the convolved function~$f$.

Carrying out the convolution~\eqref{eq:conv} requires computing the phase
factors~$\E^{\I S\! \alpha}$ and~$\E^{-\I s \alpha'}$ in each point, which can
often be done efficiently (see Appendix~\ref{sec:angles}).  However, if the
support of the convolution kernel~$K$ is sufficiently small, the local geometry
is close to flat, and~$\alpha \approx \alpha'$.  In that case, the phase
factors reduce to unity if we choose a convolution with~$S = s$ that does not
change the spin weight of the function.

\begin{figure}%
\centering%
\includegraphics[scale=.85]{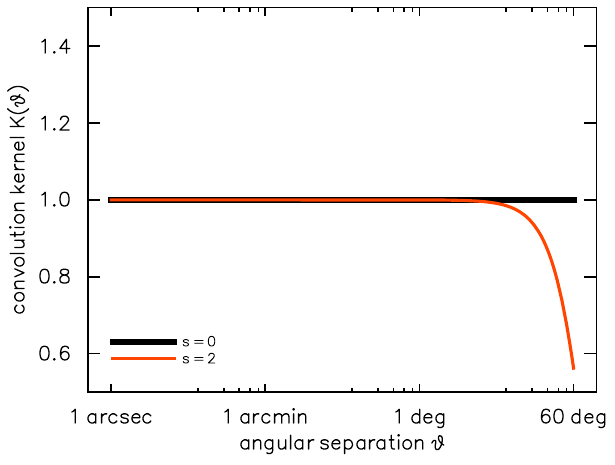}%
\caption{%
    The convolution kernel~\eqref{eq:Kf} with no cut-off for spin weight~$s=0$
    (\emph{black}) and spin weight~$s = 2$ (\emph{red}).  For sufficiently
    small angles, the convolution kernel becomes indistinguishable from a flat
    spherical disc even in the spin-weighted case.
}%
\label{fig:Kf}%
\end{figure}

\begin{figure}%
\centering%
\includegraphics[scale=.85]{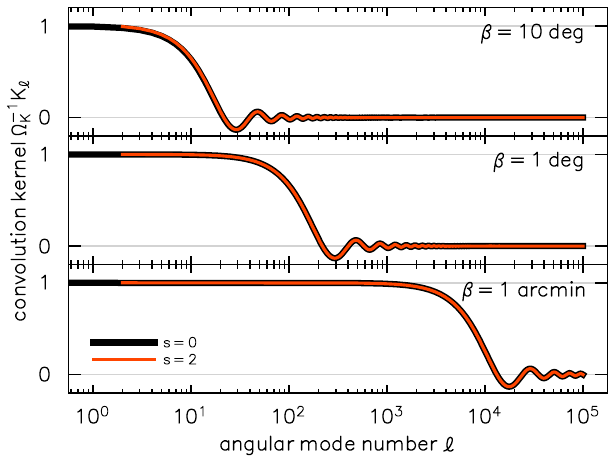}%
\caption{%
    Normalised coefficients of the convolution kernel~\eqref{eq:Kf} for spin
    weights~$s = 0$ (\emph{black}), $s = 2$ (\emph{red}), and kernel
    sizes~$\beta = 10$~degrees (\emph{top}), $\beta = 1$~degree
    (\emph{middle}), $\beta = 1$~arcmin (\emph{bottom}).  There is excellent
    agreement between the coefficients for~$s = 2$ and~$s = 0$, except in the
    case of a large kernel ($\beta = 10$~degrees) at large angular scales ($l <
    10$).
}%
\label{fig:Kl}%
\end{figure}

It remains to find a tractable convolution kernel~$K$.  For a function~$f$
with~$s = 0$, the natural choice is a spherical disc of some chosen angular
size~$\beta > 0$.  However, for~$s \ne 0$, the same kernel does not produce
analytically tractable coefficients~\eqref{eq:Kl-Kf} for the convolution
theorem.  We hence propose a modified convolution kernel that works for any
spin weight~$S = s \ge 0$, and reduces to a spherical disc if~$s = 0$,
\begin{equation}
\label{eq:Kf}
    K(\theta)
    = \begin{dcases}
        \bigl[\cos\tfrac{\theta}{2}\bigr]^{2s} &
        \text{if $\theta \le \beta$,} \\
        0 &
        \text{otherwise,}
    \end{dcases}
\end{equation}
where~$\beta$ is the angular size (i.e., radius) of the kernel.  The effective
area of the convolution kernel is\footnote{%
    Not to be confused with the cosmological curvature matter density.}
\begin{equation}
\label{eq:Kf-norm}
    \Omega_K
    = 2\pi \int_{0}^{\pi} \! K(\theta) \sin(\theta) \, \D \theta
    = 4\pi \, \frac{1 - \bigl[\cos\frac{\beta}{2}\bigr]^{2s+2}}{s+1} \,.
\end{equation}
The choice of kernel~\eqref{eq:Kf} is firstly motivated by the fact that there
is a known expression for its coefficients~\eqref{eq:Kl-Kf},\footnote{%
    The convolution kernel~\eqref{eq:Kf} and coefficients~\eqref{eq:Kl} follow
    from the integral~(4.11.9) of \citet{1988qtam.book.....V}, which in fact
    yields a more general, spin-changing convolution kernel with~$S \ne s$.}
\begin{equation}
\label{eq:Kl}
    K_l
    = \begin{dcases}
        \frac{
            4\pi \, \bigl[\sin\tfrac{\beta}{2}\bigr] \,
                \bigl[\cos\tfrac{\beta}{2}\bigr]^{2s+1} \, d^l_{ss+1}(\beta)
        }{
            \sqrt{l \, (l+1) - s \, (s+1)}
        } &
        \text{if $l > s$,} \\
        4\pi \, \frac{1 - \bigl[\cos\frac{\beta }{2}\bigr]^{4s+2}}{2s+1} &
        \text{if $l = s$.}
    \end{dcases}
\end{equation}
Secondly, for small kernel sizes~$\beta \lesssim 1$~degree, the convolution
kernel~\eqref{eq:Kf} is essentially a flat spherical disc even when the spin
weight is~$s = 2$ (Fig.~\ref{fig:Kf}), in which case the
coefficients~\eqref{eq:Kl} for~$s = 0$ and~$s = 2$ become essentially the same
(Fig.~\ref{fig:Kl}).  This makes the specific kernel~\eqref{eq:Kf} a good
practical choice for maps when the resolution is below the degree-scale.

In summary, the convolution~\eqref{eq:conv} means that we can create finite
resolution maps of the point-mass like number density~\eqref{eq:n} or
field~\eqref{eq:f} by picking a sampling scheme and for each grid point summing
each observed point with the weight~$K(\theta)$ given by the convolution
kernel~$K$, omitting the phase factors in the convolution~\eqref{eq:conv} if
the resolution allows it.  For a convolution kernel such as~\eqref{eq:Kf} with
small angular size~$\beta \ll \pi$, an alternative method is to reverse the
order of operations, and find all grid points closer than~$\beta$ for each
observed point.  This can result in vast performance improvements, particularly
if the grid points can be queried efficiently, e.g., when using Cartesian or
\emph{HEALPix} grids.

Once maps are created, it suffices to compute their spherical harmonic
coefficients~$F_{lm}$, and reconstruct the coefficients~$f_{lm}$ of the
spherical function from the convolution theorem~\eqref{eq:convthm},
\begin{equation}
\label{eq:inv-convthm}
    f_{lm} = \frac{1}{K_l} \, F_{lm} \,.
\end{equation}
Naturally, this is only possible when $K_l \ne 0$, which limits the angular
mode numbers~$l$ that can be recovered for a given convolution kernel.
However, if the deconvolution~\eqref{eq:inv-convthm} is possible for all
numbers~$l \le l_{\max}$, we are readily able to compute the angular power
spectrum~\eqref{eq:cl} of~$f$ and~$f'$ from the finite-resolution maps~$F$
and~$F'$.  In cases where the deconvolution is impossible or undesirable, we
can instead use the convolution theorem~\eqref{eq:convthm} to model the effect
of the convolution on the expected angular power spectra.  In practice, this
can be done at no computational cost, by absorbing the convolution kernel into
the mixing matrix~\eqref{eq:ev-cl-mm}.

\subsection{\emph{HEALPix} pseudo-convolution}

\begin{figure}%
\centering%
\includegraphics[scale=.85]{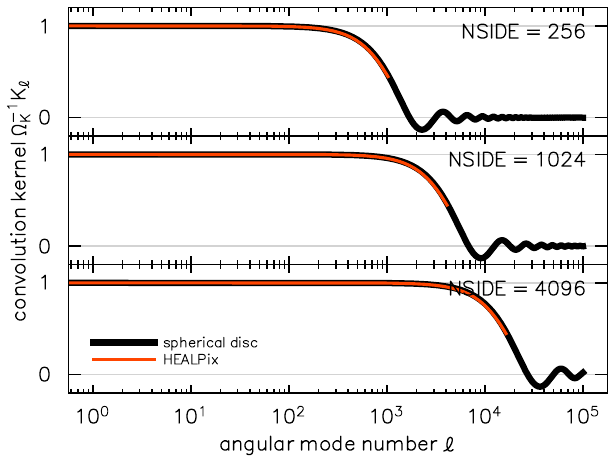}%
\caption{%
    The \emph{HEALPix} pixel window function (\emph{red}) for resolution
    parameters~$\mathtt{NSIDE} = 256$ (\emph{top}), $\mathtt{NSIDE} = 1024$
    (\emph{middle}), $\mathtt{NSIDE} = 4096$ (\emph{bottom}).  Also shown is
    the convolution kernel of a spherical disc with the same pixel area
    (\emph{black}).  The \emph{HEALPix} pixel window function is only provided
    up to~$l = 4 \, \mathtt{NSIDE}$, where it starts to fall below the kernel
    of the spherical disc.
}%
\label{fig:healpix-Kl}%
\end{figure}

Even for small kernels, the radius search required by the spherical convolution
comes at a non-negligible computational cost.  Given the number of galaxies
observed by \Euclid, this cost quickly becomes prohibitive, unless faster,
specialised algorithms can be found.  However, when using the \emph{HEALPix}
grid for sampling, we can alternatively follow the standard procedure of simply
summing the points in each \emph{HEALPix} pixel.  But even though all
\emph{HEALPix} pixels have the same area, this operation is not a true
spherical convolution, due to the slightly varying pixel shapes
\citep{2005ApJ...622..759G}.  Nevertheless, summation over~\emph{HEALPix}
pixels does obey the convolution theorem~\eqref{eq:convthm} approximately, and
the normalised coefficients~$\Omega_K^{-1} K_l$ are known as the \emph{HEALPix}
pixel window function (Fig.~\ref{fig:healpix-Kl}).  As it turns out, this
pseudo-convolution can be adequate for \Euclid analysis, which we will
demonstrate in Sect.~\ref{sec:validation}.

\begin{figure}%
\centering%
\includegraphics[scale=.85]{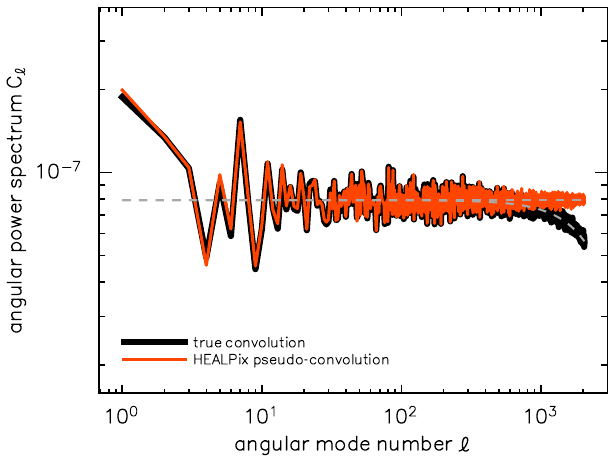}%
\caption{%
    Angular power spectrum of $1\,000\,000$ uniform random points, computed
    using the \emph{HEALPix} pseudo-convolution (\emph{red}) and a true
    convolution with spherical discs of the same area (\emph{black}).  Both
    results agree with their respective expectation (\emph{dashed}).  The
    convolution kernel is only imprinted on the true convolution.
}%
\label{fig:healpix-noise}%
\end{figure}

There is, however, one fundamental difference between a true spherical
convolution and \emph{HEALPix} pseudo-convolution.  According to the
convolution theorem~\eqref{eq:convthm}, the convolution kernel~$K_l$ is
imprinted on all spherical harmonic coefficients of a map, and consequently on
the angular power spectrum~\eqref{eq:cl},
\begin{equation}
\label{eq:cl-FF}
    C_l^{FF'} = K_l^2 \, C_l^{ff'} \,.
\end{equation}
In particular, it follows that the convolution kernel should also affect
additive bias terms such as~$A^{nn'}$ in the spectrum~\eqref{eq:cl-nn-split} of
point processes, or~$A^{ff'}$ in the expected spectrum~\eqref{eq:ev-cl-ff-noi}
of random fields.  These bias terms will therefore no longer be constant after
a true convolution.  This is not the case for the \emph{HEALPix}
pseudo-convolution (Fig.~\ref{fig:healpix-noise}), since the non-overlapping
\emph{HEALPix} pixels cannot imprint structure, such as the convolution kernel,
below the pixel scale.  In fact, a similar effect occurs when sampling a true
convolution so sparsely that the convolution kernel areas no longer overlap.

There is a practical consequence of this difference between true convolution
and \emph{HEALPix} pseudo-convolution.  For a map created by true convolution,
the deconvolution~\eqref{eq:inv-convthm} turns an additive bias term in the
angular power spectrum back into a constant.  For a \emph{HEALPix} map, the
same deconvolution of the pixel window function turns an additive bias~$A$ into
an $l$-dependent bias~$A/K_l^2$.  When comparing \emph{HEALPix} spectra and
their expectations, the additive bias must therefore either be subtracted from
the measured spectra before deconvolution of the pixel window function, or the
correct $l$-dependent bias must be used, e.g., in
expectations~\eqref{eq:ev-cl-nn}, \eqref{eq:ev-cl-dd},
and~\eqref{eq:ev-cl-ff-noi}.

\subsection{Maps}

We can now define the maps we make.  For simplicity, we always call one value
of a finite map a ``pixel'' with area~$\Omega_K$, with the understanding that
this may refer either to an actual \emph{HEALPix} pixel or to the kernel of a
true convolution.

The convolution of the number density~$n$ with definition~\eqref{eq:n} is the
map~$N$ of number counts in each pixel.  The convolution of~$\bar{n}$ is the
map~$\bar{N}$ of mean number counts; we write it as $\bar{N} = \bar{N}_0 \, V$
using the mean number of points per pixel~$\bar{N}_0 = \Omega_K \, \bar{n}_0$
and a map~$V$ that is the convolution of~$v$ divided by~$\Omega_K$.  We
call~$V$ the `visibility map', since it is the pixel-averaged equivalent of the
visibility~$v$, with pixel values between~$0$ and~$1$.  Since the
convolution~\eqref{eq:conv} is a linear operation, the
expectation~\eqref{eq:ev-n} translates from the number densities to the number
count maps, $\ev{N} = \bar{N}$.

To isolate the clustering signal in the number count map~$N$, we define a
map~$\Delta$ for the density contrast~\eqref{eq:delta},
\begin{equation}
\label{eq:Delta}
    \Delta = \frac{N - \bar{N}}{\bar{N}_0} \,.
\end{equation}
It is clear that~$\Delta$ is the convolution of~$\delta$, but divided by the
pixel area, so that the numerical values of~$\Delta$ have the correct,
intuitive scale where $-1$ means ``empty space''.  Deconvolution of~$\Delta$
must therefore be carried out with the normalised convolution
kernel~$\Omega_K^{-1} \, K_l$.

For a field~$f$ such as, e.g., cosmic shear, we compute the map~$F$ by summing
the weighted field values~$w_k \, f_k$ in each pixel, and dividing the result
by a constant mean pixel weight~$\bar{W}_0$.  Specifically, we
choose~$\bar{W}_0$ to be the mean weight divided by the mean visibility, as
computed from the maps, which makes~$\bar{W}_0$ relatively insensitive to the
survey footprint and systematics.  The map~$F$ is hence the convolution of the
function~$f$ with definition~\eqref{eq:f}, divided by~$\bar{W}_0$ to remove
explicit dependencies on pixel area and overall weight factors.  Similarly, we
compute the weight map~$W$ as the total weight in each pixel, i.e., the
convolution of~$w$ with definition~\eqref{eq:w}, divided by~$\bar{W}_0$.

Since~$\bar{W}_0$ contains a factor of the pixel area, deconvolution of~$F$
and~$W$ is carried out with the normalised convolution kernel~$\Omega_K^{-1} \,
K_l$.  Furthermore, since both~$F$ and~$W$ are scaled identically, the
resulting mixing matrix automatically applies the correct factors
of~$\bar{W}_0$ to the expected angular power spectra.  However, we do need to
account for the scaling by~$\bar{W}_0$ when computing any additive bias terms.

In particular, we do not average the field values in each pixel by dividing the
map~$F$ by the map~$W$.  For \Euclid, the resolution of our maps is such that
about half of all observed pixels contain fewer than two observed values. Using
a weighted average would simply divide out the given weights in these pixels,
resulting in an unweighted cosmic shear map \citep[see,
e.g.,][]{2011MNRAS.412...65H, 2019PASJ...71...43H, 2021JCAP...03..067N}.  This
is clear when looking at the spherical functions~$f$ and~$w$ with
definitions~\eqref{eq:f} and~\eqref{eq:w}, respectively:  dividing~$f$ by~$w$
where both are non-zero is equivalent to using unit weights in~$f$.

\section{Validation}
\label{sec:validation}

In the preceding sections, we have derived the overall theory of angular power
spectra from discrete sets of observations, their expectations, and ways to
efficiently compute spectra from maps.  We now turn to the validation of our
findings.  One part of this are the explicit assumptions that we have made
throughout;  these represent specific scientific questions that are partially
the subject of active research in their own right, and we will not investigate
their validity here.

In what follows, we validate our specific implementation of the methodology
described above.  This is a publicly available code called
\texttt{Heracles},\textsuperscript{\ref{fn:url}} developed within the \Euclid
Science Ground Segment.  It contains routines for catalogue reading,
map-making, spherical harmonic transforms, angular power spectra, and mixing
matrices.  The code can be used as a \texttt{Python} library, e.g., for data
exploration in a notebook interface, or via a standalone command-line
interface, e.g., for batch data processing.  In particular, the code also
contains an implementation of the discrete angular power spectrum methodology,
which is based on fast, non-uniform computation of spherical harmonics
\citep{2023A&A...678A.165R} as implemented in the \texttt{ducc}
package.\footnote{%
    \url{https://gitlab.mpcdf.mpg.de/mtr/ducc}}

To validate the performance of \texttt{Heracles}, we carry out the following
series of tests:
\begin{itemize}
\item [i)] We estimate the mean density of galaxies from the visible sky
    fraction in the various \Euclid data releases,
\item [ii)] we test if the phase factors in the spherical convolution can be
    neglected for map-based spectra,
\item [iii)] we assess the overall accuracy of our measurements, and
\item [iv)] we apply the methodology in a data processing setting that
    mimics the first \Euclid data release.
\end{itemize}
Where simulations are created, we generally employ the same flat $\Lambda$CDM
cosmology as \Euclid's Flagship simulation \citep{2024arXiv240513495E}, with
parameter values $\Omega_{\rm m} = 0.319$, $\Omega_{\rm b} = 0.049$, $A_{\rm s}
= 2.1 \times 10^{-9}$, $n_{\rm s} = 0.96$, and $h = 0.67$.

\subsection{Mean density estimation}

\begin{figure}%
\centering%
\includegraphics[scale=.85]{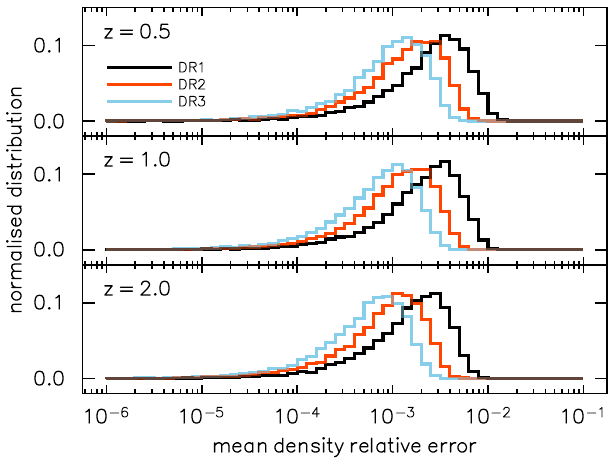}%
\caption{%
    Simulated distribution of the relative error when estimating the mean
    density~$\bar{n}$ from part of the sky, using the footprint of \Euclid DR1
    (\emph{black}), DR2 (\emph{red}), and DR3 (\emph{blue}) for redshifts $z =
    0.5$ (\emph{top}), $z = 1.0$ (\emph{middle}), and $z = 2.0$
    (\emph{bottom}).
}%
\label{fig:mean_density}%
\end{figure}

Constructing the density contrast~\eqref{eq:delta} requires knowledge of the
mean density of galaxies over the entire sky, which we must estimate from the
visible sky fraction.  If our estimate is inaccurate, we bias the angular power
spectrum~$C_l^{\delta\delta'}$ in a non-trivial manner with respect to the
expectation~\eqref{eq:ev-cl-dd}.  The problem in estimating the mean density
accurately is that the visible part of the sky might be particularly over- or
underdense compared to the true mean.  The probability of this depends on the
area of the observed sky, as well as the typical size of large-scale density
fluctuations, and hence the clustering of points.  This problem is closely
related to the integral constraint for real-space estimators.

To test the impact on \Euclid observations, we generate 10\,000 lognormal
realisations of a galaxy distribution with a linear galaxy bias
\citep{2023OJAp....6E..11T}.  To account for the redshift evolution of galaxy
clustering, we test redshifts~$z = 0.5, 1.0, 2.0$, using a redshift-dependent
bias that was fitted to the \Euclid Flagship simulation
\citep{2022A&A...662A..93E}.  Since the error in the mean density is a function
of visible sky fraction, we further use a representative footprint
\citep{2022A&A...662A.112E} for each of the three \Euclid data releases DR1
(1~year, 6\% sky coverage), DR2 (3~years, 18\% sky coverage), and DR3 (6~years,
36\% sky coverage).  The results show that the expected relative error in the
mean density is at the per mille level for all data releases and redshifts,
with a scatter that stays below the per cent level for DR2 and beyond
(Fig.~\ref{fig:mean_density}).

\subsection{Phase factors}

\begin{figure}%
\centering%
\includegraphics[scale=.85]{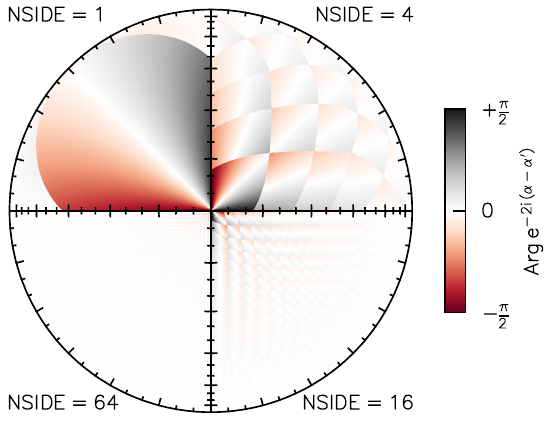}%
\caption{%
    Argument of the phase factor bias $\E^{-2\I \, (\alpha - \alpha')}$ for
    \emph{HEALPix} maps with $\mathtt{NSIDE} = 1, 4, 16, 64$, shown in
    orthographic projection with the north pole at the centre and the equator
    at the border.  The angles~$\alpha$ and~$\alpha'$ are defined in
    Appendix~\ref{sec:angles}.  For $\mathtt{NSIDE} = 64$, the resolution is at
    the degree-scale, and the phase factors are close to unity.
}%
\label{fig:healpix-phase}%
\end{figure}

To test whether or not we can neglect the phase factors in the
convolution~\eqref{eq:conv}, we need to quantify their impact on maps at the
required resolution for \Euclid.  Consider a fixed pixel located at a
position~$\U_0$.  By neglecting the phase factors, the
convolution~\eqref{eq:conv} is approximated as
\begin{equation}
\label{eq:conv-approx}
    F(\U_0) \approx \int \! f(\U') \, K(\theta) \, \D\U' \,.
\end{equation}
To cancel the phase factors, the approximation effectively applies a
position-dependent multiplicative bias~$\E^{-\I \, (S\! \alpha - s \alpha')}$
to the function~$f$ over the pixel area, where the angles~$\alpha$
and~$\alpha'$ are taken with respect to~$\U_0$.  Under a rotation of~$\gamma'$
in~$\U'$, the angle~$\alpha'$ transforms as~$\alpha' \mapsto \alpha' -
\gamma'$; as a function of~$\U'$, the bias therefore has a spin weight of~$s$.

For cosmic shear, we can make maps of this bias, using the expressions from
Appendix~\ref{sec:angles} and setting~$S = s = 2$.  The result is shown in
Fig.~\ref{fig:healpix-phase} for \emph{HEALPix} maps with resolution parameters
$\mathtt{NSIDE} = 1, 4, 16, 64$.  Parallel transport along a meridian has phase
factors of unity, so that the phase factor bias is effectively a function of
azimuthal distance from the pixel centre.  The pixel resolution starts to fall
below the degree-scale at~$\mathtt{NSIDE} = 64$, and the phase factor bias
becomes small, due to the essentially flat geometry of the pixels.

Overall, we expect no impact from neglected phase factors for \emph{HEALPix}
maps with resolution parameter $\mathtt{NSIDE} \gtrsim 1024$.  If necessary,
the phase factor bias could be mitigated even further by choosing the
coordinate system such that the poles fall into a masked region, e.g., the
galactic plane.

\subsection{Accuracy}

\begin{figure*}%
\centering%
\includegraphics[scale=.85]{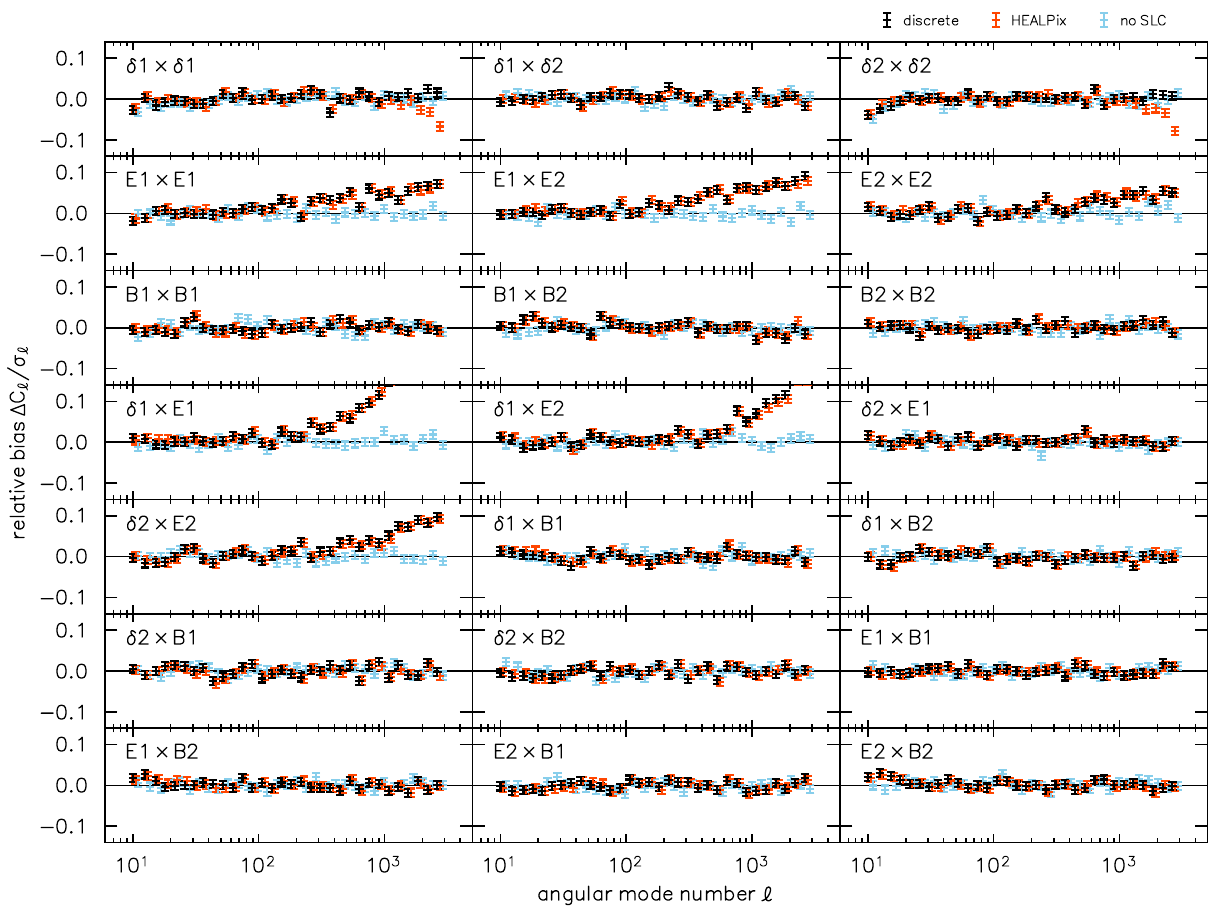}%
\caption{%
    Bias relative to statistical uncertainty between measured and expected
    angular power spectra from 10\,000 simulations with a \Euclid DR1-like
    setup.  Shown are spectra for combinations of galaxy density~$\delta$ and
    cosmic shear $E$- and $B$-mode in two tomographic bins.  The discrete
    angular power spectra show very good agreement between measurement and
    expectation (\emph{black}), except for effects that can be ascribed to
    source--lens clustering.  The same effects are visible for map-based
    spectra from \emph{HEALPix}; in addition, these also show a small residual
    in angular clustering due to the pseudo-convolution of \emph{HEALPix}
    pixels (\emph{red}).  When source--lens clustering is taken into account
    (\emph{blue}), the relative bias of the discrete spectra is consistent with
    zero at the~1\% level (\emph{error bars}).  Points of the three data sets
    are slightly offset for better visibility.
}%
\label{fig:bias}%
\end{figure*}

We now validate the results we derive in Sect.~\ref{sec:expectations} with
simulations.  To characterise the accuracy of our measurement, these require
many realisations that, for \Euclid, cover a significant fraction of the sky.
Usually, lognormal simulations would be the method of choice here; however, the
transformations involved in sampling lognormal fields are not exact
\citep{2023OJAp....6E..11T}, which adds an element of uncertainty to the
validation.  Gaussian random fields can be simulated with exactly prescribed
two-point statistics; however, for realistic angular power spectra and values
of linear galaxy bias, the realised density fields almost surely contain
regions where the number density becomes negative.  Instead, we use squared
Gaussian random fields, which is a toy model we develop in
Appendix~\ref{sec:sqnorm} that can be simulated exactly and remains physically
valid everywhere.

To accurately quantify the effects of map-based measurements, we require
simulations which are not themselves affected by pixel effects.  We therefore
simulate the fields not in real space, but via their spherical harmonic
expansion~\eqref{eq:def-f-ylm}.  We then sample points using a rejection
sampling scheme that accepts or rejects points with a probability that is
proportional to the simulated density field, evaluated in each sampled point
from the spherical harmonic expansion.  Instead of a survey footprint map, we
draw points from a spherical cap of 2500~deg$^2$, matching the anticipated area
of \Euclid's DR1, and located at the centre of the \Euclid Flagship simulation
\citep{2024arXiv240513495E}.  The weak lensing fields are subsequently
evaluated at the sampled positions from their spherical harmonic expansions,
without any intermediary interpolation.

We then generate 10\,000 realisations of these simulations.  To show results
for both auto- and cross-spectra, we simulate two Gaussian tomographic redshift
bins centred on representative redshifts of~$z = 0.5, 1.0$ with a width
of~$\sigma_z = 0.125$.  To simulate galaxy clustering, we use realistic
redshift-dependent galaxy bias \citep{2022A&A...662A..93E}, with a \Euclid-like
galaxy density of 2 galaxies per arcmin$^2$ in each tomographic bin.  For each
simulated galaxy, we store its position, as well as its observed ellipticity
from weak lensing with a random intrinsic ellipticity drawn from a hyperbolic
normal distribution \citep{2023OJAp....6E..11T}, using a per-component standard
deviation $\sigma_\epsilon = 0.26$.  Instead of using the correct weak lensing
action~\eqref{eq:shear}, we simply sum the intrinsic ellipticity and the
shear~$\gamma$ from weak lensing, since we might otherwise pick up biases due
to the reduced shear approximation \citep{2020A&A...636A..95D}.  Finally, we
give every galaxy a random shear weight from a log-uniform distribution
between~$10^{-2}$ and~$10^2$, to simulate a dynamic range that should exceed
any real shear measurement method.

For each simulation, we measure the angular power spectra of the reconstructed
density field~$\delta$, the cosmic shear $E$- and $B$-mode
(Appendix~\ref{sec:e-b-modes}), and their cross-correlations in the two
tomographic bins for angular modes up to~$l = 3000$.  In addition, we measure
the angular power spectra of the visibility~$V$ and the shear weight~$W$ in
each bin for modes up to~$l = 9000$, from which we compute the mixing
matrices~\eqref{eq:mm} truncated at~$l = 3000$ and~$l_1 = 6000$.  By
construction, our simulations are band-limited at~$l = 6000$, so that these
truncated mixing matrices should contain all non-zero entries, and yield exact
expectations for our measurements.

We thus obtain measured and expected angular power spectra for all combinations
of probes across the two tomographic bins: angular clustering, cosmic shear,
and galaxy--galaxy lensing.  To reduce noise, we average the spectra over 32
angular bins with logarithmic spacing between~$l = 10$ and~$l = 3000$.  We then
compute the mean of the bias~$\Delta C_l = C_l - \ev{C_l}$ between measurements
and expectations, which we scale relative to the standard deviation~$\sigma_l$
of each measurement over the set of realisations.

We carry out the measurements for each simulation using \emph{i)} the exact
angular power spectra computed from discrete sets of points, and \emph{ii)}
map-based angular power spectrum from \emph{HEALPix} maps with resolution
parameter $\mathtt{NSIDE} = 4096$.  The results are shown in
Fig.~\ref{fig:bias}.  For the exact, discrete spectra, we find agreement at
the~1\% level relative to the standard deviation, except for effects that can
be ascribed to source--lens clustering (i.e., Assumption~\ref{asm-4}).  For the
\emph{HEALPix}-based spectra, the results show an additional per-cent-level
bias in angular clustering at small scales.  Further testing reveals that the
small-scale \emph{HEALPix} bias has a dependency on the location of the survey
footprint, and we can hence ascribe it to the pseudo-convolution with varying
pixel shapes.  The dependence of the recovered angular power spectra on pixel
shapes is more thoroughly explored elsewhere (Hall \& Tessore in prep.).

To demonstrate that the bias in Fig.~\ref{fig:bias} is in fact source--lens
clustering, we run a second set of simulations where the positions of shears
are distributed according to an independent (but identically clustered)
realisation of large-scale structure.  When source--lens clustering is thus
taken into account, the discrete spectra show a relative bias that is
consistent with zero at the~1\% level for all probes.

Overall, we find that the \emph{HEALPix}-based measurements are only marginally
biased with respect to the exact discrete angular power spectra.  In light of
the lower computational cost, we therefore adopt this method for \Euclid's DR1
analysis, which will enable faster turnaround in the data processing.  However,
since accuracy is a function of survey area and galaxy density, this may no
longer be the case for subsequent data releases.

\subsection{Applicability to \Euclid DR1}

To demonstrate that we have a viable pipeline for \Euclid's first data release
(DR1), we process a realistic DR1-like data volume.  To do so, we select
galaxies contained in the provisional northern DR1 footprint
\citep{2022A&A...662A.112E} from the \Euclid Flagship simulation
\citep{2024arXiv240513495E}, obtained from CosmoHub \citep{2020A&C....3200391T,
2017ehep.confE.488C}.  The \Euclid data processing pipeline aims to support up
to 13 tomographic redshift bins \citep{2024arXiv240513491E}, and since the
number of spectra and mixing matrices increases quadratically with the number
of tomographic bins, we want to ensure compliance with such a setting.  Using
the simulated photometric redshifts, we therefore bin galaxies into 13
equi-populated tomographic redshift bins.  We then measure all 780 possible
auto- and cross-correlations between galaxy positions and cosmic shear $E$- and
$B$-modes.  To compare the measurements with expectations, we further compute
mixing matrices for all spectra from the simulated visibility and shear weight
maps.  For this test, we apply the \emph{HEALPix}-based methodology, with
resolution parameter~$\mathtt{NSIDE} = 4096$ and maximum angular mode number~$l
= 5000$ for all probes, which exceeds the ``optimistic'' forecast of scale cuts
\citep{2020A&A...642A.191E}.

The results are shown in Figs.~\ref{fig:fs2-a} and~\ref{fig:fs2-b}.  To compute
the expected spectra, we obtain theoretical full-sky predictions with the
\textit{Cosmology Likelihood for Observables in Euclid} code, \texttt{CLOE}
(Euclid Collaboration: Joudaki et al. in prep.), using the implemented
prescription for photometric harmonic-space observables  (Euclid Collaboration:
Cardone et al. in prep.).  We use \texttt{HMCode2020}
\citep{2021MNRAS.502.1401M} to model the non-linear matter power spectrum, as
provided in the public code \texttt{CAMB} \citep{2011PhRvD..84d3516C}.  For the
background cosmology, we use the same parameter values as in the \Euclid
Flagship simulation \citep{2024arXiv240513495E}.  To compute the observables,
we use the simulated redshift distributions~$n(z)$, and a linear galaxy bias
measured from the \Euclid Flagship simulation \citep{2022A&A...662A..93E}.
Since we are only testing the feasibility of the data processing here, we do
not perform any fine-tuning of the non-linear modelling, or take into account
systematic effects such as magnification bias or intrinsic alignments.  This is
visible, e.g., in the angular clustering at small scales~$l \gtrsim 1000$.
Nevertheless, we obtain a level of agreement between measurements and
expectations that is in line with previous results \citep{2024arXiv240513495E}.

Processing the data from DR1-like catalogues to all 780 angular power spectra
is very fast:  obtaining the spherical harmonic expansions of the fields and
weights in one tomographic bin takes around 3 minutes of wall clock time, and
all tomographic bins can be processed in parallel.  The subsequent computation
of angular power spectra from all combinations of spherical harmonic
coefficients of the fields has negligible cost.  Computation of the mixing
matrices from the spherical harmonic coefficients of the weights is a more
resource-intensive operation, taking around 75 CPU core-hours in total;
however, all mixing matrices can be computed in parallel as necessary.
Overall, we therefore expect no significant impact on \Euclid's DR1 processing
from the measurement of angular power spectra.

\section{Discussion and conclusion}
\label{sec:discussion}

We derive a complete framework to obtain exact measurements and expectations
for the angular power spectra from discrete sets of data.  Starting from an
exact, map-free formalism, we find new results such as exact, non-stochastic
expressions for the additive (``noise'') biases~\eqref{eq:A-nn}
and~\eqref{eq:A-ff} for angular clustering and random fields, respectively.
Furthermore, by explicitly tracking what assumptions enter our measurements, we
are able to separate the ``methodological'' accuracy of our results (i.e., when
all assumptions are true) from the ``true'' accuracy of the results, e.g., in
cases such as intrinsic alignments and source--lens clustering, which need to
be treated at the level of theoretical predictions.  When validating our
results on simulations, we find that discrete angular power spectra can achieve
biases of less than 1\% with respect to their standard deviation in a \Euclid
DR1-like setting.  Overall, we are therefore confident in our ability to
measure angular power spectra to the very high level of accuracy required to
achieve \Euclid's ambitious science goals.

Using the theory of spherical convolution~\eqref{eq:conv}, we can connect our
exact theory with the standard practice of measuring angular power spectra
from, e.g., \emph{HEALPix} maps.  Conceptually, this is a step away from the
picture in which the observed maps are ``noisy tracers'' of an underlying
continuous field such as, e.g., the true galaxy density or the true cosmic
shear.  In practice, the main difference between the exact map-based formalism
and standard practice is that observations are summed over one map ``pixel''
(i.e., convolution kernel) but not averaged.  This side-steps common issues
with map-based spectra, e.g., empty pixels, or the fact that pixels containing
just one single observation point are effectively unweighted after averaging.

When analysing \Euclid DR1-like simulations, we find that \emph{HEALPix}-based
spectra can perform at a level of accuracy that is comparable to the discrete
spectra.  An exception is the case of angular galaxy clustering, where the
\emph{HEALPix} pseudo-convolution, due to varying pixel shapes, introduces a
position-dependent bias at the per-cent-level for angular mode numbers~$l$ at
or above the resolution parameter~$\mathtt{NSIDE}$.  However, since this bias
is both small and mitigable by relevant analysis choices (i.e., map resolution
and scale cuts), we plan to employ the map-based methodology for \Euclid's
first data release.

We make our implementation of the methodology presented here available in the
form of a code called \texttt{Heracles}.\textsuperscript{\ref{fn:url}} This
code, originally created for 3\texttimes2pt data processing in the \Euclid
Science Ground Segment, was designed from the ground up to be user-friendly and
widely applicable to any given probe and survey, and will be maintained for
public use.

\begin{acknowledgements}
We thank Kevin Wolz and David Alonso for their collaboration during the
completion of this manuscript.  We also thank Ben Wandelt for a number of
excellent comments, and the anonymous referee for improving our manuscript.  NT
thanks Martin Reinecke for pointing out the possibility of implementing the
discrete angular power spectrum computation using \texttt{ducc}.  This work was
supported by UK Space Agency grants ST/W002574/1 and ST/X00208X/1.  AH
acknowledges support through a Royal Society University Research Fellowship.
GCH acknowledges support through the ESA research fellowship programme.  This
work has made use of CosmoHub, developed by PIC (maintained by IFAE and CIEMAT)
in collaboration with ICE-CSIC.  CosmoHub received funding from the Spanish
government (MCIN/AEI/10.13039/501100011033), the EU NextGeneration/PRTR
(PRTR-C17.I1), and the Generalitat de Catalunya.  For the purpose of open
access, the authors have applied a Creative Commons Attribution (CC BY) licence
to any Author Accepted Manuscript version arising from this submission.

\par\AckEC
\end{acknowledgements}

\bibliography{main}

\begin{thebibliography}{65}
\expandafter\ifx\csname natexlab\endcsname\relax\def\natexlab#1{#1}\fi

\bibitem[{{Abbott} {et~al.}(2022){Abbott}, {Aguena}, {Alarcon}, {Allam}, {Alves}, {Amon}, {Andrade-Oliveira}, {Annis}, {Avila}, {Bacon}, {Baxter}, {Bechtol}, {Becker}, {Bernstein}, {Bhargava}, {Birrer}, {Blazek}, {Brandao-Souza}, {Bridle}, {Brooks}, {Buckley-Geer}, {Burke}, {Camacho}, {Campos}, {Carnero Rosell}, {Carrasco Kind}, {Carretero}, {Castander}, {Cawthon}, {Chang}, {Chen}, {Chen}, {Choi}, {Conselice}, {Cordero}, {Costanzi}, {Crocce}, {da Costa}, {da Silva Pereira}, {Davis}, {Davis}, {De Vicente}, {DeRose}, {Desai}, {Di Valentino}, {Diehl}, {Dietrich}, {Dodelson}, {Doel}, {Doux}, {Drlica-Wagner}, {Eckert}, {Eifler}, {Elsner}, {Elvin-Poole}, {Everett}, {Evrard}, {Fang}, {Farahi}, {Fernandez}, {Ferrero}, {Fert{\'e}}, {Fosalba}, {Friedrich}, {Frieman}, {Garc{\'\i}a-Bellido}, {Gatti}, {Gaztanaga}, {Gerdes}, {Giannantonio}, {Giannini}, {Gruen}, {Gruendl}, {Gschwend}, {Gutierrez}, {Harrison}, {Hartley}, {Herner}, {Hinton}, {Hollowood}, {Honscheid}, {Hoyle}, {Huff}, {Huterer}, {Jain}, {James}, {Jarvis}, {Jeffrey}, {Jeltema}, {Kovacs}, {Krause}, {Kron}, {Kuehn}, {Kuropatkin}, {Lahav}, {Leget}, {Lemos}, {Liddle}, {Lidman}, {Lima}, {Lin}, {MacCrann}, {Maia}, {Marshall}, {Martini}, {McCullough}, {Melchior}, {Mena-Fern{\'a}ndez}, {Menanteau}, {Miquel}, {Mohr}, {Morgan}, {Muir}, {Myles}, {Nadathur}, {Navarro-Alsina}, {Nichol}, {Ogando}, {Omori}, {Palmese}, {Pandey}, {Park}, {Paz-Chinch{\'o}n}, {Petravick}, {Pieres}, {Plazas Malag{\'o}n}, {Porredon}, {Prat}, {Raveri}, {Rodriguez-Monroy}, {Rollins}, {Romer}, {Roodman}, {Rosenfeld}, {Ross}, {Rykoff}, {Samuroff}, {S{\'a}nchez}, {Sanchez}, {Sanchez}, {Sanchez Cid}, {Scarpine}, {Schubnell}, {Scolnic}, {Secco}, {Serrano}, {Sevilla-Noarbe}, {Sheldon}, {Shin}, {Smith}, {Soares-Santos}, {Suchyta}, {Swanson}, {Tabbutt}, {Tarle}, {Thomas}, {To}, {Troja}, {Troxel}, {Tucker}, {Tutusaus}, {Varga}, {Walker}, {Weaverdyck}, {Wechsler}, {Weller}, {Yanny}, {Yin}, {Zhang}, {Zuntz}, \& {DES Collaboration}}]{2022PhRvD.105b3520A}
{Abbott}, T.~M.~C., {Aguena}, M., {Alarcon}, A., {et~al.} 2022, \prd, 105, 023520

\bibitem[{{Alonso} {et~al.}(2019){Alonso}, {Sanchez}, {Slosar}, \& {LSST Dark Energy Science Collaboration}}]{2019MNRAS.484.4127A}
{Alonso}, D., {Sanchez}, J., {Slosar}, A., \& {LSST Dark Energy Science Collaboration}. 2019, \mnras, 484, 4127

\bibitem[{{Alsing} {et~al.}(2016){Alsing}, {Heavens}, {Jaffe}, {Kiessling}, {Wandelt}, \& {Hoffmann}}]{2016MNRAS.455.4452A}
{Alsing}, J., {Heavens}, A., {Jaffe}, A.~H., {et~al.} 2016, \mnras, 455, 4452

\bibitem[{{Baleato Lizancos} \& {White}(2024)}]{2024JCAP...05..010B}
{Baleato Lizancos}, A. \& {White}, M. 2024, JCAP, 05, 010

\bibitem[{{Bartelmann} \& {Schneider}(2001)}]{2001PhR...340..291B}
{Bartelmann}, M. \& {Schneider}, P. 2001, \physrep, 340, 291

\bibitem[{{Boyle}(2016)}]{2016JMP....57i2504B}
{Boyle}, M. 2016, Journal of Mathematical Physics, 57, 092504

\bibitem[{{Brown} {et~al.}(2005){Brown}, {Castro}, \& {Taylor}}]{2005MNRAS.360.1262B}
{Brown}, M.~L., {Castro}, P.~G., \& {Taylor}, A.~N. 2005, \mnras, 360, 1262

\bibitem[{{Carretero} {et~al.}(2017){Carretero}, {Tallada}, {Casals}, {Caubet}, {Castander}, {Blot}, {Alarc{\'o}n}, {Serrano}, {Fosalba}, {Acosta-Silva}, {Tonello}, {Torradeflot}, {Eriksen}, {Neissner}, \& {Delfino}}]{2017ehep.confE.488C}
{Carretero}, J., {Tallada}, P., {Casals}, J., {et~al.} 2017, in Proceedings of the European Physical Society Conference on High Energy Physics. 5-12 July, 488

\bibitem[{{Challinor} \& {Lewis}(2011)}]{2011PhRvD..84d3516C}
{Challinor}, A. \& {Lewis}, A. 2011, \prd, 84, 043516

\bibitem[{{Deshpande} {et~al.}(2020){Deshpande}, {Kitching}, {Cardone}, {Taylor}, {Casas}, {Camera}, {Carbone}, {Kilbinger}, {Pettorino}, {Sakr}, {Sapone}, {Tutusaus}, {Auricchio}, {Bodendorf}, {Bonino}, {Brescia}, {Capobianco}, {Carretero}, {Castellano}, {Cavuoti}, {Cledassou}, {Congedo}, {Conversi}, {Corcione}, {Cropper}, {Dubath}, {Dusini}, {Fabbian}, {Franceschi}, {Fumana}, {Garilli}, {Grupp}, {Hoekstra}, {Hormuth}, {Israel}, {Jahnke}, {Kermiche}, {Kubik}, {Kunz}, {Lacasa}, {Ligori}, {Lilje}, {Lloro}, {Maiorano}, {Marggraf}, {Massey}, {Mei}, {Meneghetti}, {Meylan}, {Moscardini}, {Padilla}, {Paltani}, {Pasian}, {Pires}, {Polenta}, {Poncet}, {Raison}, {Rhodes}, {Roncarelli}, {Saglia}, {Schneider}, {Secroun}, {Serrano}, {Sirri}, {Starck}, {Sureau}, {Taylor}, {Tereno}, {Toledo-Moreo}, {Valenziano}, {Wang}, \& {Zoubian}}]{2020A&A...636A..95D}
{Deshpande}, A.~C., {Kitching}, T.~D., {Cardone}, V.~F., {et~al.} 2020, \aap, 636, A95

\bibitem[{{Driscoll} \& {Healy}(1994)}]{1994AdApM..15..202D}
{Driscoll}, J.~R. \& {Healy}, D.~M. 1994, Advances in Applied Mathematics, 15, 202

\bibitem[{{Dupac} \& {Tauber}(2005)}]{2005A&A...430..363D}
{Dupac}, X. \& {Tauber}, J. 2005, \aap, 430, 363

\bibitem[{{Edmonds}(1960)}]{1960amqm.book.....E}
{Edmonds}, A.~R. 1960, {Angular Momentum in Quantum Mechanics} (Princeton University Press)

\bibitem[{{Euclid Collaboration: Blanchard} {et~al.}(2020){Euclid Collaboration: Blanchard}, {Camera}, {Carbone}, {Cardone}, {Casas}, {Clesse}, {Ili{\'c}}, {Kilbinger}, {Kitching}, {Kunz}, {Lacasa}, {Linder}, {Majerotto}, {Markovi{\v{c}}}, {Martinelli}, {Pettorino}, {Pourtsidou}, {Sakr}, {S{\'a}nchez}, {Sapone}, {Tutusaus}, {Yahia-Cherif}, {Yankelevich}, {Andreon}, {Aussel}, {Balaguera-Antol{\'\i}nez}, {Baldi}, {Bardelli}, {Bender}, {Biviano}, {Bonino}, {Boucaud}, {Bozzo}, {Branchini}, {Brau-Nogue}, {Brescia}, {Brinchmann}, {Burigana}, {Cabanac}, {Capobianco}, {Cappi}, {Carretero}, {Carvalho}, {Casas}, {Castander}, {Castellano}, {Cavuoti}, {Cimatti}, {Cledassou}, {Colodro-Conde}, {Congedo}, {Conselice}, {Conversi}, {Copin}, {Corcione}, {Coupon}, {Courtois}, {Cropper}, {Da Silva}, {de la Torre}, {Di Ferdinando}, {Dubath}, {Ducret}, {Duncan}, {Dupac}, {Dusini}, {Fabbian}, {Fabricius}, {Farrens}, {Fosalba}, {Fotopoulou}, {Fourmanoit}, {Frailis}, {Franceschi}, {Franzetti}, {Fumana}, {Galeotta}, {Gillard}, {Gillis}, {Giocoli}, {G{\'o}mez-Alvarez}, {Graci{\'a}-Carpio}, {Grupp}, {Guzzo}, {Hoekstra}, {Hormuth}, {Israel}, {Jahnke}, {Keihanen}, {Kermiche}, {Kirkpatrick}, {Kohley}, {Kubik}, {Kurki-Suonio}, {Ligori}, {Lilje}, {Lloro}, {Maino}, {Maiorano}, {Marggraf}, {Martinet}, {Marulli}, {Massey}, {Medinaceli}, {Mei}, {Mellier}, {Metcalf}, {Metge}, {Meylan}, {Moresco}, {Moscardini}, {Munari}, {Nichol}, {Niemi}, {Nucita}, {Padilla}, {Paltani}, {Pasian}, {Percival}, {Pires}, {Polenta}, {Poncet}, {Pozzetti}, {Racca}, {Raison}, {Renzi}, {Rhodes}, {Romelli}, {Roncarelli}, {Rossetti}, {Saglia}, {Schneider}, {Scottez}, {Secroun}, {Sirri}, {Stanco}, {Starck}, {Sureau}, {Tallada-Cresp{\'\i}}, {Tavagnacco}, {Taylor}, {Tenti}, {Tereno}, {Toledo-Moreo}, {Torradeflot}, {Valenziano}, {Vassallo}, {Verdoes Kleijn}, {Viel}, {Wang}, {Zacchei}, {Zoubian}, \& {Zucca}}]{2020A&A...642A.191E}
{Euclid Collaboration: Blanchard}, A., {Camera}, S., {Carbone}, C., {et~al.} 2020, \aap, 642, A191

\bibitem[{{Euclid Collaboration: Castander} {et~al.}(2024){Euclid Collaboration: Castander}, {Fosalba}, {Stadel}, {Potter}, {Carretero}, {Tallada-Cresp{\'\i}}, {Pozzetti}, {Bolzonella}, {Mamon}, {Blot}, {Hoffmann}, {Huertas-Company}, {Monaco}, {Gonzalez}, {De Lucia}, {Scarlata}, {Breton}, {Linke}, {Viglione}, {Li}, {Zhai}, {Baghkhani}, {Pardede}, {Neissner}, {Teyssier}, {Crocce}, {Tutusaus}, {Miller}, {Congedo}, {Biviano}, {Hirschmann}, {Pezzotta}, {Aussel}, {Hoekstra}, {Kitching}, {Percival}, {Guzzo}, {Mellier}, {Oesch}, {Bowler}, {Bruton}, {Allevato}, {Gonzalez-Perez}, {Manera}, {Avila}, {Kov{\'a}cs}, {Aghanim}, {Altieri}, {Amara}, {Amendola}, {Andreon}, {Auricchio}, {Baldi}, {Balestra}, {Bardelli}, {Bender}, {Bodendorf}, {Bonino}, {Branchini}, {Brescia}, {Brinchmann}, {Camera}, {Capobianco}, {Carbone}, {Casas}, {Castellano}, {Cavuoti}, {Cimatti}, {Conselice}, {Conversi}, {Copin}, {Corcione}, {Courbin}, {Courtois}, {Da Silva}, {Degaudenzi}, {Di Giorgio}, {Dinis}, {Douspis}, {Dubath}, {Duncan}, {Dupac}, {Dusini}, {Ealet}, {Farina}, {Farrens}, {Ferriol}, {Fotopoulou}, {Fourmanoit}, {Frailis}, {Franceschi}, {Franzetti}, {Galeotta}, {Gillard}, {Gillis}, {Giocoli}, {G{\'o}mez-Alvarez}, {Granett}, {Grazian}, {Grupp}, {Haugan}, {Holliman}, {Holmes}, {Hook}, {Hormuth}, {Hornstrup}, {Hudelot}, {Jahnke}, {Jhabvala}, {Joachimi}, {Keih{\"a}nen}, {Kermiche}, {Kiessling}, {Kilbinger}, {Kohley}, {Kubik}, {K{\"u}mmel}, {Kunz}, {Kurki-Suonio}, {Lahav}, {Laureijs}, {Le Mignant}, {Ligori}, {Lilje}, {Lindholm}, {Lloro}, {Maino}, {Maiorano}, {Mansutti}, {Marggraf}, {Markovic}, {Martinet}, {Marulli}, {Massey}, {Masters}, {Maurogordato}, {McCracken}, {Medinaceli}, {Mei}, {Melchior}, {Meneghetti}, {Merlin}, {Meylan}, {Mohr}, {Moresco}, {Moscardini}, {Munari}, {Nakajima}, {Nichol}, {Niemi}, {Padilla}, {Paech}, {Paltani}, {Pasian}, {Peacock}, {Pedersen}, {Pettorino}, {Pires}, {Polenta}, {Poncet}, {Popa}, {Raison}, {Rebolo}, {Renzi}, {Rhodes}, {Riccio}, {Romelli}, {Roncarelli}, {Rosset}, {Rossetti}, {Saglia}, {Sapone}, {Schirmer}, {Schneider}, {Schrabback}, {Scodeggio}, {Secroun}, {Seidel}, {Serrano}, {Sirignano}, {Sirri}, {Stanco}, {Starck}, {Taylor}, {Teplitz}, {Tereno}, {Toledo-Moreo}, {Torradeflot}, {Tsyganov}, {Valenziano}, {Vassallo}, {Veropalumbo}, {Wang}, {Weller}, {Zacchei}, {Zamorani}, {Zerbi}, {Zoubian}, {Zucca}, {Baccigalupi}, {Bernardeau}, {Boucaud}, {Bozzo}, {Burigana}, {Calabrese}, {Casenove}, {Castignani}, {Colodro-Conde}, {Di Ferdinando}, {Escartin Vigo}, {Fabbian}, {Finelli}, {Gracia-Carpio}, {Ili{\'c}}, {Liebing}, {Marcin}, {Martinelli}, {Matthew}, {Mauri}, {P{\"o}ntinen}, {Porciani}, {Sakr}, {Scottez}, {Sefusatti}, {Steinwagner}, {Tenti}, {Viel}, {Wiesmann}, {Akrami}, {Anselmi}, {Archidiacono}, {Atrio-Barandela}, {Aubourg}, {Balaguera-Antolinez}, {Ballardini}, {Bertacca}, {Bethermin}, {Blanchard}, {B{\"o}hringer}, {Borgani}, {Bouvard}, {Cabanac}, {Calabro}, {Camacho Quevedo}, {Canas-Herrera}, {Cappi}, {Caro}, {Carvalho}, {Castro}, {Chambers}, {Contarini}, {Contini}, {Cooray}, {Costanzi}, {Cucciati}, {Davini}, {De Caro}, {de la Torre}, {Desprez}, {D{\'\i}az-S{\'a}nchez}, {Diaz}, {Di Domizio}, {Dole}, {Escoffier}, {Ezziati}, {Ferrari}, {Ferreira}, {Ferrero}, {Finoguenov}, {Fontana}, {Fornari}, {Gabarra}, {Ganga}, {Garc{\'\i}a-Bellido}, {Gasparetto}, {Gaztanaga}, {Giacomini}, {Gianotti}, {Gonzalez}, {Gozaliasl}, {Hall}, {Hartley}, {Hildebrandt}, {Hjorth}, {Holland}, {Ilbert}, {Joudaki}, {Jullo}, {Kajava}, {Kansal}, {Karagiannis}, {Kirkpatrick}, {Le Graet}, {Legrand}, {Lesgourgues}, {Liaudat}, {Loureiro}, {Macias-Perez}, {Magliocchetti}, {Mancini}, {Mannucci}, {Maoli}, {Martins}, {Maurin}, {Metcalf}, {Migliaccio}, {Miluzio}, {Mora}, {Moretti}, {Morgante}, {Nadathur}, {Nicastro}, {Walton}, {Oguri}, {Patrizii}, {Popa}, {Pourtsidou}, {Reimberg}, {Risso}, {Rocci}, {Rollins}, {Rusholme}, {Sahl{\'e}n}, {S{\'a}nchez}, {Schaye}, {Schewtschenko}, {Schneider}, {Schultheis}, {Sereno}, {Shankar}, {Shulevski}, {Silvestri}, {Simon}, {Spurio Mancini}, {Stanford}, {Tanidis}, {Tao}, {Tessore}, {Testera}, {Tewes}, {Toft}, {Tosi}, {Troja}, {Tucci}, {Valieri}, {Valiviita}, {Vergani}, {Vernizzi}, {Verza}, {Vielzeuf}, {Weaver}, {Zalesky}, {Dimauro}, {Duc}, {Fang}, {Ferguson}, {Gutierrez}, {Kova\{{\v{c}}\}i{\'c}}, {Kruk}, {Le Brun}, {Montoro}, {Murray}, {Pagano}, {Paoletti}, {Sarpa}, {Viitanen}, {Mart{\'\i}n-Fleitas}, \& {Yung}}]{2024arXiv240513495E}
{Euclid Collaboration: Castander}, F.~J., {Fosalba}, P., {Stadel}, J., {et~al.} 2024, arXiv:2405.13495

\bibitem[{{Euclid Collaboration: Lepori} {et~al.}(2022){Euclid Collaboration: Lepori}, {Tutusaus}, {Viglione}, {Bonvin}, {Camera}, {Castander}, {Durrer}, {Fosalba}, {Jelic-Cizmek}, {Kunz}, {Adamek}, {Casas}, {Martinelli}, {Sakr}, {Sapone}, {Amara}, {Auricchio}, {Bodendorf}, {Bonino}, {Branchini}, {Brescia}, {Brinchmann}, {Capobianco}, {Carbone}, {Carretero}, {Castellano}, {Cavuoti}, {Cimatti}, {Cledassou}, {Congedo}, {Conselice}, {Conversi}, {Copin}, {Corcione}, {Courbin}, {Da Silva}, {Degaudenzi}, {Douspis}, {Dubath}, {Dupac}, {Dusini}, {Ealet}, {Farrens}, {Ferriol}, {Franceschi}, {Fumana}, {Garilli}, {Gillard}, {Gillis}, {Giocoli}, {Grazian}, {Grupp}, {Guzzo}, {Haugan}, {Holmes}, {Hormuth}, {Hudelot}, {Jahnke}, {Kermiche}, {Kiessling}, {Kilbinger}, {Kitching}, {K{\"u}mmel}, {Kurki-Suonio}, {Ligori}, {Lilje}, {Lloro}, {Mansutti}, {Marggraf}, {Markovic}, {Marulli}, {Massey}, {Maurogordato}, {Melchior}, {Meneghetti}, {Merlin}, {Meylan}, {Moresco}, {Moscardini}, {Munari}, {Nakajima}, {Niemi}, {Padilla}, {Paltani}, {Pasian}, {Pedersen}, {Percival}, {Pettorino}, {Pires}, {Poncet}, {Popa}, {Pozzetti}, {Raison}, {Rhodes}, {Roncarelli}, {Rossetti}, {Saglia}, {Schneider}, {Secroun}, {Seidel}, {Serrano}, {Sirignano}, {Sirri}, {Stanco}, {Starck}, {Tallada-Cresp{\'\i}}, {Taylor}, {Tereno}, {Toledo-Moreo}, {Torradeflot}, {Valentijn}, {Valenziano}, {Wang}, {Weller}, {Zamorani}, {Zoubian}, {Andreon}, {Bardelli}, {Fabbian}, {Graci{\'a}-Carpio}, {Maino}, {Medinaceli}, {Mei}, {Renzi}, {Romelli}, {Sureau}, {Vassallo}, {Zacchei}, {Zucca}, {Baccigalupi}, {Balaguera-Antol{\'\i}nez}, {Bernardeau}, {Biviano}, {Blanchard}, {Bolzonella}, {Borgani}, {Bozzo}, {Burigana}, {Cabanac}, {Cappi}, {Carvalho}, {Castignani}, {Colodro-Conde}, {Coupon}, {Courtois}, {Cuby}, {Davini}, {de la Torre}, {Di Ferdinando}, {Farina}, {Ferreira}, {Finelli}, {Galeotta}, {Ganga}, {Garcia-Bellido}, {Gaztanaga}, {Gozaliasl}, {Hook}, {Ili{\'c}}, {Joachimi}, {Kansal}, {Keihanen}, {Kirkpatrick}, {Lindholm}, {Mainetti}, {Maoli}, {Martinet}, {Maturi}, {Metcalf}, {Monaco}, {Morgante}, {Nightingale}, {Nucita}, {Patrizii}, {Popa}, {Potter}, {Riccio}, {S{\'a}nchez}, {Schirmer}, {Schultheis}, {Scottez}, {Sefusatti}, {Tramacere}, {Valiviita}, {Viel}, \& {Hildebrandt}}]{2022A&A...662A..93E}
{Euclid Collaboration: Lepori}, F., {Tutusaus}, I., {Viglione}, C., {et~al.} 2022, \aap, 662, A93

\bibitem[{{Euclid Collaboration: Mellier} {et~al.}(2024){Euclid Collaboration: Mellier}, {Abdurro'uf}, {Acevedo Barroso}, {Ach{\'u}carro}, {Adamek}, {Adam}, {Addison}, {Aghanim}, {Aguena}, {Ajani}, {Akrami}, {Al-Bahlawan}, {Alavi}, {Albuquerque}, {Alestas}, {Alguero}, {Allaoui}, {Allen}, {Allevato}, {Alonso-Tetilla}, {Altieri}, {Alvarez-Candal}, {Amara}, {Amendola}, {Amiaux}, {Andika}, {Andreon}, {Andrews}, {Angora}, {Angulo}, {Annibali}, {Anselmi}, {Anselmi}, {Arcari}, {Archidiacono}, {Aric{\`o}}, {Arnaud}, {Arnouts}, {Asgari}, {Asorey}, {Atayde}, {Atek}, {Atrio-Barandela}, {Aubert}, {Aubourg}, {Auphan}, {Auricchio}, {Aussel}, {Aussel}, {Avelino}, {Avgoustidis}, {Avila}, {Awan}, {Azzollini}, {Baccigalupi}, {Bachelet}, {Bacon}, {Baes}, {Bagley}, {Bahr-Kalus}, {Balaguera-Antolinez}, {Balbinot}, {Balcells}, {Baldi}, {Baldry}, {Balestra}, {Ballardini}, {Ballester}, {Balogh}, {Ba{\~n}ados}, {Barbier}, {Bardelli}, {Barreiro}, {Barriere}, {Barros}, {Barthelemy}, {Bartolo}, {Basset}, {Battaglia}, {Battisti}, {Baugh}, {Baumont}, {Bazzanini}, {Beaulieu}, {Beckmann}, {Belikov}, {Bel}, {Bellagamba}, {Bella}, {Bellini}, {Benabed}, {Bender}, {Benevento}, {Bennett}, {Benson}, {Bergamini}, {Bermejo-Climent}, {Bernardeau}, {Bertacca}, {Berthe}, {Berthier}, {Bethermin}, {Beutler}, {Bevillon}, {Bhargava}, {Bhatawdekar}, {Bisigello}, {Biviano}, {Blake}, {Blanchard}, {Blazek}, {Blot}, {Bosco}, {Bodendorf}, {Boenke}, {B{\"o}hringer}, {Bolzonella}, {Bonchi}, {Bonici}, {Bonino}, {Bonino}, {Bonvin}, {Bon}, {Booth}, {Borgani}, {Borlaff}, {Borsato}, {Bosco}, {Bose}, {Botticella}, {Boucaud}, {Bouche}, {Boucher}, {Boutigny}, {Bouvard}, {Bouy}, {Bowler}, {Bozza}, {Bozzo}, {Branchini}, {Brau-Nogue}, {Brekke}, {Bremer}, {Brescia}, {Breton}, {Brinchmann}, {Brinckmann}, {Brockley-Blatt}, {Brodwin}, {Brouard}, {Brown}, {Bruton}, {Bucko}, {Buddelmeijer}, {Buenadicha}, {Buitrago}, {Burger}, {Burigana}, {Busillo}, {Busonero}, {Cabanac}, {Cabayol-Garcia}, {Cagliari}, {Caillat}, {Caillat}, {Calabrese}, {Calabro}, {Calderone}, {Calura}, {Camacho Quevedo}, {Camera}, {Campos}, {Canas-Herrera}, {Candini}, {Cantiello}, {Capobianco}, {Cappellaro}, {Cappelluti}, {Cappi}, {Caputi}, {Cara}, {Carbone}, {Cardone}, {Carella}, {Carlberg}, {Carle}, {Carminati}, {Caro}, {Carrasco}, {Carretero}, {Carrilho}, {Carron Duque}, {Carry}, {Carvalho}, {Carvalho}, {Casas}, {Casas}, {Casenove}, {Casey}, {Cassata}, {Castander}, {Castelao}, {Castellano}, {Castiblanco}, {Castignani}, {Castro}, {Cavet}, {Cavuoti}, {Chabaud}, {Chambers}, {Charles}, {Charlot}, {Chartab}, {Chary}, {Chaumeil}, {Cho}, {Chon}, {Ciancetta}, {Ciliegi}, {Cimatti}, {Cimino}, {Cioni}, {Claydon}, {Cleland}, {Cl{\'e}ment}, {Clements}, {Clerc}, {Clesse}, {Codis}, {Cogato}, {Colbert}, {Cole}, {Coles}, {Collett}, {Collins}, {Colodro-Conde}, {Colombo}, {Combes}, {Conforti}, {Congedo}, {Conseil}, {Conselice}, {Contarini}, {Contini}, {Conversi}, {Cooray}, {Copin}, {Corasaniti}, {Corcho-Caballero}, {Corcione}, {Cordes}, {Corpace}, {Correnti}, {Costanzi}, {Costille}, {Courbin}, {Courcoult Mifsud}, {Courtois}, {Cousinou}, {Covone}, {Cowell}, {Cragg}, {Cresci}, {Cristiani}, {Crocce}, {Cropper}, {E Crouzet}, {Csizi}, {Cuby}, {Cucchetti}, {Cucciati}, {Cuillandre}, {Cunha}, {Cuozzo}, {Daddi}, {D'Addona}, {Dafonte}, {Dagoneau}, {Dalessandro}, {Dalton}, {D'Amico}, {Dannerbauer}, {Danto}, {Das}, {Da Silva}, {da Silva}, {Daste}, {Davies}, {Davini}, {de Boer}, {Decarli}, {De Caro}, {Degaudenzi}, {Degni}, {de Jong}, {de la Bella}, {de la Torre}, {Delhaise}, {Delley}, {Delucchi}, {De Lucia}, {Denniston}, {De Paolis}, {De Petris}, {Derosa}, {Desai}, {Desjacques}, {Despali}, {Desprez}, {De Vicente-Albendea}, {Deville}, {Dias}, {D{\'\i}az-S{\'a}nchez}, {Diaz}, {Di Domizio}, {Diego}, {Di Ferdinando}, {Di Giorgio}, {Dimauro}, {Dinis}, {Dolag}, {Dolding}, {Dole}, {Dom{\'\i}nguez S{\'a}nchez}, {Dor{\'e}}, {Dournac}, {Douspis}, {Dreihahn}, {Droge}, {Dryer}, {Dubath}, {Duc}, {Ducret}, {Duffy}, {Dufresne}, {Duncan}, {Dupac}, {Duret}, {Durrer}, {Durret}, {Dusini}, {Ealet}, {Eggemeier}, {Eisenhardt}, {Elbaz}, {Elkhashab}, {Ellien}, {Endicott}, {Enia}, {Erben}, {Escartin Vigo}, {Escoffier}, {Escudero Sanz}, {Essert}, {Ettori}, {Ezziati}, {Fabbian}, {Fabricius}, {Fang}, {Farina}, {Farina}, {Farinelli}, {Farrens}, {Faustini}, {Feltre}, {Ferguson}, {Ferrando}, {Ferrari}, {Ferr{\'e}-Mateu}, {Ferreira}, {Ferreras}, {Ferrero}, {Ferriol}, {Ferruit}, {Filleul}, {Finelli}, {Finkelstein}, {Finoguenov}, {Fiorini}, {Flentge}, {Focardi}, {Fonseca}, {Fontana}, {Fontanot}, {Fornari}, {Fosalba}, {Fossati}, {Fotopoulou}, {Fouchez}, {Fourmanoit}, {Frailis}, {Fraix-Burnet}, {Franceschi}, {Franco}, {Franzetti}, {Freihoefer}, {Frittoli}, {Frugier}, {Frusciante}, {Fumagalli}, {Fumagalli}, {Fumana}, {Fu}, {Gabarra}, {Galeotta}, {Galluccio}, {Ganga}, {Gao}, {Garc{\'\i}a-Bellido}, {Garcia}, {Gardner}, {Garilli}, {Gaspar-Venancio}, {Gasparetto}, {Gautard}, {Gavazzi}, {Gaztanaga}, {Genolet}, {Genova Santos}, {Gentile}, {George}, {Ghaffari}, {Giacomini}, {Gianotti}, {Gibb}, {Gillard}, {Gillis}, {Ginolfi}, {Giocoli}, {Girardi}, {Giri}, {Goh}, {G{\'o}mez-Alvarez}, {Gonzalez}, {Gonzalez}, {Gonzalez}, {Gouyou Beauchamps}, {Gozaliasl}, {Gracia-Carpio}, {Grandis}, {Granett}, {Granvik}, {Grazian}, {Gregorio}, {Grenet}, {Grillo}, {Grupp}, {Gruppioni}, {Gruppuso}, {Guerbuez}, {Guerrini}, {Guidi}, {Guillard}, {Gutierrez}, {Guttridge}, {Guzzo}, {Gwyn}, {Haapala}, {Haase}, {Haddow}, {Hailey}, {Hall}, {Hall}, {Hamaus}, {Haridasu}, {Harnois-D{\'e}raps}, {Harper}, {Hartley}, {Hasinger}, {Hassani}, {Hatch}, {Haugan}, {H{\"a}u{\ss}ler}, {Heavens}, {Heisenberg}, {Helmi}, {Helou}, {Hemmati}, {Henares}, {Herent}, {Hern{\'a}ndez-Monteagudo}, {Heuberger}, {Hewett}, {Heydenreich}, {Hildebrandt}, {Hirschmann}, {Hjorth}, {Hoar}, {Hoekstra}, {Holland}, {Holliman}, {Holmes}, {Hook}, {Horeau}, {Hormuth}, {Hornstrup}, {Hosseini}, {Hu}, {Hudelot}, {Hudson}, {Huertas-Company}, {Huff}, {Hughes}, {Humphrey}, {Hunt}, {Huynh}, {Ibata}, {Ichikawa}, {Iglesias-Groth}, {Ilbert}, {Ili{\'c}}, {Ingoglia}, {Iodice}, {Israel}, {Israelsson}, {Izzo}, {Jablonka}, {Jackson}, {Jacobson}, {Jafariyazani}, {Jahnke}, {Jansen}, {Jarvis}, {Jasche}, {Jauzac}, {Jeffrey}, {Jhabvala}, {Jimenez-Teja}, {Jimenez Mu{\~n}oz}, {Joachimi}, {Johansson}, {Joudaki}, {Jullo}, {Kajava}, {Kang}, {Kannawadi}, {Kansal}, {Karagiannis}, {K{\"a}rcher}, {Kashlinsky}, {Kazandjian}, {Keck}, {Keih{\"a}nen}, {Kerins}, {Kermiche}, {Khalil}, {Kiessling}, {Kiiveri}, {Kilbinger}, {Kim}, {King}, {Kirkpatrick}, {Kitching}, {Kluge}, {Knabenhans}, {Knapen}, {Knebe}, {Kneib}, {Kohley}, {Koopmans}, {Koskinen}, {Koulouridis}, {Kou}, {Kov{\'a}cs}, {Kova\{{\v{c}}\}i{\'c}}, {Kowalczyk}, {Koyama}, {Kraljic}, {Krause}, {Kruk}, {Kubik}, {Kuchner}, {Kuijken}, {K{\"u}mmel}, {Kunz}, {Kurki-Suonio}, {Lacasa}, {Lacey}, {La Franca}, {Lagarde}, {Lahav}, {Laigle}, {La Marca}, {La Marle}, {Lamine}, {Lam}, {Lan{\c{c}}on}, {Landt}, {Langer}, {Lapi}, {Larcheveque}, {Larsen}, {Lattanzi}, {Laudisio}, {Laugier}, {Laureijs}, {Lavaux}, {Lawrenson}, {Lazanu}, {Lazeyras}, {Le Boulc'h}, {Le Brun}, {Le Brun}, {Leclercq}, {Lee}, {Le Graet}, {Legrand}, {Leirvik}, {Le Jeune}, {Lembo}, {Le Mignant}, {Lepinzan}, {Lepori}, {Lesci}, {Lesgourgues}, {Leuzzi}, {Levi}, {Liaudat}, {Libet}, {Liebing}, {Ligori}, {Lilje}, {Lin}, {Linde}, {Linder}, {Lindholm}, {Linke}, {Li}, {Liu}, {Lloro}, {Lobo}, {Lodieu}, {Lombardi}, {Lombriser}, {Lonare}, {Longo}, {L{\'o}pez-Caniego}, {Lopez Lopez}, {Alvarez}, {Loureiro}, {Loveday}, {Lusso}, {Macias-Perez}, {Maciaszek}, {Magliocchetti}, {Magnard}, {Magnier}, {Magro}, {Mahler}, {Mainetti}, {Maino}, {Maiorano}, {Maiorano}, {Malavasi}, {Mamon}, {Mancini}, {Mandelbaum}, {Manera}, {Manj{\'o}n-Garc{\'\i}a}, {Mannucci}, {Mansutti}, {Manteiga Outeiro}, {Maoli}, {Maraston}, {Marcin}, {Marcos-Arenal}, {Margalef-Bentabol}, {Marggraf}, {Marinucci}, {Marinucci}, {Markovic}, {Marleau}, {Marpaud}, {Martignac}, {Mart{\'\i}n-Fleitas}, {Martin-Moruno}, {Martin}, {Martinelli}, {Martinet}, {Martin}, {Martins}, {Marulli}, {Massari}, {Massey}, {Masters}, {Matarrese}, {Matsuoka}, {Matthew}, {Maughan}, {Mauri}, {Maurin}, {Maurogordato}, {McCarthy}, {McConnachie}, {McCracken}, {McDonald}, {McEwen}, {McPartland}, {Medinaceli}, {Mehta}, {Mei}, {Melchior}, {Melin}, {M{\'e}nard}, {Mendes}, {Mendez-Abreu}, {Meneghetti}, {Mercurio}, {Merlin}, {Metcalf}, {Meylan}, {Migliaccio}, {Mignoli}, {Miller}, {Miluzio}, {Milvang-Jensen}, {Mimoso}, {Miquel}, {Miyatake}, {Mobasher}, {Mohr}, {Monaco}, {Mongui{\'o}}, {Montoro}, {Mora}, {Moradinezhad Dizgah}, {Moresco}, {Moretti}, {Morgante}, {Morisset}, {Moriya}, {Morris}, {Mortlock}, {Moscardini}, {Mota}, {Moustakas}, {Moutard}, {M{\"u}ller}, {Munari}, {Murphree}, {Murray}, {Murray}, {Musi}, {Nadathur}, {Nagam}, {Nagao}, {Naidoo}, {Nakajima}, {Nally}, {Natoli}, {Navarro-Alsina}, {Navarro Girones}, {Neissner}, {Nersesian}, {Nesseris}, {Nguyen-Kim}, {Nicastro}, {Nichol}, {Nielbock}, {Niemi}, {Nieto}, {Nilsson}, {Noller}, {Norberg}, {Nourizonoz}, {Ntelis}, {Nucita}, {Nugent}, {Nunes}, {Nutma}, {Ocampo}, {Odier}, {Oesch}, {Oguri}, {Magalhaes Oliveira}, {Onoue}, {Oosterbroek}, {Oppizzi}, {Ordenovic}, {Osato}, {Pacaud}, {Pace}, {Padilla}, {Paech}, {Pagano}, {Page}, {Palazzi}, {Paltani}, {Pamuk}, {Pandolfi}, {Paoletti}, {Paolillo}, {Papaderos}, {Pardede}, {Parimbelli}, {Parmar}, {Partmann}, {Pasian}, {Passalacqua}, {Paterson}, {Patrizii}, {Pattison}, {Paulino-Afonso}, {Paviot}, {Peacock}, {Pearce}, {Pedersen}, {Peel}, {Peletier}, {Pellejero Ibanez}, {Pello}, {Penny}, {Percival}, {Perez-Garrido}, {Perotto}, {Pettorino}, {Pezzotta}, {Pezzuto}, {Philippon}, {Piersanti}, {Pietroni}, {Piga}, {Pilo}, {Pires}, {Pisani}, {Pizzella}, {Pizzuti}, {Plana}, {Polenta}, {Pollack}, {Poncet}, {P{\"o}ntinen}, {Pool}, {Popa}, {Popa}, {Popp}, {Porciani}, {Porth}, {Potter}, {Poulain}, {Pourtsidou}, {Pozzetti}, {Prandoni}, {Pratt}, {Prezelus}, {Prieto}, {Pugno}, {Quai}, {Quilley}, {Racca}, {Raccanelli}, {R{\'a}cz}, {Radinovi{\'c}}, {Radovich}, {Ragagnin}, {Ragnit}, {Raison},
  {Ramos-Chernenko}, {Ranc}, {Raylet}, {Rebolo}, {Refregier}, {Reimberg}, {Reiprich}, {Renk}, {Renzi}, {Retre}, {Revaz}, {Reyl{\'e}}, {Reynolds}, {Rhodes}, {Ricci}, {Ricci}, {Riccio}, {Ricken}, {Rissanen}, {Risso}, {Rix}, {Robin}, {Rocca-Volmerange}, {Rocci}, {Rodenhuis}, {Rodighiero}, {Rodriguez Monroy}, {Rollins}, {Romanello}, {Roman}, {Romelli}, {Romero-Gomez}, {Roncarelli}, {Rosati}, {Rosset}, {Rossetti}, {Roster}, {Rottgering}, {Rozas-Fern{\'a}ndez}, {Ruane}, {Rubino-Martin}, {Rudolph}, {Ruppin}, {Rusholme}, {Sacquegna}, {S{\'a}ez-Casares}, {Saga}, {Saglia}, {Sahl{\'e}n}, {Saifollahi}, {Sakr}, {Salvalaggio}, {Salvaterra}, {Salvati}, {Salvato}, {Salvignol}, {S{\'a}nchez}, {Sanchez}, {Sanders}, {Sapone}, {Saponara}, {Sarpa}, {Sarron}, {Sartori}, {Sassolas}, {Sauniere}, {Sauvage}, {Sawicki}, {Scaramella}, {Scarlata}, {Scharr{\'e}}, {Schaye}, {Schewtschenko}, {Schindler}, {Schinnerer}, {Schirmer}, {Schmidt}, {Schmidt}, {Schmidt}, {Schneider}, {Schneider}, {Schneider}, {Sch{\"o}neberg}, {Schrabback}, {Schultheis}, {Schulz}, {Schwartz}, {Sciotti}, {Scodeggio}, {Scognamiglio}, {Scott}, {Scottez}, {Secroun}, {Sefusatti}, {Seidel}, {Seiffert}, {Sellentin}, {Selwood}, {Semboloni}, {Sereno}, {Serjeant}, {Serrano}, {Shankar}, {Sharples}, {Short}, {Shulevski}, {Shuntov}, {Sias}, {Sikkema}, {Silvestri}, {Simon}, {Sirignano}, {Sirri}, {Skottfelt}, {Slezak}, {Sluse}, {Smith}, {Smith}, {Smith}, {Smit}, {Soldano}, {Solheim}, {Sorce}, {Sorrenti}, {Soubrie}, {Spinoglio}, {Spurio Mancini}, {Stadel}, {Stagnaro}, {Stanco}, {Stanford}, {Starck}, {Stassi}, {Steinwagner}, {Stern}, {Stone}, {Strada}, {Strafella}, {Stramaccioni}, {Surace}, {Sureau}, {Suyu}, {Swindells}, {Szafraniec}, {Szapudi}, {Taamoli}, {Talia}, {Tallada-Cresp{\'\i}}, {Tanidis}, {Tao}, {Tarr{\'\i}o}, {Tavagnacco}, {Taylor}, {Taylor}, {Taylor}, {Teixeira}, {Tenti}, {Teodoro Idiago}, {Teplitz}, {Tereno}, {Tessore}, {Testa}, {Testera}, {Tewes}, {Teyssier}, {Theret}, {Thizy}, {Thomas}, {Toba}, {Toft}, {Toledo-Moreo}, {Tolstoy}, {Tommasi}, {Torbaniuk}, {Torradeflot}, {Tortora}, {Tosi}, {Tosti}, {Trifoglio}, {Troja}, {Trombetti}, {Tronconi}, {Tsedrik}, {Tsyganov}, {Tucci}, {Tutusaus}, {Uhlemann}, {Ulivi}, {Urbano}, {Vacher}, {Vaillon}, {Valdes}, {Valentijn}, {Valenziano}, {Valieri}, {Valiviita}, {Van den Broeck}, {Vassallo}, {Vavrek}, {Venemans}, {Venhola}, {Ventura}, {Verdoes Kleijn}, {Vergani}, {Verma}, {Vernizzi}, {Veropalumbo}, {Verza}, {Vescovi}, {Vibert}, {Viel}, {Vielzeuf}, {Viglione}, {Viitanen}, {Villaescusa-Navarro}, {Vinciguerra}, {Visticot}, {Voggel}, {von Wietersheim-Kramsta}, {Vriend}, {Wachter}, {Walmsley}, {Walth}, {Walton}, {Walton}, {Wander}, {Wang}, {Wang}, {Weaver}, {Weller}, {Whalen}, {Wiesmann}, {Wilde}, {Williams}, {Winther}, {Wittje}, {Wong}, {Wright}, {Yankelevich}, {Yeung}, {Youles}, {Yung}, {Zacchei}, {Zalesky}, {Zamorani}, {Zamorano Vitorelli}, {Zanoni Marc}, {Zennaro}, {Zerbi}, {Zinchenko}, {Zoubian}, {Zucca}, \& {Zumalacarregui}}]{2024arXiv240513491E}
{Euclid Collaboration: Mellier}, Y., {Abdurro'uf}, {Acevedo Barroso}, J.~A., {et~al.} 2024, arXiv:2405.13491

\bibitem[{{Euclid Collaboration: Scaramella} {et~al.}(2022){Euclid Collaboration: Scaramella}, {Amiaux}, {Mellier}, {Burigana}, {Carvalho}, {Cuillandre}, {Da Silva}, {Derosa}, {Dinis}, {Maiorano}, {Maris}, {Tereno}, {Laureijs}, {Boenke}, {Buenadicha}, {Dupac}, {Gaspar Venancio}, {G{\'o}mez-{\'A}lvarez}, {Hoar}, {Lorenzo Alvarez}, {Racca}, {Saavedra-Criado}, {Schwartz}, {Vavrek}, {Schirmer}, {Aussel}, {Azzollini}, {Cardone}, {Cropper}, {Ealet}, {Garilli}, {Gillard}, {Granett}, {Guzzo}, {Hoekstra}, {Jahnke}, {Kitching}, {Maciaszek}, {Meneghetti}, {Miller}, {Nakajima}, {Niemi}, {Pasian}, {Percival}, {Pottinger}, {Sauvage}, {Scodeggio}, {Wachter}, {Zacchei}, {Aghanim}, {Amara}, {Auphan}, {Auricchio}, {Awan}, {Balestra}, {Bender}, {Bodendorf}, {Bonino}, {Branchini}, {Brau-Nogue}, {Brescia}, {Candini}, {Capobianco}, {Carbone}, {Carlberg}, {Carretero}, {Casas}, {Castander}, {Castellano}, {Cavuoti}, {Cimatti}, {Cledassou}, {Congedo}, {Conselice}, {Conversi}, {Copin}, {Corcione}, {Costille}, {Courbin}, {Degaudenzi}, {Douspis}, {Dubath}, {Duncan}, {Dusini}, {Farrens}, {Ferriol}, {Fosalba}, {Fourmanoit}, {Frailis}, {Franceschi}, {Franzetti}, {Fumana}, {Gillis}, {Giocoli}, {Grazian}, {Grupp}, {Haugan}, {Holmes}, {Hormuth}, {Hudelot}, {Kermiche}, {Kiessling}, {Kilbinger}, {Kohley}, {Kubik}, {K{\"u}mmel}, {Kunz}, {Kurki-Suonio}, {Lahav}, {Ligori}, {Lilje}, {Lloro}, {Mansutti}, {Marggraf}, {Markovic}, {Marulli}, {Massey}, {Maurogordato}, {Melchior}, {Merlin}, {Meylan}, {Mohr}, {Moresco}, {Morin}, {Moscardini}, {Munari}, {Nichol}, {Padilla}, {Paltani}, {Peacock}, {Pedersen}, {Pettorino}, {Pires}, {Poncet}, {Popa}, {Pozzetti}, {Raison}, {Rebolo}, {Rhodes}, {Rix}, {Roncarelli}, {Rossetti}, {Saglia}, {Schneider}, {Schrabback}, {Secroun}, {Seidel}, {Serrano}, {Sirignano}, {Sirri}, {Skottfelt}, {Stanco}, {Starck}, {Tallada-Cresp{\'\i}}, {Tavagnacco}, {Taylor}, {Teplitz}, {Toledo-Moreo}, {Torradeflot}, {Trifoglio}, {Valentijn}, {Valenziano}, {Verdoes Kleijn}, {Wang}, {Welikala}, {Weller}, {Wetzstein}, {Zamorani}, {Zoubian}, {Andreon}, {Baldi}, {Bardelli}, {Boucaud}, {Camera}, {Di Ferdinando}, {Fabbian}, {Farinelli}, {Galeotta}, {Graci{\'a}-Carpio}, {Maino}, {Medinaceli}, {Mei}, {Neissner}, {Polenta}, {Renzi}, {Romelli}, {Rosset}, {Sureau}, {Tenti}, {Vassallo}, {Zucca}, {Baccigalupi}, {Balaguera-Antol{\'\i}nez}, {Battaglia}, {Biviano}, {Borgani}, {Bozzo}, {Cabanac}, {Cappi}, {Casas}, {Castignani}, {Colodro-Conde}, {Coupon}, {Courtois}, {Cuby}, {de la Torre}, {Desai}, {Dole}, {Fabricius}, {Farina}, {Ferreira}, {Finelli}, {Flose-Reimberg}, {Fotopoulou}, {Ganga}, {Gozaliasl}, {Hook}, {Keihanen}, {Kirkpatrick}, {Liebing}, {Lindholm}, {Mainetti}, {Martinelli}, {Martinet}, {Maturi}, {McCracken}, {Metcalf}, {Morgante}, {Nightingale}, {Nucita}, {Patrizii}, {Potter}, {Riccio}, {S{\'a}nchez}, {Sapone}, {Schewtschenko}, {Schultheis}, {Scottez}, {Teyssier}, {Tutusaus}, {Valiviita}, {Viel}, {Vriend}, \& {Whittaker}}]{2022A&A...662A.112E}
{Euclid Collaboration: Scaramella}, R., {Amiaux}, J., {Mellier}, Y., {et~al.} 2022, \aap, 662, A112

\bibitem[{{G{\'o}rski} {et~al.}(2005){G{\'o}rski}, {Hivon}, {Banday}, {Wandelt}, {Hansen}, {Reinecke}, \& {Bartelmann}}]{2005ApJ...622..759G}
{G{\'o}rski}, K.~M., {Hivon}, E., {Banday}, A.~J., {et~al.} 2005, \apj, 622, 759

\bibitem[{{Heavens} \& {Taylor}(1995)}]{1995MNRAS.275..483H}
{Heavens}, A.~F. \& {Taylor}, A.~N. 1995, \mnras, 275, 483

\bibitem[{{Heymans} {et~al.}(2021){Heymans}, {Tr{\"o}ster}, {Asgari}, {Blake}, {Hildebrandt}, {Joachimi}, {Kuijken}, {Lin}, {S{\'a}nchez}, {van den Busch}, {Wright}, {Amon}, {Bilicki}, {de Jong}, {Crocce}, {Dvornik}, {Erben}, {Fortuna}, {Getman}, {Giblin}, {Glazebrook}, {Hoekstra}, {Joudaki}, {Kannawadi}, {K{\"o}hlinger}, {Lidman}, {Miller}, {Napolitano}, {Parkinson}, {Schneider}, {Shan}, {Valentijn}, {Verdoes Kleijn}, \& {Wolf}}]{2021A&A...646A.140H}
{Heymans}, C., {Tr{\"o}ster}, T., {Asgari}, M., {et~al.} 2021, \aap, 646, A140

\bibitem[{{Hikage} {et~al.}(2019){Hikage}, {Oguri}, {Hamana}, {More}, {Mandelbaum}, {Takada}, {K{\"o}hlinger}, {Miyatake}, {Nishizawa}, {Aihara}, {Armstrong}, {Bosch}, {Coupon}, {Ducout}, {Ho}, {Hsieh}, {Komiyama}, {Lanusse}, {Leauthaud}, {Lupton}, {Medezinski}, {Mineo}, {Miyama}, {Miyazaki}, {Murata}, {Murayama}, {Shirasaki}, {Sif{\'o}n}, {Simet}, {Speagle}, {Spergel}, {Strauss}, {Sugiyama}, {Tanaka}, {Utsumi}, {Wang}, \& {Yamada}}]{2019PASJ...71...43H}
{Hikage}, C., {Oguri}, M., {Hamana}, T., {et~al.} 2019, \pasj, 71, 43

\bibitem[{{Hikage} {et~al.}(2011){Hikage}, {Takada}, {Hamana}, \& {Spergel}}]{2011MNRAS.412...65H}
{Hikage}, C., {Takada}, M., {Hamana}, T., \& {Spergel}, D. 2011, \mnras, 412, 65

\bibitem[{{Hilbert} {et~al.}(2011){Hilbert}, {Hartlap}, \& {Schneider}}]{2011A&A...536A..85H}
{Hilbert}, S., {Hartlap}, J., \& {Schneider}, P. 2011, \aap, 536, A85

\bibitem[{{Hivon} {et~al.}(2002){Hivon}, {G{\'o}rski}, {Netterfield}, {Crill}, {Prunet}, \& {Hansen}}]{2002ApJ...567....2H}
{Hivon}, E., {G{\'o}rski}, K.~M., {Netterfield}, C.~B., {et~al.} 2002, \apj, 567, 2

\bibitem[{{Huffenberger} \& {Wandelt}(2010)}]{2010ApJS..189..255H}
{Huffenberger}, K.~M. \& {Wandelt}, B.~D. 2010, \apjs, 189, 255

\bibitem[{{Joachimi} {et~al.}(2015){Joachimi}, {Cacciato}, {Kitching}, {Leonard}, {Mandelbaum}, {Sch{\"a}fer}, {Sif{\'o}n}, {Hoekstra}, {Kiessling}, {Kirk}, \& {Rassat}}]{2015SSRv..193....1J}
{Joachimi}, B., {Cacciato}, M., {Kitching}, T.~D., {et~al.} 2015, \ssr, 193, 1

\bibitem[{{Joachimi} {et~al.}(2021){Joachimi}, {Lin}, {Asgari}, {Tr{\"o}ster}, {Heymans}, {Hildebrandt}, {K{\"o}hlinger}, {S{\'a}nchez}, {Wright}, {Bilicki}, {Blake}, {van den Busch}, {Crocce}, {Dvornik}, {Erben}, {Getman}, {Giblin}, {Hoekstra}, {Kannawadi}, {Kuijken}, {Napolitano}, {Schneider}, {Scoccimarro}, {Sellentin}, {Shan}, {von Wietersheim-Kramsta}, \& {Zuntz}}]{2021A&A...646A.129J}
{Joachimi}, B., {Lin}, C.~A., {Asgari}, M., {et~al.} 2021, \aap, 646, A129

\bibitem[{{Johnston} {et~al.}(2021){Johnston}, {Wright}, {Joachimi}, {Bilicki}, {Elisa Chisari}, {Dvornik}, {Erben}, {Giblin}, {Heymans}, {Hildebrandt}, {Hoekstra}, {Joudaki}, \& {Vakili}}]{2021A&A...648A..98J}
{Johnston}, H., {Wright}, A.~H., {Joachimi}, B., {et~al.} 2021, \aap, 648, A98

\bibitem[{{Kamionkowski} {et~al.}(1997){Kamionkowski}, {Kosowsky}, \& {Stebbins}}]{1997PhRvD..55.7368K}
{Kamionkowski}, M., {Kosowsky}, A., \& {Stebbins}, A. 1997, \prd, 55, 7368

\bibitem[{{Kerscher} {et~al.}(2000){Kerscher}, {Szapudi}, \& {Szalay}}]{2000ApJ...535L..13K}
{Kerscher}, M., {Szapudi}, I., \& {Szalay}, A.~S. 2000, \apjl, 535, L13

\bibitem[{{Kiessling} {et~al.}(2015){Kiessling}, {Cacciato}, {Joachimi}, {Kirk}, {Kitching}, {Leonard}, {Mandelbaum}, {Sch{\"a}fer}, {Sif{\'o}n}, {Brown}, \& {Rassat}}]{2015SSRv..193...67K}
{Kiessling}, A., {Cacciato}, M., {Joachimi}, B., {et~al.} 2015, \ssr, 193, 67

\bibitem[{{Kirk} {et~al.}(2015){Kirk}, {Brown}, {Hoekstra}, {Joachimi}, {Kitching}, {Mandelbaum}, {Sif{\'o}n}, {Cacciato}, {Choi}, {Kiessling}, {Leonard}, {Rassat}, \& {Sch{\"a}fer}}]{2015SSRv..193..139K}
{Kirk}, D., {Brown}, M.~L., {Hoekstra}, H., {et~al.} 2015, \ssr, 193, 139

\bibitem[{{Kitching} {et~al.}(2021){Kitching}, {Deshpande}, \& {Taylor}}]{2021OJAp....4E..17K}
{Kitching}, T., {Deshpande}, A., \& {Taylor}, P. 2021, The Open Journal of Astrophysics, 4, 17

\bibitem[{{Kitching} \& {Deshpande}(2022)}]{2022OJAp....5E...6K}
{Kitching}, T.~D. \& {Deshpande}, A.~C. 2022, The Open Journal of Astrophysics, 5, 6

\bibitem[{{Landy} \& {Szalay}(1993)}]{1993ApJ...412...64L}
{Landy}, S.~D. \& {Szalay}, A.~S. 1993, \apj, 412, 64

\bibitem[{{Laureijs} {et~al.}(2011){Laureijs}, {Amiaux}, {Arduini}, {Augu{\`e}res}, {Brinchmann}, {Cole}, {Cropper}, {Dabin}, {Duvet}, {Ealet}, {Garilli}, {Gondoin}, {Guzzo}, {Hoar}, {Hoekstra}, {Holmes}, {Kitching}, {Maciaszek}, {Mellier}, {Pasian}, {Percival}, {Rhodes}, {Saavedra Criado}, {Sauvage}, {Scaramella}, {Valenziano}, {Warren}, {Bender}, {Castander}, {Cimatti}, {Le F{\`e}vre}, {Kurki-Suonio}, {Levi}, {Lilje}, {Meylan}, {Nichol}, {Pedersen}, {Popa}, {Rebolo Lopez}, {Rix}, {Rottgering}, {Zeilinger}, {Grupp}, {Hudelot}, {Massey}, {Meneghetti}, {Miller}, {Paltani}, {Paulin-Henriksson}, {Pires}, {Saxton}, {Schrabback}, {Seidel}, {Walsh}, {Aghanim}, {Amendola}, {Bartlett}, {Baccigalupi}, {Beaulieu}, {Benabed}, {Cuby}, {Elbaz}, {Fosalba}, {Gavazzi}, {Helmi}, {Hook}, {Irwin}, {Kneib}, {Kunz}, {Mannucci}, {Moscardini}, {Tao}, {Teyssier}, {Weller}, {Zamorani}, {Zapatero Osorio}, {Boulade}, {Foumond}, {Di Giorgio}, {Guttridge}, {James}, {Kemp}, {Martignac}, {Spencer}, {Walton}, {Bl{\"u}mchen}, {Bonoli}, {Bortoletto}, {Cerna}, {Corcione}, {Fabron}, {Jahnke}, {Ligori}, {Madrid}, {Martin}, {Morgante}, {Pamplona}, {Prieto}, {Riva}, {Toledo}, {Trifoglio}, {Zerbi}, {Abdalla}, {Douspis}, {Grenet}, {Borgani}, {Bouwens}, {Courbin}, {Delouis}, {Dubath}, {Fontana}, {Frailis}, {Grazian}, {Koppenh{\"o}fer}, {Mansutti}, {Melchior}, {Mignoli}, {Mohr}, {Neissner}, {Noddle}, {Poncet}, {Scodeggio}, {Serrano}, {Shane}, {Starck}, {Surace}, {Taylor}, {Verdoes-Kleijn}, {Vuerli}, {Williams}, {Zacchei}, {Altieri}, {Escudero Sanz}, {Kohley}, {Oosterbroek}, {Astier}, {Bacon}, {Bardelli}, {Baugh}, {Bellagamba}, {Benoist}, {Bianchi}, {Biviano}, {Branchini}, {Carbone}, {Cardone}, {Clements}, {Colombi}, {Conselice}, {Cresci}, {Deacon}, {Dunlop}, {Fedeli}, {Fontanot}, {Franzetti}, {Giocoli}, {Garcia-Bellido}, {Gow}, {Heavens}, {Hewett}, {Heymans}, {Holland}, {Huang}, {Ilbert}, {Joachimi}, {Jennins}, {Kerins}, {Kiessling}, {Kirk}, {Kotak}, {Krause}, {Lahav}, {van Leeuwen}, {Lesgourgues}, {Lombardi}, {Magliocchetti}, {Maguire}, {Majerotto}, {Maoli}, {Marulli}, {Maurogordato}, {McCracken}, {McLure}, {Melchiorri}, {Merson}, {Moresco}, {Nonino}, {Norberg}, {Peacock}, {Pello}, {Penny}, {Pettorino}, {Di Porto}, {Pozzetti}, {Quercellini}, {Radovich}, {Rassat}, {Roche}, {Ronayette}, {Rossetti}, {Sartoris}, {Schneider}, {Semboloni}, {Serjeant}, {Simpson}, {Skordis}, {Smadja}, {Smartt}, {Spano}, {Spiro}, {Sullivan}, {Tilquin}, {Trotta}, {Verde}, {Wang}, {Williger}, {Zhao}, {Zoubian}, \& {Zucca}}]{2011arXiv1110.3193L}
{Laureijs}, R., {Amiaux}, J., {Arduini}, S., {et~al.} 2011, arXiv:1110.3193

\bibitem[{{Linke} {et~al.}(2024){Linke}, {Unruh}, {Wittje}, {Schrabback}, {Grandis}, {Asgari}, {Dvornik}, {Hildebrandt}, {Hoekstra}, {Joachimi}, {Reischke}, {van den Busch}, {Wright}, {Schneider}, {Aghanim}, {Altieri}, {Amara}, {Andreon}, {Auricchio}, {Baccigalupi}, {Baldi}, {Bardelli}, {Bonino}, {Branchini}, {Brescia}, {Brinchmann}, {Camera}, {Capobianco}, {Carbone}, {Cardone}, {Carretero}, {Casas}, {Castander}, {Castellano}, {Cavuoti}, {Cimatti}, {Congedo}, {Conselice}, {Conversi}, {Copin}, {Courbin}, {Courtois}, {Da Silva}, {Degaudenzi}, {Dinis}, {Douspis}, {Dubath}, {Dupac}, {Dusini}, {Farina}, {Farrens}, {Ferriol}, {Fosalba}, {Frailis}, {Franceschi}, {Fumana}, {Galeotta}, {Gillis}, {Giocoli}, {Grazian}, {Grupp}, {Guzzo}, {Haugan}, {Holmes}, {Hook}, {Hormuth}, {Hornstrup}, {Hudelot}, {Jahnke}, {Keih{\"a}nen}, {Kermiche}, {Kiessling}, {Kilbinger}, {Kitching}, {Kubik}, {Kuijken}, {K{\"u}mmel}, {Kunz}, {Kurki-Suonio}, {Ligori}, {Lilje}, {Lindholm}, {Lloro}, {Maino}, {Maiorano}, {Mansutti}, {Marggraf}, {Markovic}, {Martinet}, {Marulli}, {Massey}, {McCracken}, {Medinaceli}, {Mei}, {Mellier}, {Meneghetti}, {Merlin}, {Meylan}, {Moresco}, {Moscardini}, {Munari}, {Nakajima}, {Nichol}, {Niemi}, {Nightingale}, {Padilla}, {Paltani}, {Pasian}, {Pedersen}, {Pettorino}, {Pires}, {Polenta}, {Poncet}, {Popa}, {Raison}, {Rebolo}, {Renzi}, {Rhodes}, {Riccio}, {Romelli}, {Roncarelli}, {Saglia}, {Sakr}, {Sapone}, {Sartoris}, {Schirmer}, {Secroun}, {Seidel}, {Serrano}, {Sirignano}, {Sirri}, {Stanco}, {Starck}, {Tallada-Cresp{\'\i}}, {Taylor}, {Tereno}, {Toledo-Moreo}, {Torradeflot}, {Tutusaus}, {Valenziano}, {Vassallo}, {Verdoes Kleijn}, {Veropalumbo}, {Wang}, {Weller}, {Zamorani}, {Zucca}, {Burigana}, {Pezzotta}, {Porciani}, {Scottez}, {Viel}, \& {Le Brun}}]{2024arXiv240709810L}
{Linke}, L., {Unruh}, S., {Wittje}, A., {et~al.} 2024, arXiv:2407.09810

\bibitem[{{Loureiro} {et~al.}(2023){Loureiro}, {Whiteway}, {Sellentin}, {Silva Lafaurie}, {Jaffe}, \& {Heavens}}]{2023OJAp....6E...6L}
{Loureiro}, A., {Whiteway}, L., {Sellentin}, E., {et~al.} 2023, The Open Journal of Astrophysics, 6, 6

\bibitem[{{Maraio} {et~al.}(2023){Maraio}, {Hall}, \& {Taylor}}]{2023MNRAS.520.4836M}
{Maraio}, A., {Hall}, A., \& {Taylor}, A. 2023, \mnras, 520, 4836

\bibitem[{{McEwen} {et~al.}(2015){McEwen}, {Leistedt}, {B{\"u}ttner}, {Peiris}, \& {Wiaux}}]{2015arXiv150906749M}
{McEwen}, J.~D., {Leistedt}, B., {B{\"u}ttner}, M., {Peiris}, H.~V., \& {Wiaux}, Y. 2015, arXiv:1509.06749

\bibitem[{{McEwen} \& {Wiaux}(2011)}]{2011ITSP...59.5876M}
{McEwen}, J.~D. \& {Wiaux}, Y. 2011, IEEE Transactions on Signal Processing, 59, 5876

\bibitem[{{Mead} {et~al.}(2021){Mead}, {Brieden}, {Tr{\"o}ster}, \& {Heymans}}]{2021MNRAS.502.1401M}
{Mead}, A.~J., {Brieden}, S., {Tr{\"o}ster}, T., \& {Heymans}, C. 2021, \mnras, 502, 1401

\bibitem[{{More} {et~al.}(2023){More}, {Sugiyama}, {Miyatake}, {Rau}, {Shirasaki}, {Li}, {Nishizawa}, {Osato}, {Zhang}, {Takada}, {Hamana}, {Takahashi}, {Dalal}, {Mandelbaum}, {Strauss}, {Kobayashi}, {Nishimichi}, {Oguri}, {Luo}, {Kannawadi}, {Hsieh}, {Armstrong}, {Bosch}, {Komiyama}, {Lupton}, {Lust}, {MacArthur}, {Miyazaki}, {Murayama}, {Okura}, {Price}, {Tait}, {Tanaka}, \& {Wang}}]{2023PhRvD.108l3520M}
{More}, S., {Sugiyama}, S., {Miyatake}, H., {et~al.} 2023, \prd, 108, 123520

\bibitem[{{Nicola} {et~al.}(2021){Nicola}, {Garc{\'\i}a-Garc{\'\i}a}, {Alonso}, {Dunkley}, {Ferreira}, {Slosar}, \& {Spergel}}]{2021JCAP...03..067N}
{Nicola}, A., {Garc{\'\i}a-Garc{\'\i}a}, C., {Alonso}, D., {et~al.} 2021, JCAP, 03, 067

\bibitem[{{Peebles}(1973)}]{1973ApJ...185..413P}
{Peebles}, P.~J.~E. 1973, \apj, 185, 413

\bibitem[{{Percival} {et~al.}(2004){Percival}, {Burkey}, {Heavens}, {Taylor}, {Cole}, {Peacock}, {Baugh}, {Bland-Hawthorn}, {Bridges}, {Cannon}, {Colless}, {Collins}, {Couch}, {Dalton}, {De Propris}, {Driver}, {Efstathiou}, {Ellis}, {Frenk}, {Glazebrook}, {Jackson}, {Lahav}, {Lewis}, {Lumsden}, {Maddox}, {Norberg}, {Peterson}, {Sutherland}, \& {Taylor}}]{2004MNRAS.353.1201P}
{Percival}, W.~J., {Burkey}, D., {Heavens}, A., {et~al.} 2004, \mnras, 353, 1201

\bibitem[{{Reinecke} {et~al.}(2023){Reinecke}, {Belkner}, \& {Carron}}]{2023A&A...678A.165R}
{Reinecke}, M., {Belkner}, S., \& {Carron}, J. 2023, \aap, 678, A165

\bibitem[{{Rodr{\'\i}guez-Monroy} {et~al.}(2022){Rodr{\'\i}guez-Monroy}, {Weaverdyck}, {Elvin-Poole}, {Crocce}, {Carnero Rosell}, {Andrade-Oliveira}, {Avila}, {Bechtol}, {Bernstein}, {Blazek}, {Camacho}, {Cawthon}, {De Vicente}, {DeRose}, {Dodelson}, {Everett}, {Fang}, {Ferrero}, {Fert{\'e}}, {Friedrich}, {Gaztanaga}, {Giannini}, {Gruendl}, {Hartley}, {Herner}, {Huff}, {Jarvis}, {Krause}, {MacCrann}, {Mena-Fern{\'a}ndez}, {Muir}, {Pandey}, {Park}, {Porredon}, {Prat}, {Rosenfeld}, {Ross}, {Rozo}, {Rykoff}, {Sanchez}, {Sanchez Cid}, {Sevilla-Noarbe}, {Tabbutt}, {To}, {Wagoner}, {Wechsler}, {Aguena}, {Allam}, {Amon}, {Annis}, {Bacon}, {Baxter}, {Bertin}, {Bhargava}, {Brooks}, {Burke}, {Carrasco Kind}, {Carretero}, {Castander}, {Choi}, {Conselice}, {Costanzi}, {da Costa}, {Pereira}, {Desai}, {Diehl}, {Flaugher}, {Fosalba}, {Frieman}, {Garc{\'\i}a-Bellido}, {Giannantonio}, {Gruen}, {Gschwend}, {Gutierrez}, {Hinton}, {Hollowood}, {Honscheid}, {Huterer}, {Jain}, {James}, {Kuehn}, {Kuropatkin}, {Lima}, {Maia}, {March}, {Marshall}, {Melchior}, {Menanteau}, {Miller}, {Miquel}, {Mohr}, {Morgan}, {Palmese}, {Paz-Chinch{\'o}n}, {Pieres}, {Plazas Malag{\'o}n}, {Roodman}, {Scarpine}, {Serrano}, {Smith}, {Soares-Santos}, {Suchyta}, {Tarle}, {Thomas}, {Varga}, \& {DES Collaboration}}]{2022MNRAS.511.2665R}
{Rodr{\'\i}guez-Monroy}, M., {Weaverdyck}, N., {Elvin-Poole}, J., {et~al.} 2022, \mnras, 511, 2665

\bibitem[{{Schneider} {et~al.}(2010){Schneider}, {Eifler}, \& {Krause}}]{2010A&A...520A.116S}
{Schneider}, P., {Eifler}, T., \& {Krause}, E. 2010, \aap, 520, A116

\bibitem[{{Schneider} {et~al.}(2002){Schneider}, {van Waerbeke}, {Kilbinger}, \& {Mellier}}]{2002A&A...396....1S}
{Schneider}, P., {van Waerbeke}, L., {Kilbinger}, M., \& {Mellier}, Y. 2002, \aap, 396, 1

\bibitem[{{Seitz} \& {Schneider}(1997)}]{1997A&A...318..687S}
{Seitz}, C. \& {Schneider}, P. 1997, \aap, 318, 687

\bibitem[{{Sellentin} {et~al.}(2023){Sellentin}, {Loureiro}, {Whiteway}, {Lafaurie}, {Balan}, {Olamaie}, {Jaffe}, \& {Heavens}}]{2023OJAp....6E..31S}
{Sellentin}, E., {Loureiro}, A., {Whiteway}, L., {et~al.} 2023, The Open Journal of Astrophysics, 6, 31

\bibitem[{{Singh} {et~al.}(2017){Singh}, {Mandelbaum}, {Seljak}, {Slosar}, \& {Vazquez Gonzalez}}]{2017MNRAS.471.3827S}
{Singh}, S., {Mandelbaum}, R., {Seljak}, U., {Slosar}, A., \& {Vazquez Gonzalez}, J. 2017, \mnras, 471, 3827

\bibitem[{{Tadros} {et~al.}(1999){Tadros}, {Ballinger}, {Taylor}, {Heavens}, {Efstathiou}, {Saunders}, {Frenk}, {Keeble}, {McMahon}, {Maddox}, {Oliver}, {Rowan-Robinson}, {Sutherland}, \& {White}}]{1999MNRAS.305..527T}
{Tadros}, H., {Ballinger}, W.~E., {Taylor}, A.~N., {et~al.} 1999, \mnras, 305, 527

\bibitem[{{Tallada} {et~al.}(2020){Tallada}, {Carretero}, {Casals}, {Acosta-Silva}, {Serrano}, {Caubet}, {Castander}, {C{\'e}sar}, {Crocce}, {Delfino}, {Eriksen}, {Fosalba}, {Gazta{\~n}aga}, {Merino}, {Neissner}, \& {Tonello}}]{2020A&C....3200391T}
{Tallada}, P., {Carretero}, J., {Casals}, J., {et~al.} 2020, Astronomy and Computing, 32, 100391

\bibitem[{{Tegmark}(1997)}]{1997PhRvD..55.5895T}
{Tegmark}, M. 1997, \prd, 55, 5895

\bibitem[{{Tegmark} \& {de Oliveira-Costa}(2001)}]{2001PhRvD..64f3001T}
{Tegmark}, M. \& {de Oliveira-Costa}, A. 2001, \prd, 64, 063001

\bibitem[{{Tessore} {et~al.}(2023){Tessore}, {Loureiro}, {Joachimi}, {von Wietersheim-Kramsta}, \& {Jeffrey}}]{2023OJAp....6E..11T}
{Tessore}, N., {Loureiro}, A., {Joachimi}, B., {von Wietersheim-Kramsta}, M., \& {Jeffrey}, N. 2023, The Open Journal of Astrophysics, 6, 11

\bibitem[{{Troxel} \& {Ishak}(2015)}]{2015PhR...558....1T}
{Troxel}, M.~A. \& {Ishak}, M. 2015, \physrep, 558, 1

\bibitem[{{Varshalovich} {et~al.}(1988){Varshalovich}, {Moskalev}, \& {Khersonskii}}]{1988qtam.book.....V}
{Varshalovich}, D.~A., {Moskalev}, A.~N., \& {Khersonskii}, V.~K. 1988, {Quantum Theory of Angular Momentum} (World Scientific)

\bibitem[{{Wandelt} \& {G{\'o}rski}(2001)}]{2001PhRvD..63l3002W}
{Wandelt}, B.~D. \& {G{\'o}rski}, K.~M. 2001, \prd, 63, 123002

\bibitem[{{Wandelt} {et~al.}(2001){Wandelt}, {Hivon}, \& {G{\'o}rski}}]{2001PhRvD..64h3003W}
{Wandelt}, B.~D., {Hivon}, E., \& {G{\'o}rski}, K.~M. 2001, \prd, 64, 083003

\bibitem[{{Wolz} {et~al.}(2024){Wolz}, {Alonso}, \& {Nicola}}]{2024arXiv240721013W}
{Wolz}, K., {Alonso}, D., \& {Nicola}, A. 2024, arXiv:2407.21013

\bibitem[{{Zaldarriaga} \& {Seljak}(1997)}]{1997PhRvD..55.1830Z}
{Zaldarriaga}, M. \& {Seljak}, U. 1997, \prd, 55, 1830

\end{thebibliography}

\begin{appendix}

\onecolumn

\section{Additional figures}

\begin{figure*}[h!]%
\centering%
\includegraphics[scale=.6]{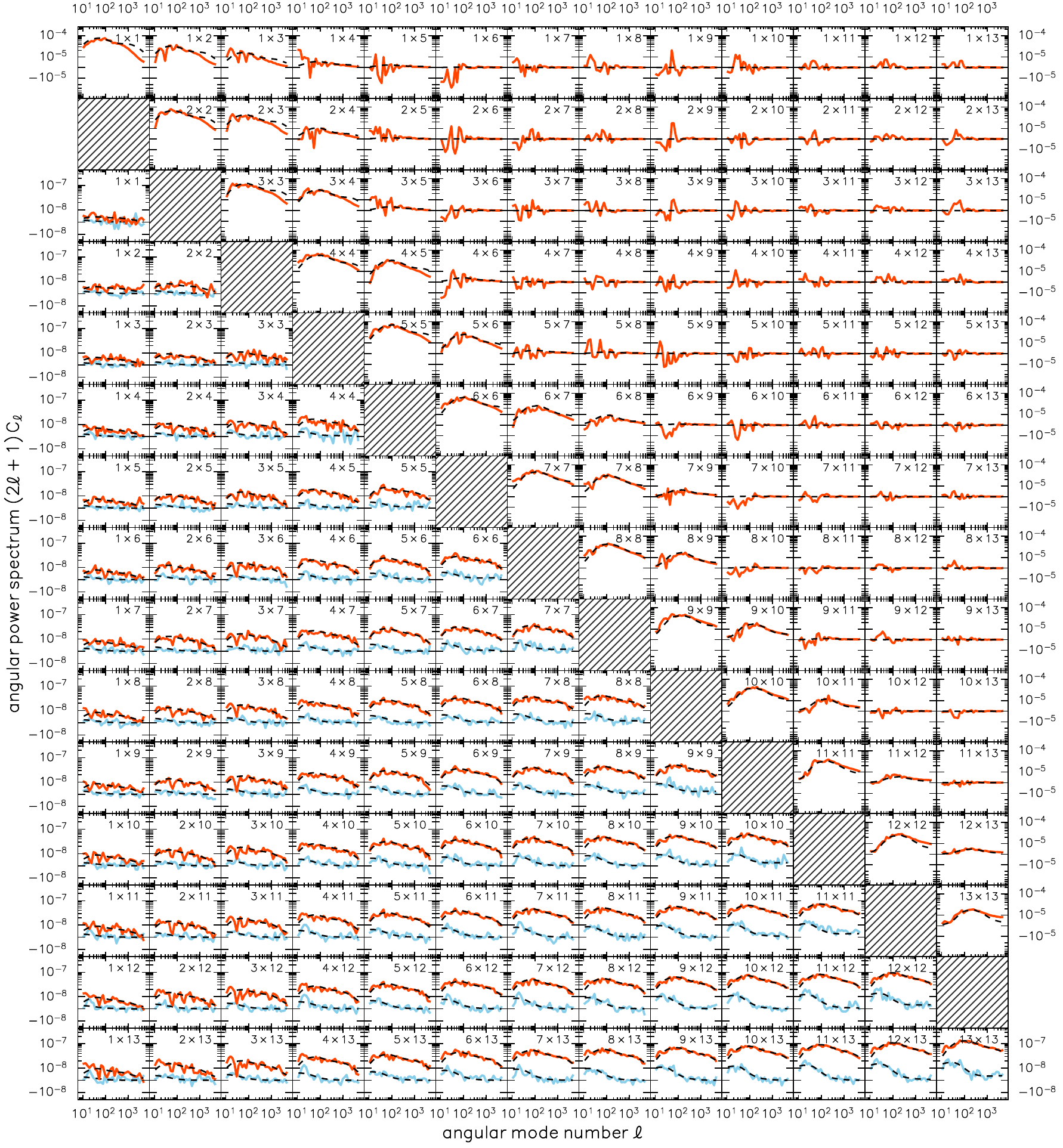}%
\caption{%
    Angular power spectra (\emph{red}) for angular clustering (\emph{upper
    triangle}) and cosmic shear (\emph{lower triangle}) in the \Euclid Flagship
    simulation with a DR1-like footprint.  For cosmic shear, the $B$-mode
    spectrum due to mode mixing is shown in \emph{blue}.  Also shown is the
    expectation for each spectrum (\emph{dashed}), as computed from the
    respective mixing matrices.  All spectra are binned into 32 angular bins
    with logarithmic spacing between~$l = 10$ and~$l = 5000$.  The $y$-axis
    changes to linear scaling when passing through the origin.
}%
\label{fig:fs2-a}%
\end{figure*}

\begin{figure*}[p!]%
\centering%
\includegraphics[scale=.6]{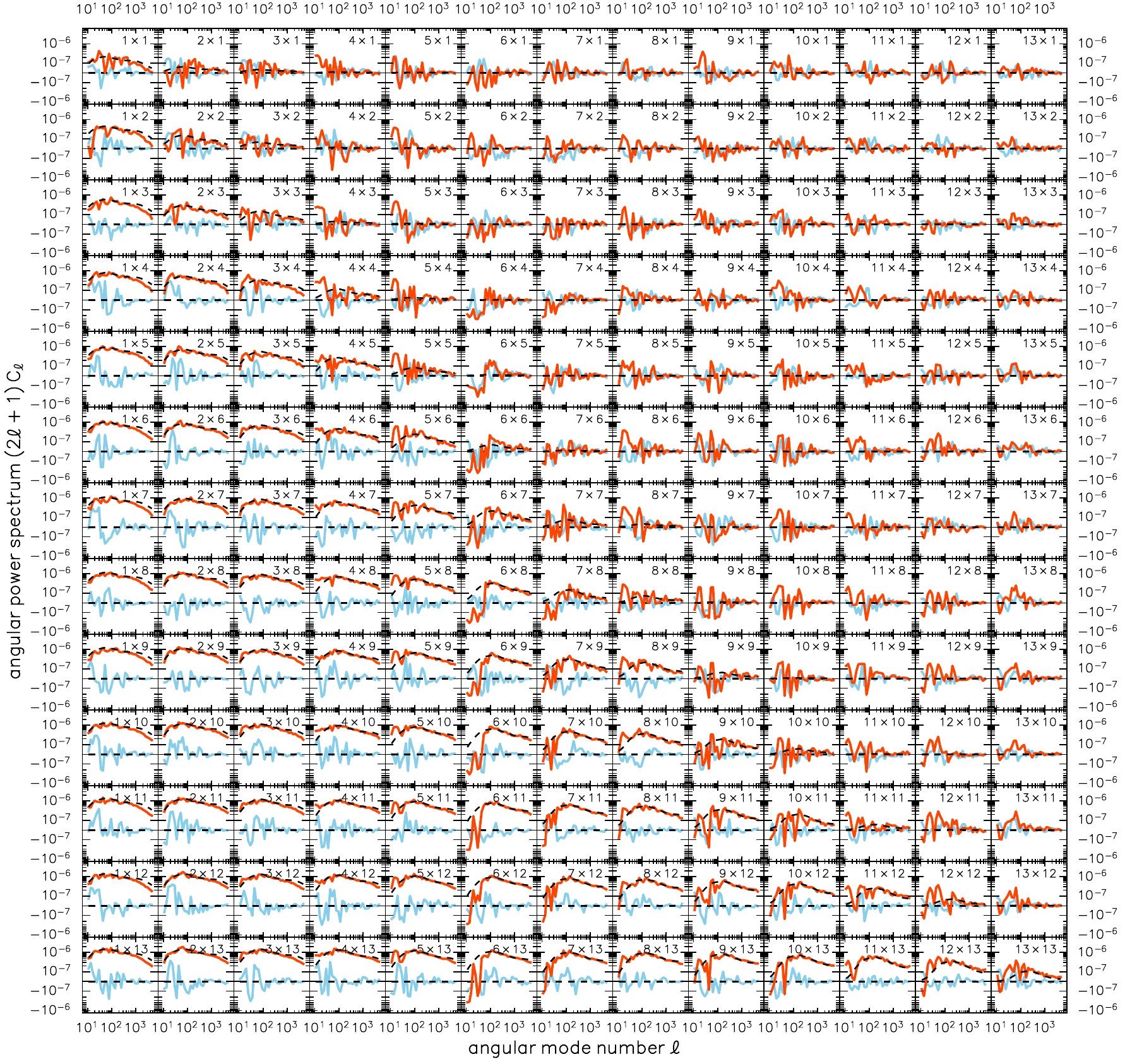}%
\caption{%
    Same as Fig.~\ref{fig:fs2-a} for galaxy--galaxy lensing.
}%
\label{fig:fs2-b}%
\end{figure*}

\twocolumn

\section{Alternative estimators for angular clustering}
\label{sec:alt-ang-clus}

In Sect.~\ref{sec:point-processes}, we measure the two-point statistics for
point processes using the particular choice of density
contrast~\eqref{eq:delta}.  Here, we consider a number of alternative choices.

Firstly, we can trivially replace the mean number density~$\bar{n}$ by a
catalogue of random points (``randoms'').  Given the definition~\eqref{eq:nbar}
of the mean number density, these randoms must be distributed according to the
visibility~$v$ \citep{2024JCAP...05..010B}.

Secondly, in full correspondence to real-space methods, we can directly
construct estimators of the angular power spectrum~$\w_l$ from the
expectation~\eqref{eq:ev-cl-nn}.  For example, using a formal
inverse~$(M^{\bar{n}\bar{n}'})^{-1}$ of the mixing matrix, we can construct the
estimator
\begin{equation}
\label{eq:est-nat}
    \hat{\w}_l^{\mathrm{N}}
    = \sum_{l'} (M^{\bar{n}\bar{n}'})^{-1}_{ll'} \, \Bigl[
        C_{l'}^{nn'}
        - C_{l'}^{\bar{n}\bar{n}'}
        - A^{nn'}
    \Bigr] \,.
\end{equation}
In the taxonomy of \citet{2000ApJ...535L..13K}, this corresponds to the
``natural'' real-space estimator $(DD - RR)/RR$.  Furthermore, by
expectation~\eqref{eq:ev-n}, we have $\ev{C_l^{n\bar{n}'}} =
C_l^{\bar{n}\bar{n}'}$, and we can hence construct a more advanced estimator
\begin{equation}
\label{eq:est-ls}
    \hat{\w}_l^{\mathrm{LS}}
    = \sum_{l'} (M^{\bar{n}\bar{n}'})^{-1}_{ll'} \, \Bigl[
        C_{l'}^{nn'}
        - C_{l'}^{n\bar{n}'}
        - C_{l'}^{\bar{n}n'}
        + C_{l'}^{\bar{n}\bar{n}'}
        - A^{nn'}
    \Bigr] \,.
\end{equation}
This is the harmonic-space equivalent of the \citet{1993ApJ...412...64L}
estimator $(DD - DR - RD + RR)/RR$.

The estimators~\eqref{eq:est-nat} and~\eqref{eq:est-ls} both rely on inversion
of the mixing matrix.  We can similarly construct a partial-sky variant of the
natural estimator in harmonic space,
\begin{equation}
\label{eq:est-nat-part}
    \tilde{\w}_l^{\mathrm{N}}
    = C_l^{nn'}
    - C_l^{\bar{n}\bar{n}'}
    - A^{nn'} \,,
\end{equation}
as well as a partial-sky variant of the harmonic-space Landy-Szalay estimator,
\begin{equation}
\label{eq:est-ls-part}
    \tilde{\w}_l^{\mathrm{LS}}
    = C_l^{nn'}
    - C_l^{n\bar{n}'}
    - C_l^{\bar{n}n'}
    + C_l^{\bar{n}\bar{n}'}
    - A^{nn'} \,.
\end{equation}
The respective expectation of both partial-sky estimators is the product of
mixing matrix and full-sky expectation.  In particular, the partial-sky
Landy--Szalay estimator~\eqref{eq:est-ls-part} is essentially the same as the
measured angular power spectrum~\eqref{eq:cl-dd} of the density
contrast~\eqref{eq:delta}.

Lastly, to see why we normalise the density contrast~\eqref{eq:delta} by a
constant~$\bar{n}_0$, consider an alternative definition of the density
contrast with an arbitrary normalisation function~$q$,
\begin{equation}
\label{eq:delta-alt}
    \delta_q(\U)
    = \frac{n(\U) - \bar{n}(\U)}{\bar{n}_0 \, q(\U)} \,.
\end{equation}
It follows from the definitions of the number density~\eqref{eq:n} and mean
number density~\eqref{eq:nbar} that~$\delta_q$ is equivalent to~$\delta$ under
a change of weights~$w_k \mapsto w_k/q(\U_k)$ and, consequently, a change of
visibility~$v(\U) \mapsto v(\U)/q(\U)$.  Defining the density
contrast~\eqref{eq:delta} with a different normalisation therefore effectively
replaces the given set of weights with a different set of weights.

\section{Relative orientation on the sphere}
\label{sec:angles}

\begin{figure}%
\centering%
\includegraphics[scale=1.0]{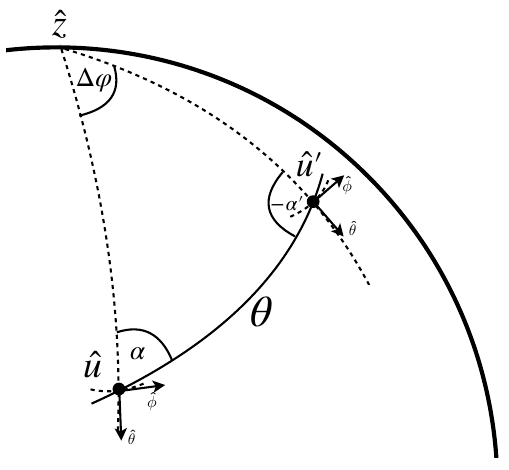}%
\caption{%
    The angles $\alpha, \theta, \alpha'$ that describe the relative orientation
    between two points~$\U$ and~$\U'$ on the sphere can be obtained from the
    spherical triangle between the north pole, $\U$, and~$\U'$, with~$\Delta
    \varphi = \varphi - \varphi'$.
}%
\label{fig:triangle}%
\end{figure}

To obtain the angles~$\alpha$ and~$\alpha'$ that describe the relative
orientation of points~$\U$ and~$\U'$ on the sphere, it suffices to solve the
spherical triangle shown in Fig.~\ref{fig:triangle} (for more information, see
Hall \& Tessore in prep.),
\begin{align}
    \alpha &= \arctan \frac{
        \sin \vartheta'\sin (\varphi - \varphi')
    }{
        \sin \vartheta \cos \vartheta'-\cos \vartheta \sin \vartheta'\cos
        (\varphi - \varphi')
    } \,, \\[4pt]
    \alpha' &= -\arctan \frac{
        \sin \vartheta \sin (\varphi - \varphi')
    }{
        \sin \vartheta'\cos \vartheta  - \cos \vartheta'\sin \vartheta \cos
        (\varphi - \varphi')
    } \,.
\end{align}
The respective numerators and denominators are written here such that their
signs yield the correct quadrant for the inverse tangent.

The same angles can be expressed in terms of the components of the unit
vectors~$\U = \{x, y, z\}$ and~$\U' = \{x', y', z'\}$ as
\begin{align}
    \alpha &= \arctan \frac{yx' - xy'}{z' - z \cos\theta} \,, \\[4pt]
    \alpha' &= -\arctan \frac{yx' - xy'}{z - z' \cos\theta} \,,
\end{align}
with~$\cos\theta = \U \cdot \U'$.  This form is often useful in applications
where points are available as vectors, since it requires no additional
trigonometric operations.  In fact, for $\alpha = \arctan(q/p)$, we find a
familiar expression for the spin-$2$ phase factors that appear, e.g., in the
spherical harmonic addition theorem~\eqref{eq:addthm},
\begin{equation}
    \E^{2\I\alpha}
    = \frac{p^2 - q^2 + 2 \, \I \, pq}{p^2 + q^2} \,.
\end{equation}
The phase factors can hence be computed entirely in terms of the vector
components of~$\U$ and~$\U'$.

\section{Decomposition into \texorpdfstring{$E$}{E}- and
\texorpdfstring{$B$}{B}-modes}
\label{sec:e-b-modes}

If the spherical function~$f$ is a complex-valued random field, the two-point
statistics of~$f$ and a second, not necessarily distinct, field~$f'$ are not
fully characterised by the expected angular correlations~\eqref{eq:ev-ff-cf}
alone.  Like for any complex random variable, we also require the associated
pseudo-correlation~$\ev{C^{f^*\!f'}(\theta)}$, i.e., the correlation of the
complex conjugated random field~$f^*$ and~$f'$,
\begin{equation}
    \ev{f(\U) \, f'(\U')}
    = \E^{\I s \alpha} \, \ev{C^{f^*\!f'}(\theta)} \, \E^{\I s' \alpha'} \,,
\end{equation}
where we have used the fact that the spin weight of~$f^*$ is~$-s$.  The same
information is contained in the pseudo-spectrum\footnote{%
    The prefix ``pseudo-'' is used here in the statistical sense, and not to be
    confused with meaning ``partial sky'', for which it is unfortunately
    sometimes also used.}
\begin{equation}
    \ev{(f^*)^*_{lm} \, f'_{l'm'}}
    = \delta^{\rm K}_{ll'} \, \delta^{\rm K}_{mm'} \, \ev{C_l^{f^*\!f'}} \,,
\end{equation}
which is merely expectation~\eqref{eq:ev-flm-prod} applied to~$f^*$ and~$f'$.

Instead of using spectra and pseudo-spectra, it is often more convenient to
work with a different decomposition of the harmonic-space two-point statistics,
namely that into $E$- and $B$-modes \citep{1997PhRvD..55.1830Z}.  For a
spherical function~$f$ with spin weight~$s$, the respective~$E$- and $B$-modes
are defined as linear combinations of the spherical harmonic coefficients
of~$f$ and~$f^*$,
\begin{align}
\label{eq:elm}
    E_{lm}
    &= -\frac{f_{lm} + (-1)^s \, (f^*)_{lm}}{2} \,, \\
\label{eq:blm}
    B_{lm}
    &= -\frac{f_{lm} - (-1)^s \, (f^*)_{lm}}{2\I} \,,
\end{align}
where the overall negative sign is the convention adopted by \emph{HEALPix}.
The $E$- and $B$-mode spectra are then obtained by using~$E_{lm}$ and~$B_{lm}$
in the angular power spectrum~\eqref{eq:cl}.  Since the
coefficients~\eqref{eq:elm} and~\eqref{eq:blm} are linear combinations
of~$f_{lm}$ and~$(f^*)_{lm}$, it is clear that the resulting $E$- and $B$-mode
spectra are linear combinations of the spectra~$C_l^{ff'}$, $C_l^{f^*\!f'}$,
etc. of the constituent fields and their complex conjugates.

To compute, e.g., the expectation for partial-sky $E$- and $B$-mode spectra, it
therefore suffices to apply the mixing matrix formalism described above to the
individual spectra of the fields, and express the result in terms of the
full-sky $E$- and $B$-mode spectra.  In doing so, one finds that the mixing
matrices also introduce mixing between $E$- and $B$-modes
\citep{2005MNRAS.360.1262B}.

\section{Squared normal fields}
\label{sec:sqnorm}

In this section, we consider a Gaussian random field~$X$ on the sphere that is
transformed into a random field~$Y$ by an arbitrary function~$t$,
\begin{equation}
    Y(\U) = t\bigl(X(\U)\bigr) \,.
\end{equation}
As shown by \citet{2023OJAp....6E..11T}, a band-limited angular power
spectrum~$G_l$ for~$X$ does not generally result in a band-limited angular
power spectrum~$C_l$ for~$Y$.  In practice, we are not generally able to
construct a spectrum~$G_l$ that, after transformation, reproduces a desired
spectrum~$C_l$ exactly.  For validation, we now try and identify a special case
where that is possible.  In particular, we look for a transformation with the
following two characteristics:
\begin{itemize}
\item[i)] The transformed field~$Y$ is bounded from below, so that we can
    simulate density contrasts~$\delta$ for angular clustering that respect
    the physical bound~$\delta \ge -1$.
\item[ii)] The transformed field~$Y$ has a band-limited spectrum.
\end{itemize}
Lognormal fields satisfy the first criterium, but lack a strictly band-limited
spectrum.  However, it turns out that both criteria are fulfilled by squaring a
Gaussian random field.

\begin{figure}%
\centering%
\includegraphics[scale=.85]{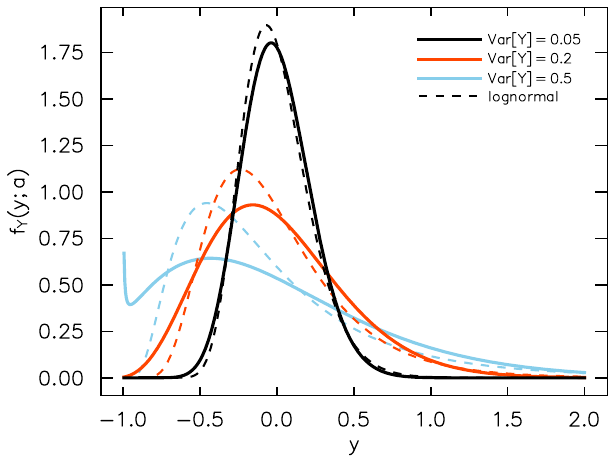}%
\caption{%
    Shape of squared normal distributions (\emph{solid}) and lognormal
    distributions with the same variance (\emph{dashed}).  For small variances,
    the distribution is approximately normal (\emph{black}).  For large
    variances, the distribution saturates at its lower bound (\emph{blue}).  In
    the intermediate regime, the distribution has a roughly lognormal shape
    (\emph{red}).
}%
\label{fig:sqnorm}%
\end{figure}

Let~$X$ be a normal random variable with zero mean and variance~$\sigma^2 \le
1$, and let~$a = \sqrt{1 - \sigma^2}$.  Define the random variable~$Y$ as a
quadratic transformation of~$X$, resulting in a scaled and shifted
non-central~$\chi^2$ random variable with 1 degree of freedom,
\begin{equation}
\label{eq:sqnorm}
    Y = \lambda \, [(X - a)^2 - 1] \,,
\end{equation}
where~$\lambda > 0$ is the scale parameter of the distribution, which also
fixes the minimum value of~$Y$.\footnote{%
    For that reason, the scale parameter~$\lambda$ of a lognormal random
    variable is commonly called the ``shift'' parameter
    \citep{2023OJAp....6E..11T}.}
A straightforward calculation shows that~$Y$ has zero mean and variance
\begin{equation}
    \ev{Y^2}
    = 2 \lambda^2 \, \sigma^2 \, (2 - \sigma^2) \,.
\end{equation}
Inserting~$\sigma^2 = 1 - a^2$, the variance can equivalently be expressed in
terms of~$a$,
\begin{equation}
    \ev{Y^2}
    = 2 \lambda^2 \, (1 - a^4) \,,
\end{equation}
and the value of~$a$ can hence be obtained from the variance of the transformed
random variable,
\begin{equation}
    a
    = \Biggl(
        1 - \frac{\ev{Y^2}}{2 \lambda^2}
    \Biggr)^{\frac{1}{4}} \,.
\end{equation}
The transformation~\eqref{eq:sqnorm} is therefore readily obtained in either
direction.  Setting~$\lambda = 1$, the standardised probability distribution
function of~$Y$ is
\begin{equation}
    f_Y(y; a)
    = \frac{
        \exp\Bigl(-\frac{a^2 + y + 1}{2 \, (1-a^2)}\Bigr)
        \cosh\Bigl(\frac{a \sqrt{y+1}}{1-a^2}\Bigr)
    }{
        \sqrt{2\pi \, (1-a^2) \, (y+1)}
    } \,.
\end{equation}
The distribution approaches normality for small variances, and becomes more
skewed as the variance increases, similar to the lognormal distribution
(Fig.~\ref{fig:sqnorm}).

We then apply the transformation~\eqref{eq:sqnorm} pointwise to a pair~$X$
and~$X'$ of jointly homogeneous Gaussian random fields on the sphere.  By
expectation~\eqref{eq:ev-ff-cf}, there is a correlation function~$G$ such
that~$\ev{X(\U) \, X'(\U')} = G(\theta)$.  It can be shown that the transformed
fields~$Y$ and~$Y'$ are also jointly homogeneous \citep{2023OJAp....6E..11T},
and there is hence an angular correlation function~$C$ such that~$\ev{Y(\U) \,
Y'(\U')} = C(\theta)$.  Using the transformation~\eqref{eq:sqnorm}, we can
compute~$C(\theta)$ in terms of~$G(\theta)$,
\begin{equation}
\label{eq:sqnorm-ct-gt}
    C(\theta)
    = 2 \lambda \lambda' \, G(\theta) \, \bigl(G(\theta) + 2aa'\bigr) \,.
\end{equation}
By completing the square, we also obtain the inverse relation,
\begin{equation}
\label{eq:sqnorm-gt-ct}
    G(\theta)
    = \sqrt{
        \frac{C(\theta)}{2 \lambda \lambda'}
        + (aa')^2
    } - aa' \,.
\end{equation}
Furthermore, relation~\eqref{eq:sqnorm-ct-gt} is readily transformed to
harmonic space using expectation~\eqref{eq:ev-cl-prod},
\begin{equation}
\begin{split}
    C_l
    &= 2\lambda\lambda'
        \sum_{l_1l_2} \frac{(2l_1+1)(2l_2+1)}{4\pi} \,
            \threej{l_1,0}{l_2,0}{l,0}^2 \, G_{l_1} G_{l_2}
    \\ & \quad
    + 4 \, \lambda\lambda' \, aa' \, G_l \,.
\end{split}
\end{equation}
Since the triangle condition~$l \le l_1 + l_2$ is imposed on the sum by the
Wigner $3j$ symbols, it follows that~$C_l$ is indeed band-limited if~$G_l$ is
band-limited, at twice the angular mode number.  In this case, the non-linear
solver for~$G_l$ proposed by \citet{2023OJAp....6E..11T} can produce an
essentially exact transformation.

For our simulations, the variance of each random field, and hence~$a$, is
determined by its spectrum.  It remains to fix the value of~$\lambda$.  For
angular clustering, we simulate the density contrast~$\delta$, for which we
set~$\lambda = 1$.  For cosmic shear, we simulate the convergence
field~$\kappa$, for which we set~$\lambda$ using the fitting formula
of~\citet{2011A&A...536A..85H}.

\end{appendix}

\label{LastPage}
\end{document}